\documentclass[preprint,12pt]{elsarticle}

\usepackage{amssymb}
\usepackage{bbding}
\usepackage{amsmath}
\usepackage{tikz}
\usetikzlibrary{arrows.meta, positioning, shapes.geometric}
\usepackage{graphicx}
\usepackage{dcolumn}
\usepackage{bm}
\usepackage{xcolor}
\usepackage{enumitem}
\usepackage{booktabs}
\usepackage{soul}
\usepackage{makecell}
\usepackage{float}
\usepackage{hyperref}

\usepackage{dutchcal}

\usepackage{graphicx}
\usepackage{xcolor}
\usepackage{geometry}
\usepackage{tikz}
\usetikzlibrary{arrows.meta, positioning, shapes.geometric}
\usepackage{tikz}
\usetikzlibrary{shapes.geometric, arrows.meta, positioning}
\usepackage{booktabs}

\tikzstyle{startstop} = [rectangle, rounded corners, minimum width=3.5cm, minimum height=1.2cm,align=center, draw=black, fill=blue!10]
\tikzstyle{process} = [rectangle, minimum width=3.5cm, minimum height=1.2cm, text centered, draw=black, fill=orange!10]
\tikzstyle{decision} = [diamond, draw, fill=blue!10,
    text width=5em, align=center, inner sep=1pt]
\tikzstyle{arrow} = [thick,->,>=stealth]

\newcommand{\diff}{\mathop{}\!\mathrm{d}}
\newcommand{\ddt}[1]{\frac{\diff#1}{\diff x^0}}

\definecolor{og}{RGB}{34, 139, 34}

\def\l{\left}
\def\r{\right}

\DeclareMathOperator{\arctanh}{arctanh}

\def\ampt{{\sc a\kern-.05em m\kern-.05em p\kern-.05em t\kern-.05em}}

\def\music{{\sc m\kern-.05em u\kern-.05em s\kern-.05em i\kern-.05em c\kern-.05em}}

\def\vhlle{v{\sc h\kern-.05em l\kern-.05em l\kern-.05em e\kern-.05em}}

\def\vishnu{{\sc v\kern-.05em i\kern-.05em s\kern-.05em h\kern-.05em n\kern-.05em u\kern-.05em}}

\def\clvisc{{\sc c\kern-.05em l\kern-.05em v\kern-.05em i\kern-.05em s\kern-.05em c\kern-.05em}}

\def\ekrt{{\sc e\kern-.05em k\kern-.05em r\kern-.05em t\kern-.05em}}

\def\mcdipper{{\sc m\kern-.05em c\kern-.05em d\kern-.05em i\kern-.05em p\kern-.05em p\kern-.05em e\kern-.05em r\kern-.05em}}

\def\kompost{{\sc k\kern-.05em ø\kern-.05em m\kern-.05em p\kern-.05em ø\kern-.05em s\kern-.05em t\kern-.05em}}

\def\ipglasma{{\sc i\kern-.05em p\kern-.05em -g\kern-.05em l\kern-.05em a\kern-.05em s\kern-.05em m\kern-.05em a\kern-.05em}}

\def\mcglauber{{\sc m\kern-.05em c\kern-.05em g\kern-.05em l\kern-.05em a\kern-.05em u\kern-.05em b\kern-.05em e\kern-.05em r\kern-.05em}}

\def\mckln{{\sc m\kern-.05em c\kern-.05em k\kern-.05em l\kern-.05em n\kern-.05em}}

\def\nexspherio{{\sc n\kern-.05em e\kern-.05em x\kern-.05em s\kern-.05em p\kern-.05em h\kern-.05em e\kern-.05em r\kern-.05em i\kern-.05em o\kern-.05em}}

\def\vusphydro{{\sc v\kern-.05em -U\kern-.05em S\kern-.05em P\kern-.05em h\kern-.05em y\kern-.05em d\kern-.05em r\kern-.05em o\kern-.05em}}

\def\NuclearConfectionery{{\sc N\kern-.05em u\kern-.05em c\kern-.05em l\kern-.05em e\kern-.05em a\kern-.05em r\kern-.05em C\kern-.05em o\kern-.05em n\kern-.05em f\kern-.05em e\kern-.05em c\kern-.05em t\kern-.05em i\kern-.05em o\kern-.05em n\kern-.05em e\kern-.05em r\kern-.05em y\kern-.05em}}
\def\iccing{{\sc i\kern-.05em c\kern-.05em c\kern-.05em i\kern-.05em n\kern-.05em g\kern-.05em}}
\def\ccake{{\sc c\kern-.05em c\kern-.05em a\kern-.05em k\kern-.05em e\kern-.05em}}
\def\frosting{{\sc f\kern-.05em r\kern-.05em o\kern-.05em s\kern-.05em t\kern-.05em i\kern-.05em n\kern-.05em g\kern-.05em}}
\def\melt{{\sc m\kern-.05em e\kern-.05em l\kern-.05em t\kern-.05em}}
\def\smsh{{\sc s\kern-.05em m\kern-.05em a\kern-.05em s\kern-.05em h\kern-.05em}}
\def\trento{{\sc t\kern-.05em \lower.4ex\hbox{r}\kern-.025em e\kern-.05em n\kern-.05em t\kern-.09em}o}

\newcommand{\kokkos}{\texttt{Kokkos}}
\newcommand{\cabana}{\texttt{Cabana}}

\journal{Computer Physics Communications}

\begin{document}

\begin{frontmatter}



\title{
\texorpdfstring{\NuclearConfectionery{}}{NuclearConfectionery}:\\ Multi-stage Simulation Framework for Modeling\\Relativistic Heavy-ion Collisions} 



\author[1,2]{Kevin P. Pala}

\author[2]{Surkhab Kaur Virk}
\author[2,3]{Dekrayat Almaalol}
\author[2]{Isabella Danhoni}
\author[2]{Nanxi Yao}
\author[2]{Isaac Long}
\author[2]{Willian Serenone}
\author[2]{Jordi Salinas San Mart\'in}
\author[4]{Alayna A. Yared}
\author[4]{Christopher Plumberg}
\author[5]{Fernando Gardim}
\author[2]{Jacquelyn Noronha-Hostler}

\address[1]{Instituto de F\'{i}sica, Universidade de S\~ao Paulo, Rua do Mat\~ao 1371, 05508-090 S\~ao Paulo-SP, Brazil}
\address[2]{The Grainger College of Engineering, Illinois Center for Advanced Studies of the Universe, Department of Physics, University of Illinois at Urbana-Champaign, Urbana, IL 61801, USA}
\address[3]{Center for Nuclear Theory, Department of Physics and Astronomy, Stony Brook University, Stony Brook, NY 11794, USA}
\address[4]{Natural Science Division, Pepperdine University, Malibu, CA 90263, USA}
\address[5]{Instituto de Ci\^encia e Tecnologia, Universidade Federal de Alfenas, 37715-400 Po\c cos de Caldas, MG, Brazil}

\date{\today}


\begin{abstract}
\noindent
We present the \NuclearConfectionery{}, a modular framework for simulating the full dynamical evolution of relativistic heavy-ion collisions. Its core hydrodynamic module, \ccake{}~2.0, represents a major advance over previous SPH-based relativistic hydrodynamic codes. \ccake{}~2.0 simultaneously evolves energy–momentum and multiple conserved charges (B, S, Q) with a four-dimensional equation of state, and can be run in either Cartesian or hyperbolic coordinates, enabling consistent simulations from the RHIC Beam Energy Scan to LHC energies. We have implemented a particlization module that supports global BSQ charge conservation on the freeze-out surface; the resulting hadron ensemble is then propagated through a hadronic transport afterburner.  A source term is included on the equations of motion to couple jets to the fluid, allowing simultaneous bulk and hard-probe evolution or, alternatively, for stopped baryons at low beam energies. The framework offers flexible choices of equations of motion (Israel--Stewart, DNMR, ADNH) and transport coefficients, along with GPU-ready performance via \kokkos/\cabana, offline equation of state inversion for 4D tables, and containerized portability. We validate the code with semi-analytical benchmarks (including BSQ Gubser and Landau--Khalatnikov solutions) and extensive convergence studies. The \NuclearConfectionery{} provides a user-friendly, high-performance, open-source tool for event-by-event simulations across collision energies, offering flexibility to study QCD matter at both vanishing and finite densities.

\end{abstract}


\end{frontmatter}


\date{\today}

\newpage 


\begin{center}
{\bf PROGRAM SUMMARY}
\end{center}

\begin{small}
\noindent
{\em Manuscript Title: }\NuclearConfectionery{}: Multi-stage Simulation Framework for Modeling\\Relativistic Heavy-ion Collisions \\
{\em Authors:} Kevin P. Pala, Surkhab Kaur Virk, Dekrayat Almaalol, Isabella Danhoni, Nanxi Yao, Isaac Long, Willian Serenone, Jordi Salinas San Mart\'in, Alayna A. Yared, Christopher Plumberg, Fernando Gardim, Jacquelyn Noronha-Hostler\\
{\em Program Title:}  \NuclearConfectionery{}\\
{\em Journal Reference:}                                                \\ 
{\em Catalogue identifier:}                                             \\ 
{\em Licensing provisions:} none                                        \\ 
{\em Programming language:} \texttt{C++}, \texttt{python}, \texttt{Bash}, \texttt{SQLite}  \\
{\em Computer:} Laptop, desktop, cluster                                \\ 
{\em Operating system:} Tested on GNU/Linux Ubuntu. To operate on macOS systems with Apple Silicon (M-series) processors, a Linux virtual machine (e.g., Multipass) is required due to incompatibilities between the available Clang compiler and the current version of the code.                           \\
{\em RAM: } Depending on the resolution, centrality, and neighbor count, typical (3+1)D SPH simulations require $\sim$ 1–48 GB of RAM. High-resolution (3+1)D Au+Au runs with tens of millions of particles sit at the upper end, while small (2+1)D Pb+Pb tests use $\sim$ 1 GB.                       \\ 
{\em Number of processors used:} Serial (CPU) and Parallel (CPU \& GPU)  
{\em External routines/libraries:} GNU Scientific Library (GSL), HDF5, Numpy, Eigen, YAML\\
{\em Nature of problem:}
  Heavy-ion collisions  necessitate efficient, user-friendly dynamical simulations that can be run across a range of beam energies in order to extract information from experimental data. The systems are very small, relativistic, short lived, begin far-from-equilibrium, and expand freely in a vacuum.\\
{\em Solution method:}
Use smoothed particle hydrodynamics to describe the fluid dynamic equations of motion. Here we include shear and bulk viscosity and a full diffusion matrix (necessary for heavy-ion collisions). Adopt a generalized coordinate system that employs hyperbolic coordinates at high beam energies and Cartesian coordinates at low beam energies. A source term is included in the fluid dynamic simulations to include jets for high energies and dynamical initialization for low energies. The code is written for both CPUs and GPUS with YAML files and Docker containerization. Causality constraints are incorporated as well.  \\
{\em Restrictions:} none\\ 
{\em Unusual features:} none \\ 
{\em Additional comments:} none\\ 
{\em Running time:} See Tabs.\ \ref{table:runtimes}-\ref{table:runtimesRHIC} where estimated run times come from a single central event.  We note that since central events have the longest run times with the most particles, these should all be considered upper bounds. 

\begin{table}[ht!]
\centering
\begin{tabular}{|c|c|c|c|}
\hline
Configuration & \makecell{O+O\\0--5\%}&  \makecell{Xe+Xe\\0--5\%}  &  \makecell{Pb+Pb\\0--5\%}  \\ 
\hline
initial state (\trento{})  & $<$ 1 s & $<$ 1 s & $<$ 1 s  \\
(1 event, $\Delta x=\Delta y=0.06$ fm) & & & \\ \hline
initial state (\iccing{})  & 15 s & $\sim$ 50 s & $\sim$ 60 s  \\
(1 event) & & & \\ \hline
(2+1)D hydrodynamics (\ccake{}) & 0.05 h & 0.4 h & 0.3 h   \\
(1 event,  $h=0.3$ fm) & & & \\ \hline
Cooper--Frye freeze-out sampler & 4 s &25 s & 39 s\\ 
$N_{\left\{s\right\}}$ samples:  & 500 & 100 & 50 \\ \hline
hadronic transport (\smsh{}) & 84 s & 3.5 h & 5.3 h \\  \hline
Total (1 event) &  0.08h & 3.92 h & 5.62 h \\  \hline
\end{tabular}
\caption{Summary of runtimes in the \NuclearConfectionery{} package for varying colliding systems. Collisions are simulated at $\sqrt{s_\mathrm{NN}} = 5.02$ TeV for all system configurations considered.  Here we use $N_{\left\{s\right\}}=500$ samples of the freeze-out hypersurface per event where each sample is a unique run in \smsh{}.  The times shown for freeze-out and hadronic transport are both calculated for all samples.}
\label{table:runtimes}
\end{table}

\begin{table}[ht!]
\centering
\begin{tabular}{|c|c|c|c|c|}
\hline
\makecell{Configuration}  &
\makecell{Au+Au\\$200$ GeV} &
\makecell{Au+Au\\$19.6$ GeV} &
\makecell{Au+Au\\$19.6$ GeV}&
\makecell{Au+Au\\$19.6$ GeV} \\
\hline
{initial state (\ampt{})}  & 
$\mathcal{O}(10^2)$ s & 
$\mathcal{O}(10^2)$ s& 
$\mathcal{O}(10^2)$ s& 
$\mathcal{O}(10^2)$ s\\
(1 event, $\Delta x=\Delta y=0.1$ fm) &  $\Delta \eta=0.15$ & $\Delta \eta=0.15$ & $\Delta \eta=0.3$ & $\Delta \eta=0.3$\\ \hline
(3+1)D  (\ccake{}) & 19 h  & 9 h & 3 h & 1.8 h \\
(1 event,  $h=0.3$ fm) & (online EoS) & (online EoS) & (online EoS) & (offline EoS) \\ \hline
Freeze-out sampler & $832$ s & $230$ s & $240$ s  & $240$ s \\ 
$N_{\left\{s\right\}}$ samples: & 500 & 500 & 500 & 500\\ \hline
hadronic transport (\smsh{}) & 
 3 h& 20 min  & 20 min & 20 min \\  
\hline
Total (1 event) & 22.25 h & 9.4 h & 3.5 h & 2 h \\  \hline
\end{tabular}%
\caption{Summary of runtimes in the \NuclearConfectionery{} package for varying beam energies and $\Delta\eta$. Collisions are simulated using hyperbolic coordinates with finite BSQ charges (no diffusion) for all beam energies considered. We also compare \ccake{} simulations with the online vs. offline EoS inverter. Here we use $N_{\left\{s\right\}}=500$ samples of the freeze-out hypersurface per event where each sample is a unique run in \smsh{}.  The times shown for freeze-out and hadronic transport are both calculated for all samples. 
}
\label{table:runtimesRHIC}
\end{table}

\end{small}

\newpage


\newpage
\tableofcontents

\newpage

\section{Introduction}
\noindent
The study of relativistic heavy-ion collisions provides unique insight into the properties of strongly interacting matter under extreme conditions in a far-from-equilibrium environment. 
Quantum Chromodynamics (QCD) is the quantum field theory that describes the strong force and it can be used to calculate QCD's equilibrium properties like the equation of state (EoS) at vanishing net-baryon densities $n_B$ and determine its phases of matter.  
Due to limitations of our approach to calculate QCD on the lattice, we cannot use first-principle methods to solve for QCD in dynamical systems \cite{Troyer:2004ge}, especially when the system is far from equilibrium. 
Heavy-ion collisions are  far-from-equilibrium, dynamical systems   that cannot be used as a mechanism to directly connect the QCD EoS to experimental data. 
Rather, one relies on multi-stage frameworks \cite{Petersen:2008dd,Heinz:2013th,Goes-Hirayama:2025nls} that are centered around relativistic viscous hydrodynamics where the QCD EoS is an input, but then this framework makes the direct connection to experimental data \cite{Alba:2017hhe,Monnai:2021kgu,Spieles:2020zaa}. 

A challenge for these frameworks is to be flexible enough to tackle the wide range of phenomena that appear within heavy-ion collisions, since the dynamics of the system vary strongly with the center-of-mass beam energy, $\sqrt{s_{NN}}$. At the highest center-of-mass energies $\sqrt{s_{NN}} \sim 0.2$--$6$ TeV reached at the Large Hadron Collider (LHC) and the sPHENIX experiment \cite{sPHENIX:2017lqb} at the Relativistic Heavy-Ion Collider (RHIC), where the net-baryon density is close to zero ($n_B \sim 0$), QCD predicts a transition from a Quark-Gluon Plasma (QGP) to a Hadron Resonance Gas (HRG). Lattice QCD calculations indicate that, in this regime, the transition is a smooth crossover occurring at a temperature of approximately $T_\mathrm{had} \sim 155$ MeV \cite{Aoki:2006we,Borsanyi:2010cj,Borsanyi:2013bia,HotQCD:2014kol}. Probes such as high-momentum jets, which are sensitive to the QGP at very small distance scales \cite{Connors:2017ptx,Cunqueiro:2021wls,Cao:2020wlm}, or extremely small colliding systems (e.g., Oxygen–Oxygen) \cite{Citron:2018lsq}, provide valuable insight into the far-from-equilibrium behavior of QCD.

At lower collision energies $\sqrt{s_{NN}}\sim 2$--$200$ GeV, significantly higher net-baryon densities are reached. In this domain, the QCD phase structure is expected to differ markedly: theoretical considerations suggest the existence of a critical endpoint \cite{Stephanov:1998dy} and a first-order phase transition line at finite $n_B$, although their exact locations remain unknown \cite{STAR:2020tga,Dexheimer:2020zzs,Lovato:2022vgq,Sorensen:2023zkk,MUSES:2023hyz}. Ongoing experimental efforts, including the RHIC Beam Energy Scan (BES) program \cite{STAR:2020tga,STAR:2025zdq}, NA61/SHINE at the SPS \cite{NA61SHINE:2017fne}, and the future CBM experiment at FAIR \cite{CBM:2016kpk}, are actively
exploring these regions of the phase diagram.

Because quarks and hadrons carry multiple conserved charges, i.e., baryon number (B), strangeness (S), and electric charge (Q), any calculations at finite $n_B$ necessarily involve the inclusion of a 4D EoS due to local fluctuations of strangeness number density $n_S$ and electric charge $n_Q$ \cite{Plumberg:2024leb}. 
For a 4D EoS the inversion problem exists because the EoS is normally calculated in terms of chemical potentials $\vec{\mu}=\left\{\mu_B,\mu_S,\mu_Q\right\}$ such that the pressure is provided in terms of $p(T,\vec{\mu})=p(T,\mu_B,\mu_S,\mu_Q)$, but the natural hydrodynamic variables are in terms of entropy density $s$ and densities, i.e., $p(s,\vec{n})=p(s,n_B,n_S,n_Q)$.
Furthermore, the inclusion of a 4D EoS also implies that an entire BSQ diffusion matrix is required to describe how charges diffuse through the QGP \cite{Greif:2017byw,Fotakis:2019nbq,Fotakis:2021diq,Fotakis:2022usk,Fotakis:2024hmz} and that at the point of particlization (when the fluid is converted into hadrons) one must conserve BSQ globally for a given event  (some also enforce local conservation \cite{Oliinychenko:2019zfk}). Since these improvements significantly slow down dynamical frameworks, it is necessary to implement optimizations, including flexible coordinate systems, offline EoS inverters, and extensive GPU parallelization.

Calculations at high $\sqrt{s_{NN}}$ have their own challenges, especially once one considers small system sizes. At high $\sqrt{s_{NN}}$, jets are produced more often and can provide insight into the microscopic interactions of the QGP \cite{Connors:2017ptx,Cunqueiro:2021wls,Cao:2020wlm}. Jets do not follow hydrodynamic modes but rather their dynamics are described by some external module to hydrodynamics (but they rely the local hydrodynamic variables as they pass through-- and deposit energy onto the fluid).   Ideally, one has jets fully-coupled to the system wherein they can ``dump'' energy into the hydrodynamic framework through a source term, but also the hydrodynamic background influences the dynamics of the jet itself \cite{Tachibana:2017syd}. Jet calculations require 3+1D simulations as well and can be enormously computationally expensive in order to obtain reasonable statistics. Furthermore, especially in small systems, there is the possibility that fluid dynamics may break down, leading  to acausal or unstable fluid cells \cite{Bemfica:2020xym,Plumberg:2021bme,Chiu:2021muk}. 
Thus, they also benefit from the same optimization and parallelization mentioned above. 

Significant advances have been made within heavy-ion collisions to dynamical frameworks to explore a range of beam energies. However, most of these advances have been made within grid-based codes (e.g., \music{} \cite{Denicol:2018wdp}, \vhlle{} \cite{Karpenko:2013wva}, \vishnu{} \cite{Du:2019obx}, \clvisc{} \cite{Wu:2021fjf}). 
One can also solve the relativistic viscous hydrodynamic equations using a Lagrangian method known as Smoothed Particle Hydrodynamics (SPH) that comes originally from the astrophysical community \cite{Monaghan:1992rr,Rosswog:2009sr}. The SPH approach offers several advantages for relativistic heavy-ion collision simulations. Since the equations of motion are solved only in regions where the fluid is present, the method efficiently adapts to the rapidly expanding and dilute medium, with the smoothing length $h$ setting the smallest resolved scale. By following fluid elements in space and time, SPH provides a Lagrangian formulation in which time derivatives are treated exactly through the $F=MA$ structure of the equations of motion \cite{Denicol:2008rj,Denicol:2009yyt}, thus avoiding common instabilities of grid-based schemes \cite{Gavassino:2025bsn}. Furthermore, because each SPH particle carries energy-momentum and all conserved charges, SPH ensures exact local conservation of baryon number, strangeness, and electric charge in the absence of diffusion, while at the same time facilitating the implementation of realistic multi-dimensional equations of state and a natural treatment of large event-by-event fluctuations in the initial conditions.

The first SPH code used within heavy-ion collisions was an ideal 3+1D code called \nexspherio{} \cite{Hama:2004rr}. Later on, a 2+1D code that incorporated shear and bulk viscosity was developed called \vusphydro{} \cite{Noronha-Hostler:2013gga,Noronha-Hostler:2014dqa}.  The \vusphydro{} code was a precursor to \ccake{}~1.0 \cite{Plumberg:2024leb}, which extended the framework to include BSQ conserved charges with a 4D EoS and also included several numerical improvements and optimizations. 
However, frameworks that cover the entire $\sqrt{s_{NN}}$ range possible for heavy-ion collisions are rare and must include multiple conserved charges, a 4D EoS, and dynamical jet-fluid coupling. These upgrades remain a significant challenge both in terms of numerical implementation but also in terms of slow-down of runtime.
Additionally, most numerical methods in the field have been developed for grid-based codes, often rendering the numerical algorithms available not portable to SPH codes and must be re-derived with SPH in mind. 
Thus, dynamical frameworks also necessitate optimization as well as heavy parallelization. 

Thus, we present \ccake{}~2.0, the next generation of the SPH-based relativistic viscous hydrodynamics code \vusphydro{}. \ccake{}~2.0 introduces several key advances:
\begin{itemize}
    \item Simultaneous evolution of energy--momentum and conserved charge densities ($n_B$, $n_S$, $n_Q$) in full 3+1D with a four-dimensional equation of state.
    \item Flexibility to run in either Cartesian or hyperbolic coordinates, enabling applications from low-energy RHIC BES collisions to top-energy LHC events.
    \item Dynamical source term framework for coupling jets and baryon stopping to the hydrodynamic evolution.
    \item Multiple formulations of viscous hydrodynamics (Israel--Stewart, DNMR, ADNH) with consistent transport coefficients.
    \item GPU-ready performance through \kokkos/\cabana, with an offline EoS inversion scheme that allows efficient use of multidimensional EoS tables.
    \item Global BSQ conservation of particle sampling, coupled to a hadronic  afterburner \smsh{} ~\cite{SMASH:2016zqf}.
    \item Built-in functionality to calculate causality and stability, eccentricities, and perform a variety of semi-analytical checks. Data analysis scripts are included to calculate a variety of experimental observables with statistical error bars. 
\end{itemize}

These developments directly address long-standing challenges in modeling relativistic heavy-ion collisions across a broad range of energies and conditions. In addition to describing the theoretical and numerical implementation, we validate \ccake{}~2.0 against semi-analytic solutions (e.g., BSQ generalizations of Gubser and Landau--Khalatnikov flows) and perform extensive convergence tests. Together with its integration into the \NuclearConfectionery{} framework, \ccake{}~2.0 provides a modular, portable, and high-performance platform for precision studies of QCD matter at both vanishing and finite densities.

This paper is organized around the various stages of a hybrid approach to simulate heavy-ion collisions and which comprise the \NuclearConfectionery{} framework. In Sec.~\ref{Sec:Observables} we discuss the event-by-event fluctuations and experimental observables that can be studied within this framework.
In Sec.~\ref{Sec:FrameworkOverview}, we provide a more detailed overview of the framework, its structure, and the challenges it is intended to address.
In Sec.~\ref{Sec:InitialState}, we discuss the initial state.
In Sec.~\ref{Sec:Hydrodynamics}, we discuss the hydrodynamics.
In Sec.~\ref{Sec:Particlization}, we discuss the particlization.
Lastly, in Sec.~\ref{Sec:Conclusions} we present our conclusions and outlook.

\subsection*{Notation}
\begin{itemize}
    \item \textbf{Metric and units.--} 
        We adopt the mostly-minus metric signature $\{+,-,-,-\}$ and natural units $\hbar = c = 1$.
    \item \textbf{Derivatives.--} 
        The covariant derivative of a vector $A^\nu$ is defined as
        $
            \nabla_\mu A^\nu = \partial_\mu A^\nu + \Gamma^{\nu}_{\mu\lambda} A^\lambda,
        $
        where $\Gamma^{\nu}_{\mu\lambda}$ are the Christoffel symbols of the second kind,
        \begin{equation*}
           \Gamma^{\nu}_{\mu\lambda} = \tfrac{1}{2} g^{\nu\rho}
            \left( 
                \partial_\mu g_{\rho\lambda} + 
                \partial_\lambda g_{\rho\mu} - 
                \partial_\rho g_{\mu\lambda}
            \right). 
        \end{equation*}
        The comoving (or convective) derivative is defined as
            $D \equiv u^\mu \nabla_\mu.$
        The time component of the covariant derivative can be expanded explicitly in terms of partial derivatives and the Christoffel symbols as
          $  \nabla_0 A^\mu 
            = \frac{{\rm d}A^\mu}{{\rm d}x^0} 
              - v^i \partial_i A^\mu 
              + \Gamma^\mu_{0\sigma} A^\sigma,$
        where $v^i = u^i / u^0$ are the spatial components of the fluid velocity.
        To calculate partial derivatives in time, one can write the comoving derivative as $  D = u^0 \partial_0 + u^i \partial_i 
              = u^0 \frac{{\rm d}}{{\rm d}x^0},$
        where ${\rm d}/{\rm d}x^0$ denotes the total derivative with respect to coordinate time. Then,
        $\partial_0 = \frac{{\rm d}}{{\rm d}x^0} + v^i \partial_i.$
    \item \textbf{Hydrodynamic derivatives.--} 
        The spatial projector operator is defined as
            $\Delta_{\alpha\beta} = g_{\alpha\beta} + u_{\alpha} u_{\beta},$
        and the spatially projected covariant derivative as
            $\nabla^{\langle\mu\rangle} = \Delta^{\mu\nu} \nabla_\nu.$
        The expansion rate is defined as
            $\theta = \nabla_\mu u^\mu,$
        and the velocity-shear tensor as
        $\sigma^{\mu\nu} = \Delta^{\mu\nu\alpha\beta} \nabla_{\alpha} u_{\beta},$
        where the symmetric, traceless projection operator is
           $ \Delta^{\mu\nu\alpha\beta} 
            \equiv 
            \tfrac{1}{2}
            \left(
                \Delta^{\mu\alpha} \Delta^{\nu\beta} 
                + \Delta^{\mu\beta} \Delta^{\nu\alpha}
            \right)
            - \tfrac{1}{3}\,
            \Delta^{\mu\nu} \Delta^{\alpha\beta}.$
\end{itemize}

\section{Event-by-event fluctuations and experimental observables}%
\label{Sec:Observables}
\noindent
We quantify the flow harmonics in a single event through flow vectors $V_n=v_ne^{-in\Psi_n}$ where $v_n$ is the magnitude of the Fourier harmonics and $\Psi_n$ is the event-plane angle (the orientation of the geometrical shape) where $n=2$ is for elliptical flow and $n=3$ is triangular flow. 
Because the orientation of $\Psi_n$ is random, we construct observables where it drops out. 
For instance, a two-particle correlation can be constructed via
\begin{equation}
    V_nV_n^* = v_n^2\left\{\cos \left[n\left(\Psi_n-\Psi_n\right)\right]+i\sin \left[n\left(\Psi_n-\Psi_n\right)\right]\right\}=v_n^2,
\end{equation}
where the observable is rotationally invariant. 
One can construct observables that look at correlations between event planes of different orders \cite{ATLAS:2014ndd}, but these still do not depend on the orientation of a single $\Psi_n$, just how they relate to each other. 

To make direct comparisons with experimental data, simulations run an initial state $i$ through the entire heavy-ion framework from start to finish to obtain a contribution to observable $\mathcal{O}_i$ with a corresponding multiplicity $M_i$. 
Each run from start to finish is called an ``event'' and is analogous to a single collision in an experiment. 
However, in the current experimental set-up there is too much error in reconstructing observables to compare a single event to a single event in a simulation. 
Thus, to reduce error, observables are averaged over many events in the following manner. 
First, one sorts events into centrality bins such as $1\%$ bins. In experiments this is based on multiplicity in theoretical frameworks this is typical done by the initial total entropy $\mathcal{S}_0$ of the event (before running hydrodynamics). 
Then, one runs millions of initial conditions, sorting by $\mathcal{S}_0$ to determine the bounds of the centrality classes---essentially a dictionary between ranges of $\mathcal{S}_0$ and centrality $\%$ is determined. 
Actual simulations, though, are run with significantly fewer events (in the order of tens of thousands) because of their long computational times (minutes to hours), which are then sorted into centrality classes based on the aforementioned dictionary.

Next, one gathers all the events in a given centrality bin of, say  $1\%$ bins. 
The exact method of calculating the observable in a given bin is
\begin{equation}
    \langle \mathcal{O}\rangle =\frac{\sum_i^{N_{\left\{\mathrm{ev}\right\}}}w_i \mathcal{O}_i}{\sum_i^{N_{\left\{\mathrm{ev}\right\}}}w_i},
\end{equation}
where $w_i$ is some weight for event $i$. 
For most observables, the multiplicity $M_i$ is used as the weight. 
However, because observables must be constructed to be rotationally invariant (in other words, the value of $\Psi_n$ should be irrelevant), then one generally correlates even numbers of particles.\footnote{For a better understanding of the flow observables and details of constructing them, see \cite{Bilandzic:2010jr,Luzum:2013yya,Schukraft:2012ah}} 
To avoid double counting particles, the weights for a 2-particle correlation are
\begin{equation}
    w_i\left\{2\right\}=M_i(M_i-1),
\end{equation}
where the curly brackets indicate the number of particles correlated. 
For a 4-particle correlation, then
\begin{equation}
    w_i\left\{4\right\}=M_i(M_i-1)(M_i-2)(M_i-3).
\end{equation}
This technique of weighting the observables is known as ``multiplicity weighing'' and is not used universally. 
It is generally preferred by the LHC experiments but not always used at RHIC (where they choose instead $w_i=1$). 
However, it can make a difference when calculating the observables themselves, see e.g. \cite{Gardim:2016nrr}, such that one must be aware of exactly how each experimental collaboration calculates their observables. 

After obtaining an observable calculated in $1\%$ bins (or some other small bin width, but we will use $1\%$ widths as the example here), we then recombine them into larger bins using the following method. 
One uses the weighing averaging technique again,
\begin{equation}\label{eqn:observable_weighted}
    \langle \mathcal{O}_\mathrm{comb}\rangle =\frac{\sum_j^\mathrm{width} w_j \mathcal{O}_j}{\sum_j^\mathrm{width}w_j},
\end{equation}
where the widths of the bins are often widths of 5\%, 10\%, or 20\%, depending on the available statistics for a given observable. 
The effects of multiplicity weighing and centrality re-binning are generally negligible for 2-particle correlations but can have up to a 5\% effect for 4-particle correlations \cite{Betz:2016ayq}. 

Because these observables are calculated with a finite number of statistics, i.e., $N_{\left\{\mathrm{ev}\right\}}<\infty$, we need to calculate the statistical error on the observable.  
Let us define the average of an observable over all events, removing a single event $j$, as $\langle \hat{\mathcal{O}}\rangle_j $, i.e., 
\begin{equation}
    \langle \hat{\mathcal{O}}\rangle_j =\frac{\sum_{i\neq j}^{N_{\left\{\mathrm{ev}\right\}}}w_i \mathcal{O}_i}{\sum_{i\neq j}^{N_{\left\{\mathrm{ev}\right\}}}w_i}.
\end{equation}
Then, the average across events $\langle \hat{\mathcal{O}}\rangle$ is
\begin{equation}
    \langle \hat{\mathcal{O}}\rangle=\frac{1}{N_{\left\{\mathrm{ev}\right\}}}\sum_{i}^{N_{\left\{\mathrm{ev}\right\}}}\langle\hat{\mathcal{O}}\rangle_i,
\end{equation}
from which we can now calculate the jackknife variance,
\begin{equation}
    \sigma^2(\langle {\mathcal{O}}\rangle)=\frac{N_{\left\{\mathrm{e}v\right\}}-1}{N_{\left\{\mathrm{ev}\right\}}}\sum_{j=1}^{N_{\left\{\mathrm{ev}\right\}}}\left( \langle \hat{\mathcal{O}}\rangle_j- \langle \hat{\mathcal{O}}\rangle\right)^2,
\end{equation}
where the one-standard deviation error is then the square root of this. Here we use jackknife to calculate an \emph{entire observable} since that avoids issues with not properly handling correlated error. For instance, if the observable is a fraction of $f=A/B$, it is ill-advised to calculate the error on A and B separately and then use error propagation, because the correlated error will then overestimate the error bars. Instead, we calculate the error on the entire fraction at once. 
To see this more clearly, the correct method for error propagation of this function is
\begin{equation}
    \sigma_f\sim |f|\sqrt{\left(\frac{\sigma_A}{A}\right)^2+\left(\frac{\sigma_B}{B}\right)^2-2\left(\frac{\sigma_{AB}}{AB}\right)^2},
\end{equation}
and since the correlated error $-2\left(\sigma_{AB}/AB\right)^2$ is not accounted for in standard error propagation, not including this term overestimates the error. 

We point out the methodology for calculating observables because these are inherently Big Data quantities that require high statistics.
To obtain these statistics a large number of initial conditions must be run within simulations to generate the final output contribution from a given event. 
This final output from a given event is then averaged over to calculate these global quantities. 
For this procedure to be possible, the simulation framework must be very efficient and user friendly. 
In the following, we discuss how we calculate specific observables. 

\subsection{Multiplicity}%
\label{Sec:Observables:Multiplicity}
\noindent
Experiments measure charged particles in momentum space using the coordinates: transverse momentum $p_T$, transverse angle $\phi$, and (pseudo)-rapidity $y$ ($\eta$). Here we will just always assume $p_T,\phi,\eta$ (but it is straightforward to convert between $y$ and $\eta$, if needed). These charged particles are sometimes labeled as certain ground stable particles, e.g., pions or even lighter resonances like $\phi$-mesons. When the particle type is identified we refer to them as PID (identified particle). 

The easiest observable to calculate in theoretical simulations is always the multiplicity, which we use as a catch-all for simply counting up either i.) all the particles produced in a single event, ii.) all the particles within some kinematic cuts, iii.) all the particles of certain type, e.g., pions (with or without some kinematic cuts),  iv.) all the particles carrying certain conserved charges, e.g., the total number of electrically charged hadrons is approximately the sum of pions, kaons, and protons. One can also look at more differential quantities such as the spectra $dN/(2\pi p_T)dp_Tdy$ that is typically the $p_T$-dependence of the multiplicity at mid-rapidity or $dN/dy$ that is the rapidity dependence of the multiplicity. 

For a single event $\left\{\mathrm{ev}\right\}$, the multiplicity $M_{\left\{\mathrm{ev}\right\}}$ is defined as
\begin{equation}
    M_i=\frac{1}{N_{\left\{s\right\}}}\sum_{\left\{s\right\}}^{N_{\left\{s\right\}}}\sum_{P_{\left\{s\right\}}=\pi,K,p,\dots} P_{\left\{s\right\}}\Theta\left(p_T^\mathrm{min}\leq p_T(P_{\left\{s\right\}})\leq p_T^\mathrm{max}\right)\Theta\left(\eta^\mathrm{min}\leq \eta(P_{\left\{s\right\}})\leq \eta^\mathrm{max}\right),
\end{equation}
where $N_{\left\{s\right\}}$ is the number of samples for a given event $\left\{\mathrm{ev}\right\}$ (see later Sec.~\ref{Sec:Particlization}). 
Here the Heaviside $\Theta$ functions  are defined with the chosen kinematic cuts and $P_{\left\{s\right\}}$ may be summed over all charged particles or a subset. Additionally, $p_T(P_{\left\{s\right\}})$ is the transverse momentum of the measured particle $P_{\left\{s\right\}}$ and $\eta(P_{\left\{s\right\}})$ is its corresponding pseudo-rapidity. In experiments, multiplicity is determined through histogramming each event. 
The total multiplicity is calculated slightly differently than what was shown in Eq.~\ref{eqn:observable_weighted} because the multiplicity weight would just cancel. Thus, the total multiplicity (regardless if it is PID, includes kinemetic cuts, etc.) is calculated as a straightforward average via
\begin{equation}
    \langle N\rangle = \frac{1}{N_{\left\{\mathrm{ev}\right\}}} \sum_{\left\{\mathrm{ev}\right\}}^{N_{\left\{\mathrm{ev}\right\}}} M_{\left\{\mathrm{ev}\right\}}.
\end{equation}

\subsection{Flow}%
\label{Sec:Observables:Flow}
\noindent
One can construct a large number of flow observables that are rotationally invariant and relevant for extracting information from heavy-ion collision data. 
Let us begin with multi-particle cumulants for integrated flow (within in some kinematic range of $p_T,\eta$). 
To preserve rotational invariance, we always correlate even number of particles.  Let us first define moments of the flow distribution:
\begin{equation}
   \langle v_n^m\rangle \equiv  \langle \left(V_n V_n^*\right)^{m/2}\rangle,
\end{equation}
where $m=2,4,6,\dots$ is the number of particles we are correlating. 
However, experimentalists do not measure the moments, but rather the cumulants such that the relationship between moments and cumulants up to 6-th order is
\begin{eqnarray}
    v_n\left\{2\right\}&=&\sqrt{\langle v_n^2\rangle},\\
    v_n\left\{4\right\}&=&\sqrt[4]{2\langle v_n^2\rangle^2-\langle v_n^4\rangle},\\
    v_n\left\{6\right\}&=&\sqrt[6]{\langle v_n^6\rangle-9\langle v_n^4\rangle\langle v_n^2\rangle+12 \langle v_n^2\rangle^3},\\
    v_n\left\{8\right\}&=&\sqrt[8]{\langle v_n^8\rangle-16\langle v_n^6\rangle\langle v_n^2\rangle-18 \langle v_n^4\rangle^2-144 \langle v_n^2\rangle^4},
\end{eqnarray}
where higher-order cumulants can be calculated as well, but they require enormous statistics. 
The averaging in the moments always includes multiplicity weighing, which is why multiplicity weighing plays a larger role in high-order cumulants. 
At the moment, we have not yet included the 6 and 8 particle cumulants but we plan to incorporate them in a future work. 

Up until now, we have considered only an integrated flow vector that does not depend on $p_T,\eta$. However, it is possible to define a differential flow vector $V_n(p_T,\eta,P)$ where $P$ is the particle type. Given that flow observables are always correlating at least 2 particles, we call the particle that is part of the differential flow vector a particle of interest, or POI. Then, POI are often correlated with reference particles, i.e., $V_n$ that is taken from the integrated, all charged particle flow.
For example, to build a 2-particle correlation for $p_T$-dependent flow (we could have chosen $\eta$ or $P$ analogously) we arrive at
\begin{equation}
    \langle V_n(p_T)V_n^*\rangle =\langle v_n(p_T)v_n\cos n\left(\psi_n-\psi_n(p_T)\right)\rangle.
\end{equation}
However, in this case we also choose to normalize with $v_n\left\{2\right\}$ to cancel the magnitude of the integrated flow harmonics, i.e., 
\begin{equation}\label{eqn:differential_flow}
    v_n(p_T)\left\{2\right\}=\frac{\langle v_n(p_T)v_n\cos n\left(\psi_n-\psi_n(p_T)\right)\rangle }{v_n\left\{2\right\}},
\end{equation}
and we can generically write the same formula for any POI ($p_T,\eta,P$),
\begin{equation}
    v_n^{\rm 1POI}\left\{2\right\}=\frac{\langle v_n^{\rm 1POI}v_n\cos n\left(\psi_n-\psi_n^{\rm 1POI}\right)\rangle }{v_n\left\{2\right\}}.
\end{equation}
It is also possible to correlate two particles of interest, such that the cosine phase drops and one obtains
\begin{equation}
    v_n^{\rm 2POI}\left\{2\right\}=\sqrt{\langle v_n^{2}\rangle^{\rm 2POI}}.
\end{equation}

One can also correlate 4+ particles where one then has to choose to correlate 1POI, 3 reference; 2 POI, 2 reference; 3POI, 1 reference; or 4POI. We will not work out all these possibilities here but rather point an interested reader to \cite{Moravcova:2020wnf,Holtermann:2023vwr,Nielsen:2023znu,Holtermann:2024vdw,Gardim:2024nyz,Nielsen:2025pkz} 
One can also build other type of flow cumulants \cite{Bilandzic:2010jr,Heinz:2013th} such as symmetric cumulants, event plane correlations, event-shape engineering \cite{Schukraft:2012ah}, and $p_T$-flow harmonic correlations. 

\subsection{Jet observables}%
\label{Sec:Observables:Jets}
\noindent
Hard scatterings between either quarks or gluons in the initial state lead to back-to-back jet formation.  Over time these jets interact through the medium via gluon-radiation or scatterings, which leads to the jets losing energy.  During this process the jets fragment (split apart), such that the jets are no longer point like particles but rather many particles clumped together in a jet cone with some radius $R_\mathrm{jet}$.
Eventually, they decouple from the medium at some point and stop interacting (although they may still fragment outside of the medium).
How jets interact with the QGP (in the medium) vs. outside the QGP (in the vacuum) are not the same.  
One way of seeing this is to compare the number of jets in Pb+Pb (ion-ion collisions or $A$+$A$) to the number of jets measured in proton-proton (or $pp$) collisions. 
We can think of an $A$+$A$ collision as containing $N_\mathrm{coll}$ number of $pp$ collision (otherwise known as the number of binary collisions). Central $A$+$A$ collisions have a larger $N_\mathrm{coll}$ compare to peripheral collisions where $N_\mathrm{coll}$ is much smaller. 
Note that $N_\mathrm{coll}$ scales with the number of participating nucleons (as in ones that collide at least once) but it can be larger than that because some nucleons undergo multiple collisions. 

Thus, it is useful to define the nuclear modification factor, $R_{AA}(p_T)$, that determines how many final state high  $p_T$ particles appear in $A$+$A$ vs. $pp$ collisions, i.e.,
\begin{equation}
    R_{AA}(p_T)=\frac{d^2N/dp_Td\eta}{N_\mathrm{coll}\,d^2N/dp_Td\eta},
\end{equation}
where we normalize the denominator by $N_\mathrm{coll}$. The way this ratio is defined, we can now quantitatively determine if $A$+$A$ collisions are just $pp$ collisions scaled up by $N_\mathrm{coll}$.  If $R_{AA}(p_T)=1$ there is no difference between $A$+$A$ and $N_\mathrm{coll}$ number $pp$ collisions, whereas if $R_{AA}(p_T)<1$ then this implies that jets lose energy in the dense medium\footnote{One can also consider $R_{AA}(p_T)>1$ that may occur at certain $p_T$ ranges, which imply that $A$+$A$ collisions have more particles at specific $p_T$. However, this possibility is only relevant for very small systems and/or extremely large $p_T$ beyond the regime that it is relevant for the fluid-medium interactions that we focus on here.}. 

Jets are relatively rare probes of the QGP.  At high $\sqrt{s_{NN}}$ we may at most have a handful of jet pairs in a given event and at low $\sqrt{s_{NN}}$ the high-$p_T$ end of the spectra falls off dramatically such that either jets are no longer produced or they are produced at such low $p_T$ that they lose all or most of their energy in the medium. 
Thus, in experimental measurements $R_{AA}$ is always measured as a global observable, averaged over many events. However, it can be measured in specific centrality classes or by separating out leading vs. sub-leading jets \cite{ATLAS:2022zbu}. 
Early jet measurements typically focused on just all charged (single) particle $R_{AA}$ \cite{ATLAS:2015qmb,CMS:2016xef,ALICE:2018vuu} whereas more recent jet measurements with the high-statistics era of LHC \cite{ATLAS:2018gwx,CMS:2021vui,ALICE:2023waz} and the new sPHENIX detector now allow for jet $R_{AA}$ measurements using jet-finding algorithms.
Many jet observables can be constructed \cite{Connors:2017ptx,Cunqueiro:2021wls} but we focus here on the $R_{AA}$ and differential flow. 

One can also calculate azimuthal anisotropies of high-$p_T$ all charged particles or jets.  Because jets are rare probes, these measurements are typically done with just 1POI found in the jet or using the highest-$p_T$ charged hadron, but it has been proposed for 2+ POI of measurements to be made in \cite{Holtermann:2023vwr,Holtermann:2024vdw} that could lead to further insights into jet fluctuations.
Azimuthal anisotropies of jets are always soft-hard correlations and follow the same formula as was already shown in Eq.~(\ref{eqn:differential_flow}), see \cite{Noronha-Hostler:2016eow,Betz:2016ayq} for more details. 

In theoretical calculations, we do not have the same statistics as experiments due to how computationally expensive simulations of jet-medium interactions are. 
Instead, we typically use a trick to obtain high-statistic observables without running millions nor billions of events, as is what is measured in the experiments. 
We run on the order of hundreds to thousands of fluid dynamic simulations and then overlay $\mathcal{O}(10^5$--$10^6)$ jet pairs on top.  
These jet pairs are not coupled to the medium and are completely independent, but they do use the information of local fluid cells to determine how much energy is lost for each jet. 
Thus, including $N_{\left\{\mathrm{jets}\right\}}=600,000$ jets on top of of $N_{\left\{\mathrm{ev}\right\}}=100$ events, is roughly equivalent (in terms of statistics) to
\begin{equation}
    \frac{N_{\left\{\mathrm{jets}\right\}}}{2}N_{\left\{\mathrm{ev}\right\}}\sim 3\cdot 10^7
\end{equation}
pairs of jets found in the experiments. 

Using such an approach, we can then reconstruct a differential $R^r_{AA,\left\{\mathrm{ev}\right\}}(p_T,\phi,\eta)$ where $r=q,g$ for quark or gluon within each event (similar to how we calculate the spectra). 
We emphasize here that this $R^r_{AA,\left\{\mathrm{ev}\right\}}(p_T,\phi,\eta)$ is \emph{only} a theoretical quantity and can only be compared to experiments after averaging over many events. 
From this quantity, we can sum over all quark and gluon jets and include all possible fragmentation functions to calculate the $R_{AA,\left\{\mathrm{ev}\right\}}(p_T)$ of a specific PID,
\begin{equation}
    R_{AA,\left\{\mathrm{ev}\right\}}^\mathrm{PID}(p_T,\phi,\eta) = \frac{\sum_{r=q,g} \int_{z_\mathrm{min}}^{1} \frac{dz}{z} d\sigma_r (\frac{p_\mathrm{PID}}{z}) R_{AA}^{r}(\frac{p_\mathrm{PID}}{z}, \phi) D_{r\rightarrow \mathrm{PID}} (z, \frac{p_\pi}{z})}{\sum_{r=q,g} \int_{z_\mathrm{min}}^{1} \frac{dz}{z} d\sigma_r (\frac{p_\mathrm{PID}}{z}) D_{r\rightarrow \mathrm{PID}} (z, \frac{p_\pi}{z})}.
\end{equation}
Here the momentum fraction is $z$ where $z_\mathrm{min}=p_\mathrm{PID}/p_\mathrm{max}$, the fragmentation functions are $D_{r\rightarrow \mathrm{PID}} (z, \frac{p_\pi}{z})$, and $d\sigma_r (\frac{p_\mathrm{PID}}{z})$ are cross sections of your
specified parton type that can be obtained from PYTHIA.

We can now calculate our total nuclear modification factor that can be compared to experiment, i.e., 
\begin{equation}
     R_{AA,\left\{\mathrm{ev}\right\}}^\mathrm{PID}(p_T)= \frac{1}{2\pi}\int_0^{2\pi} \int_{\eta_\mathrm{min}}^{\eta_\mathrm{max}} d\phi d\eta \,R_{AA,\left\{\mathrm{ev}\right\}}^\mathrm{PID}(p_T,\phi,\eta),
\end{equation}
that must be summed over all events, i.e.,
\begin{equation}
     R_{AA}^\mathrm{PID}(p_T)= \sum_{\left\{\mathrm{ev}\right\}} R_{AA,\left\{\mathrm{ev}\right\}}^\mathrm{PID}(p_T).
\end{equation}
Furthermore, in a given event $\left\{\mathrm{ev}\right\}$, we can then define an event plane angle,
\begin{equation}\label{hardPlane}
    \psi_{n,\left\{\mathrm{ev}\right\}}^\mathrm{hard}(p_T)=\frac{1}{n}\arctan\left(\frac{\int_0^{2\pi}d\phi\sin(n\phi)R_{AA,\left\{\mathrm{ev}\right\}}(p_T, \phi)}{\int_0^{2\pi}d\phi\cos(n\phi)R_{AA,\left\{\mathrm{ev}\right\}}(p_T, \phi)}\right),
\end{equation}
which is only a theoretical quantity to lead to the hard sector $v_n$. These $v_n$ are easy to get from the reaction plane,
\begin{equation}\label{hardFlow}
    v_{n,\left\{\mathrm{ev}\right\}}^\mathrm{hard}(p_T)=\frac{\frac{1}{2\pi}\int_0^{2\pi}d\phi\cos[n\phi-n\psi_{n,\left\{\mathrm{ev}\right\}}^\mathrm{hard}(p_T)]R_{AA,\left\{\mathrm{ev}\right\}}(p_T, \phi)}{R_{AA,\left\{\mathrm{ev}\right\}}(p_T)}\, .
\end{equation}
To calculate the observable related to high-$p_T$ azimuthal anisotropies we then assume this hard flow is the POI such that $v_{n,\left\{\mathrm{ev}\right\}}^\mathrm{hard}(p_T)$ can be substituted into Eq.~\ref{eqn:differential_flow}.

\section{Framework overview}%
\label{Sec:FrameworkOverview}
\noindent
In the following, we provide an overview of the framework needed to simulate heavy-ion collisions. Many of these details are important to understand the computational challenges involved in heavy-ion collisions, the need for efficient codes, and the details involved in the choice of the coordinate system and the dimensionality.

Let us begin with a brief overview of heavy-ion collision experiments. In experiments, one takes a beam of atoms of some element and removes all the electrons (hence it becomes an ion) such that only the nuclei  remain in the beams. These nuclei then are accelerated to nearly the speed of light within a collider. At a specific point in the experiment, the beams are allowed to cross such that the nuclei collide. Every time a single nucleus-nucleus collision occurs it is called an event where each event is slightly different than the last because it depends both on quantum fluctuations within each nucleus as well as the impact parameter of the collision itself. Here, nuclei collide millions or billions of times, producing many events. After each collision, the remnants of that collision free stream into the detector where there are large magnetic fields that separate out the charged, ground-state\footnote{By the time particles are measured, the time scales are many orders of magnitude longer than QCD decay rates such that no hadron resonances (like the $\rho$ meson) remain. That being said, it is often possible to work backwards to a vertex to identify when a resonance previously existed. } particles that are later measured experimentally (the energy and momentum of these particles is measured). While each event is measured separately, the events are always grouped together and sorted in different manners, and then a final global observable (averaged over many events) is what is reported as the experimental data. 

In order to provide a detailed simulation of these collisions, we must consider all the stages sequentially in a \emph{simulation chain} until the final remnant is produced. However, we do not need to take into account the final free streaming/magnetic field effects because they are already accounted for within the experimental analysis and because particles in heavy-ion collisions reach a point of ``freeze-out'' at the very end of the collision where both particle yields and kinematics remain unchanged. Freeze-out occurs because the system is rapidly expanding and at some point it becomes so dilute that particles no longer interact and cannot change kinematics. We also need to perform ``event-by-event'' calculations, which implies that each single collision of nuclei has unique initial conditions that need to be propagated through the entire simulation chain and a final observable for that specific event is calculated. This process must be repeated many times and one averages the final observable over all events to obtain the global observable. 

\begin{figure}[ht!]
    \centering
    \includegraphics[width=\linewidth]{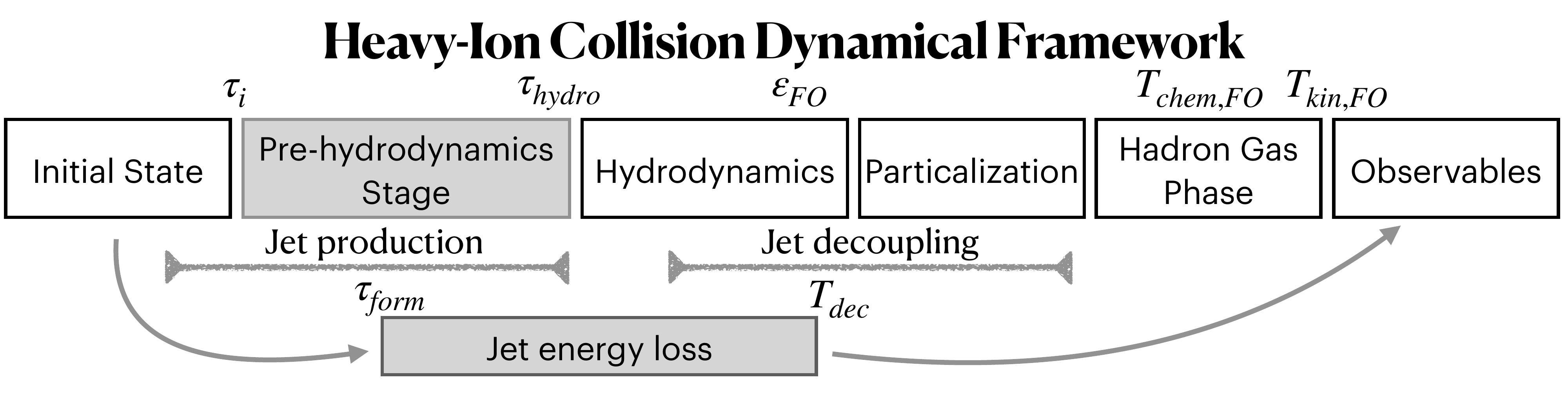}
    \caption{Heavy-ion collisions dynamical simulation framework. The boxes in white are modules that are always run whereas the boxes in gray are optional. Each initial state is an event that is indexed using $\left\{\mathrm{ev}\right\}$ and run through the entire simulation chain, where the final output is a set of observables $\vec{\mathcal{O}}_{\left\{\mathrm{ev}\right\}}$ for that particular event. The time scales, energy density, or temperatures that are relevant characteristic scales used to switch between modules are indicated above and explained in the text. Note that the particlization and observables modules have no internal dynamics such that they do not have a characteristic scale. Since this is only for a single event, the Analysis Script module is not shown.
    }
    \label{fig:HICframework}
\end{figure}

The necessary stages of a heavy-ion collision simulation (across any beam energy) are the initial stage, pre-hydrodynamic stage, hydrodynamics, particlization, and hadron gas phase, where a workflow for these stages is shown in Fig.~\ref{fig:HICframework}. These are the generic stages that must always be present, however, other codes and modules may still be necessary to make direct comparisons to data. For instance, when jets are present, they may be described by various types of energy loss pictures as they interact with the medium or split apart. Furthermore, at the end of the simulation of a given event one must first calculate the observables of a given event. After all events have completed, one then calculates global observables across all events---carefully taking into account statistical uncertainties, only then can one take these final global observables and compare directly to experimental data. 

\subsection{Modules within the Framework}%
\label{Sec:FrameworkOverview:PhysicsOverview}
\begin{itemize}
    \item 
    {\bf Initial State.--} The first stage corresponds to the construction of an \emph{initial state}, which represents the system when the nuclei first collide. 
    All initial state models generate a field of some thermodynamic quantity (i.e., energy density or entropy density) $\varepsilon(\vec{r})$, but certain models also generate the full energy-momentum tensor $T^{\mu\nu}_i$ (both in and out of equilibrium) where ``$i$'' indicates that this is the initial condition and possibly also the initial charge currents $N_{q,i}^\mu$ where $q=B,S,Q$.
    We define $\tau_i$ to be the proper time at which the initial state has been fully constructed, but we caution that some initial states are also provided in Cartesian coordinates at $t_i$ instead.
    \item 
    {\bf Pre-hydrodynamics Stage.--}  Following the initial state, there is often a \emph{pre-hydrodynamics stage}. Because the initial state may be so far from equilibrium that hydrodynamics would be acausal \cite{Plumberg:2021bme,Domingues:2024pom}, one can use a stage in between for a very short period in which the system approaches the applicability of hydrodynamics. This duration of the stage depends on the collision energy and the system size as well as the particular initial state, and pre-hydrodynamical model itself \cite{daSilva:2022xwu}. 
    We define the time scale for the end of the pre-equilibrium scale as $\tau_\mathrm{hydro}$.
    \item 
    {\bf Hydrodynamics.--}  Up next in the chain, the energy density field, charge density field, or the full $T^{\mu\nu}_i$ and $J^{\mu}_{q,i}$'s, are fed into \emph{relativistic viscous hydrodynamics}. For relativistic systems, the viscous equations of motion are not the same as those of Navier--Stokes written covariantly; it is known that relativistic extensions of Navier--Stokes lead to acausal behavior and instabilities \cite{Hiscock:1985zz}.  
    Rather, new viscous equations of motion were derived first in a phenomenological manner from the entropy current from Israel--Stewart \cite{Israel:1979wp}, later from the 14-moments method in kinetic theory (DNMR) \cite{Denicol:2012cn}, more recently using a generalized-frame in (BDNK) \cite{Bemfica:2017wps,Bemfica:2019knx,Kovtun:2019hdm}, and \cite{Almaalol:2022pjc}  recovered the 14-moments terms using the entropy current method (ADNH). 
    Here provide users with options to select either ideal, Israel--Stewart, DNMR, or ADNH equations of motion.  We do not consider BDNK because the equations of motion   would require second-order time derivatives, which we have not yet worked out in SPH. 
    \item
    {\bf Particlization.--} Generally,  the hydrodynamic stage is correlated with the Quark--Gluon Plasma phase (or in other words, it describes the dynamics of deconfined quarks and gluons). At some point these quarks and gluons join together into bound states---known as hadrons; the process of forming bound states is called \emph{hadronization}. 
    In simulations, we use Cooper--Frye \cite{Cooper:1974mv} to describe hadronization (or called \emph{particlization} since we generally switch to particles slightly after hadronization occurs) wherein contributions both from a thermal distribution and out-of-equilibrium corrections are used to determine the type of hadrons and their momentum that are produced. 
    We can define either a fixed temperature $T_\mathrm{had}$ or fixed energy density $\varepsilon_\mathrm{had}$ (more convenient at finite charge densities) at which the particles are converted into hadrons\footnote{In reality the transition from quarks into hadrons is a cross-over \cite{Aoki:2006we}, so it should be a range of temperatures to be completely accurate but there are challenges including this into the code at this time.}.
    One can either use a semi-analytical approach to obtain a continuous spectrum of hadrons or use Monte Carlo sampling to obtain the exact number of hadrons in the system while conserving BSQ charges exactly, we include both options in \ccake{}~2.0 (\ccake{}~1.0 only allowed for a continuous spectra of hadrons).
    \item
    {\bf Hadron Gas Phase.--} Once hadrons are formed they interact and decay, which is the \emph{hadron gas phase}. In simulations one can either include only direct decays (which can consider all orders of interactions as in $1\to2$ body decays but also $1\to3$ or even $1\to4$ decays) or one can use the Boltzmann equation to describe both decays and interactions but only $1\to2$ and $2\leftrightarrow2$ interactions are currently possible, if one ensures detailed balance. 
    The gas of hadrons is expanding over time such that eventually \emph{chemical freeze-out } is reached at some temperature $T_{\mathrm{chem},\,\mathrm{FO}}$ wherein the particles no longer undergo chemical changing processes, i.e., the particle yields are fixed. 
    However, for a time they can still undergo kinetic interactions that shuffle around the momenta of the particles. 
    Eventually, the gas becomes so dilute such that kinetic interactions also stop at $T_{\mathrm{kin},\,\mathrm{FO}}$, which is known as \emph{kinetic freeze-out}. 
    The simulations of this stage are called \emph{hadronic afterburners}, which cover all the necessary interactions and kinematics from the point of particlization until the final particle momenta are detected.
    \item
    {\bf Jet energy loss.--} During the early stages of the collision, hard scatterings may occur that produce back-to-back jets. The point where the jets begin interacting with the medium is the formation time $\tau_\mathrm{form}$.  These jets are bumped around by the medium (i.e., the relevant dynamical module, normally the hydrodynamic phase but it may be other modules as well) and lose energy over time. Eventually, the jets decouple from the medium at $T_\mathrm{dec}$ that most likely occurs before the system hadronizes because the jets are traveling much faster than the medium, however some groups also consider the possibility of jet-hadron interactions \cite{Dorau:2019ozd,Antiporda:2021hpk,McLaughlin:2021dlq,Bahder:2024jpa} (this is currently not available in our framework). 
    In principle, a user could couple their favorite energy loss model to the \NuclearConfectionery{} but at the moment we have implemented a parameterized energy loss model, roughly based on the  BBMG model from \cite{Betz:2014cza}. We allow for two options when it comes to jet energy loss 1.) back-to-back jets can be fully coupled to the system where they lose energy that is fed into \ccake{}~2.0 using source terms and that background also affects the jet energy loss, or 2.) the jets are only sensitive to the background of the medium (source terms are not used) but then one can oversample a single event with many jets to more efficiently calculate jet observables.
    \item 
    {\bf Observables.--} For each given event $\left\{\mathrm{ev}\right\}$ we must calculate its contribution to experimental observables.  There are many possibly experimental observables that one can calculate, but most fit into the category of 1.) yields or spectra, 2.) collective flow, or 3.) jet observables, which will be described in detail later on. Thus, for a given event $\left\{\mathrm{ev}\right\}$ we calculate an ensemble of its contribution to experimental observables $\vec{\mathcal{O}}_{\left\{\mathrm{ev}\right\}}$.
    \item
    {\bf Analysis Scripts.--} After many events have been run $N_{\left\{\mathrm{ev}\right\}}$, all of their resulting experimental observable contributions $\mathcal{O}_{\left\{\mathrm{ev}\right\}}$ are fed into a final analysis script. This script sorts events into centrality bins, takes global averages into account (considering multiplicity weighing), and calculates statistical uncertainties using jackknife error resampling.  The results from the \emph{Analysis Scripts} module is what can be compared directly to experimental data.
\end{itemize}

\subsection{Challenges within heavy-ion collision simulation frameworks}%
\noindent
Here we discuss the main challenges within heavy-ion collision simulation frameworks that we will be addressing in this paper.

\subsubsection*{Challenge \texorpdfstring{\#}{﹟}1: choice of coordinate system}%
\label{Sec:FrameworkOverview:PhysicsOverview:Challenge1}
\noindent
The nature of the system in heavy-ion collisions depends strongly on the colliding center-of-mass energy $\sqrt{s_{NN}}$ per nucleon.  In order to achieve the extremely high temperature $T$ needed to surpass $T>T_\mathrm{had}$, $\sqrt{s_{NN}}$ must be extremely large. 
However, $\sqrt{s_{NN}}$ is directly related to the speed of the incoming nuclei, such that at these high energies the nuclei experience large Lorentz
\begin{equation}
    \gamma=    \frac{1}{\sqrt{1-\left(\frac{v}{c}\right)^2}}
\end{equation}
factors (where $v$ is the velocity of the incoming nuclei and $c$ is the speed of light\footnote{Since $c=1$, we normally drop it from our equations but we point this out here to make it clear this quantity is dimensionless.}) through
\begin{equation}
    \gamma =\frac{\sqrt{s_{NN}}}{2m_p}.
\end{equation}
Then the radius of the nuclei along the direction of the beam $R_\mathrm{beam}$ is extremely Lorentz contracted,
\begin{equation}\label{eqn:Rbeam}
    R_\mathrm{beam}=\frac{2R_0m_p}{\sqrt{s_{NN}}},
\end{equation}
where $R_0$ is the standard radius for a given nucleus at rest. 
Using top energies at the LHC and collisions of heavy nuclei ($R_0\sim 6$ fm), the radius along the beam is just $R_\mathrm{beam}=2\cdot 10^{-3}$ fm, nearly 4 orders of magnitude smaller than along the transverse plane. 
Such a system can be well described in just the transverse plane (using hyperbolic coordinates that assume the systems moves at the speed of light along the beam direction and is invariant across rapidity). 
Most hydrodynamic simulations at high $\sqrt{s_{NN}}$ take advantage of this simpler system and use hyperbolic coordinates (see Sec.~\ref{Sec:InitialState:CoordinateSystem} on coordinate systems for further details) such that relativistic viscous hydrodynamics codes are normally run in 2+1 dimensions (2+1D). 

In contrast, at lower $\sqrt{s_{NN}}$ the nuclei become fully 3D such that 3+1D simulations are required. 
The lowest beam energy run in collider mode at RHIC is $\sqrt{s_{NN}}=7.7$ GeV, which leads to a $R_\mathrm{beam}=1.5$ fm which is the same order of magnitude as the transverse plane. 
In fact, even lower $\sqrt{s_{NN}}$ are possible in fixed-target colliders, although the applicability of hydrodynamics (a key component in our simulation framework) is questionable \cite{Inghirami:2022afu}.
At sufficiently low beam energies, fully 3+1D simulations in Cartesian coordinates become necessary, since the system is less strongly affected by Lorentz contraction. 

There are a few challenges that arise with the change of dimensionality as $\sqrt{s_{NN}}$ decreases: 1.) frameworks must be able to easily run both 2+1D and 3+1D simulations, 2.) preferably they also have the option of a generalized coordinate system such that Cartesian coordinates are possible at low beam energies, 3.) source terms allow one to switch on hydrodynamics earlier \cite{Shen:2017bsr} because the nuclei take longer to pass through each other \cite{Auvinen:2013sba}. 

\subsubsection*{Challenge \texorpdfstring{\#}{﹟}2: jet-medium coupling}%
\label{Sec:FrameworkOverview:PhysicsOverview:Challenge2}
\noindent
Another challenging feature of heavy-ion collisions at high $\sqrt{s_{NN}}$ is properly taking into account of jet-medium interactions in simulations.  
These jets with large transverse momentum are described by entirely different physics compared to the soft modes (at low $p_T$) within a fluid dynamics picture. 
Thus, one requires a source term within the fluid where the jets can dump energy into the fluid at each time step, but then their behavior is described by the energy loss picture of jets that  depend on the local fluid background (see e.g., \cite{Andrade:2013poa,Andrade:2014swa,Okai:2017ofp,Tachibana:2017syd,Chen:2017zte,Luo:2018pto}). 
In such a picture, the jets equations and the hydrodynamic equations must be solved simultaneously to accurate described how each affects the other. 
Furthermore, while soft physics at high $\sqrt{s_{NN}}$ can be reasonably well-described by 2+1D simulations, jets require 3+1D simulations since jets can be orientated in any direction.

Thus, the challenges involved in the proper description of jet-medium coupling are 1.) implementing source terms into the hydrodynamic simulations, 2.) translating the direction of the jets (often in Cartesian coordinates) into hyperbolic coordinates, 3.) interpolating over the fluid medium to obtain thermodynamic variables at a specific location, since the code is not grid-based, 4.) obtaining high-enough statistics given that jets are rare probes, such that back-to-back jets may produced once or only a several times in a single event, 4.) potential causality violations in fluid dynamic simulations when large amounts of energy are dumped within an event. 

\subsubsection*{Challenge \texorpdfstring{\#}{﹟}3: multi-dimensional equation of state}%
\label{Sec:FrameworkOverview:PhysicsOverview:Challenge3}
\noindent
The QCD EoS is inherently multidimensional because anytime a (anti-)quark is present, it carries multiple conserved charges. 
For instance, a strange quark (or hadron) in the fluid would carry not just baryon number (B), but strangeness (S) and electric charge (Q). 
These BSQ conserved charges require multiple changes within simulation frameworks. For instance, this implies that the EoS must vary not only in $T$ but also in all three chemical potentials $\left\{\mu_B,\mu_S,\mu_Q\right\}$ such that it is 4D \cite{Noronha-Hostler:2019ayj,Monnai:2024pvy} (and the EoS must be invertible since hydrodynamics runs in $s$ and $\vec{n}$, not in $T$ and $\vec{\mu}$). 
There are then two options: an ``online inverter'' that uses a 4D interpolation/root-finding algorithm on top of a $(T,\vec{\mu})$-EoS to find its location for the natural hydrodynamic variables while the hydrodynamic code is running or an ``offline inverter'' that inverts a 4D table once before the simulation framework begins such that the \ccake{} only needs to interpolate. 
Furthermore, the hydrodynamic equations of motion must include BSQ charge conservation at the minimum \cite{Karpenko:2013wva,Plumberg:2024leb} (in fact, BSQ diffusion may also be relevant but provides further challenges, see \cite{Greif:2017byw,Fotakis:2019nbq,Fotakis:2022usk}). 
These new equations of motion also have stability requirements that can be checked \cite{Almaalol:2022pjc} that include EoS and out-of-equilibrium contributions.
Finally, the conversion from a fluid into particles (freeze-out) is altered by the presence of these new conserved charges such that each hydrodynamic event must globally conserve BSQ charges with sampling over a hypersurface \cite{Oliinychenko:2019zfk,Oliinychenko:2020cmr} and that their contributions must be included in the distribution function \cite{Denicol:2018wdp,Plumberg:2024leb}. 
For specific combinations of $\left\{T,\mu_B,\mu_S,\mu_Q\right\}$ and particle mass at freeze-out it may not be possible to produce certain mesons (bosons) because the Bose-Einstein distribution would diverge (see discussion in \cite{Plumberg:2024leb}).

The challenges associated with the multidimensional EoS are then 1.) find an efficient and accurate method to invert a 4D EoS, 2.) a method to handle out of bound fluid cells, due to limitations in the reach of the 4D EoS, 3.) the implementation of BSQ conserved charges and diffusion within the equations of motion (there are different possible options for equations of motion so we allow the user to decide), 4.) exact BSQ global charge conservation when sampling particles, 5.) the inclusion of ``$\delta f$'' corrects for BSQ diffusion\footnote{As of writing this paper, we are only aware of the diagonal terms that have been derived in \cite{Denicol:2018wdp} at the point of particlization, off-diagonal terms have not yet been included.}, 6.) method for handling divergent meson contributions when sampling particles, 7.) revising analysis scripts to study identified particle observables à la \cite{Gardim:2024nyz,Sousa:2025oqf}.

\subsubsection*{Challenge \texorpdfstring{\#}{﹟}4: numerical efficiency and optimization}%
\label{Sec:FrameworkOverview:PhysicsOverview:Challenge4}
\noindent
The other three challenges identified above, necessitate significant computational improvements to \ccake{} in order to efficiently perform event-by-event simulations that will obtain high-enough statistics to reasonably compare to data. 
For instance, 3+1D simulations of \ccake{} on a single core could take up to 2 weeks, which is not remotely feasible if one requires tens of thousands of events per $\sqrt{s_{NN}}$. 
However, we also found that the most efficient method for running 2+1D event-by-event simulations is on a single core, but while running multiple events in parallel.
Thus, we require a code that can both run on CPUs and on GPUs (but on a single core and in parallel) for the most efficient use of our resources. 
The code should be easily portable on a variety of machines, which requires containerization so that it does not require a large number of dependencies to be installed when ran on a new machine. 
Finally, a motivation of \ccake{}~2.0 is to be eventually used to run in Bayesian analyses, which require that the free parameters can be easily identified, understood what their ranges and units are, and easily varied. 

The challenge then is to 1.) rewrite \ccake{} using \kokkos{}  and \cabana{} that allows for a simulation framework that can be run both on CPUs and GPUs, 2.) \kokkos{} and \cabana{} cannot easily handle our ``online inverter'' for the EoS such that when running in parallel or GPUs one must always use the ``offline inverter'',  3.) due to the large number of free parameters within the simulations, one wants to easily identify them in YAML files that contains their units, ranges, and allows them to be easily varied, and 4.) the code must be portable on a wide range of machines. 

\begin{figure*}[ht!]
\centering
\includegraphics[keepaspectratio, width=0.81\linewidth]{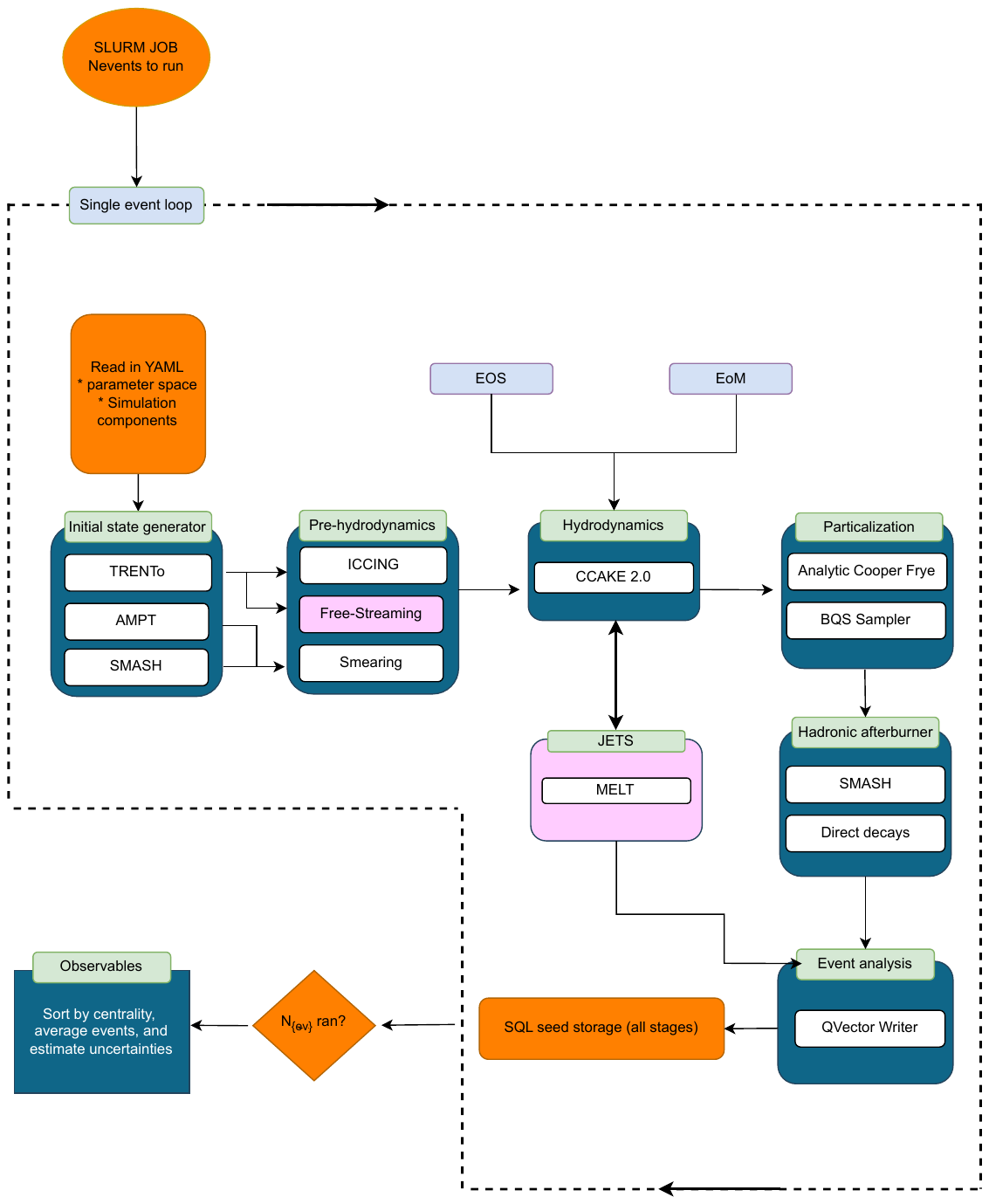}
\caption{Hybrid simulation framework: Initial Conditions, Pre-hydrodynamics 
(optional in pink), hydrodynamics (requires EoS and EoM), Particlization, 
Afterburner, and Event analysis. One option is to run Jets (in pink) coupled to 
the hydrodynamics phase that is then fed into Event Analysis afterwards. 
Example YAML files and the simulation chain can be found at 
\href{https://github.com/the-nuclear-confectionery/confectionery_chain}{\texttt{github.com/the-nuclear-confectionery/confectionery\_chain}}. }
    \label{fig:workflow}
\end{figure*}

\subsection{Framework structure}%
\label{Sec:FrameworkOverview:FrameworkStructure}
\noindent
In Fig.~\ref{fig:workflow} we demonstrate the possible workflows for the \NuclearConfectionery. The workflow begins by sending a job that contains $N_{\left\{\mathrm{ev}\right\}}$ the number of events to the cluster with a corresponding YAML file that contains all necessary input parameters.  For a given event $i$, the initial condition is produced from either \trento{}, \ampt{}, or \smsh{}. Then, an optional pre-equilibrium phase exists where a few choices exist: i.) \trento{} could be coupled to \iccing{} to initalize BSQ conserved charges or ii.) \trento{} could be couple to Free-stream to initialize a full $T^{\mu\nu}$ or iii.) \ampt{} or \smsh{} can be smeared out. The pre-equilibrium phase can also be skipped entirely and initial conditions can be fed directly into hydrodynamics. The hydrodynamics phase also requires an EoS and equations of motion (EoM).  The hydrodynamic phase has the option of being coupled to Jets through the \melt{} code. After hydrodynamics is complete, Particlization is run where either Analytical Cooper--Frye can be calculated (only possible for 2+1D simulations) or BQS Sampler can be run (possible for 2+1D or 3+1D). One particles are produced, a Hadronic afterburner is required. For the Analytical Cooper--Frye the only option is Direct decays. However, for the sampled particles from BQS Sampler \smsh{} must be run but it can either be run with Direct decays or with including the back-reactions. 
Finally, the results from the Hadronic afterburner and Jets (if run) feed into Event Analysis that calculates the set of all observables for this particular event $\vec{ \mathcal{O}}_i $. This process is repeated (either in parallel or in series, depending on the user's choice) for up to $N_{\left\{\mathrm{ev}\right\}}$.  Once all events are run then Observables calculates the global observables with statistical error bars that can be compared directly to experimental data. 

We focus most of our attention on the upgrades to the hydrodynamic evolution, which were sufficiently significant to warrant a new version to replace the original \ccake{}~1.0 \cite{Plumberg:2024leb}.
\ccake{}~2.0 includes the following upgrades: 
\begin{enumerate}
    \item Flexible dimensionality, allowing the code to be run in 1+1D, 2+1D, or 3+1D
    \item A generalized coordinate system which accommodates both hyperbolic and Cartesian coordinates using a mostly minus metric $\text{diag}(+,-,-,-)$ (although other metrics or coordinate systems may be easily added)
    \item Multiple available choices for equations of motion, including minimal Israel--Stewart (IS) \cite{Koide:2006ef}, Denicol-Niemi-Moln\'ar-Rischke (DNMR) \cite{Denicol:2012cn} that are derived from the 14 moments approach in kinetic theory, and (to appear soon) Almaalol-Dore-Noronha-Hostler (ADNH) \cite{Almaalol:2022pjc} that unifies Israel--Stewart and DNMR using the phenomenological entropy current derivation in a way which is consistent with the second law of thermodynamics. 
    \item Inclusion of source terms in the equations of motion which permit, e.g., jets which are fully coupled to the fluid \cite{Andrade:2014swa,Tachibana:2017syd,Okai:2017ofp,Chen:2017zte,Pang:2018zzo} or the dynamical inclusion of stopped baryons \cite{Denicol:2018wdp}
    \item A Module Energy Loss Treatment (melt{}) that is  jet energy loss code based on the Barbara Betz--Mikolos Gyulassy (BBMG) \cite{Betz:2014cza,Noronha-Hostler:2016eow} approach where the scaling of the path length dependence of the energy loss is parameterized and the equations are determined by dimensional analysis. The paper on melt{} is still forthcoming, but it will be available in this release. 
    \item Choice of either an offline or online inverter for a 4D EoS that allows one to map an EoS in $\left\{T,\mu_B,\mu_S,\mu_Q\right\}$ space (more natural for EoS calculations) into the natural hydrodynamic variables of entropy and number densities $\left\{s,n_B,n_S,n_Q\right\}$.
    \item Freeze-out and particlization which couple directly to our custom-made sampling routine for SPH, enabling BSQ charges to be exactly conserved globally on an event-by-event basis
    \item Options that include printing off the full $T^{\mu\nu}$ and $N_q^{\mu}$ at each time step, causality and stability constraints, internal eccentricities, energy and BSQ conservation checks, and comparisons to (semi-)analytical checks.
    \item Single event analysis module as well as a global analysis module that calculates statistical errors and sorts events into centralities. Both low and high-$p_T$ observables are included in the analyses. 
    \item Support for the \cabana{} library which ensures portability across CPUs / GPUs
    \item Convenient configuration and input using YAML files
    \item Containerization using Apptainer (just for \ccake{})
    \item Public availability of the open-source code on Github 
    \cite{ccakesite}
\end{enumerate}
%
\section{Initial State and Pre-equilibrium Stage}%
\label{Sec:InitialState}
\noindent
Here we introduce the possible options within the \NuclearConfectionery \, for initial state models and the pre-equilibrium stage. Since the pre-equilibrium stage is often closely tied to the choice in the initial state, we choose to combine their sections into one.

\subsection{Event-by-event fluctuations }
\label{Subsec:ebe}
\noindent
Every time heavy-ions collide in the laboratory, i.e., in each individual event, they produce a unique initial condition that depends on the impact parameter of the two nuclei, the positions of the nucleons (and possibly even the positions of the quarks and gluons) within the nucleus at the snapshot in time when the nuclei collide, and $\sqrt{s_{NN}}$. 
Because in the experiment, millions or even billions of events occur and they are averaged over to obtain a single experimental observable, in simulations we must run ``event-by-event'' simulations where $10^3$--$10^5$ different initial conditions are run through the entire  \NuclearConfectionery{} workflow, producing each a contribution to different observables from that event. 
At this point in time, the average shape (quantified by the eccentricity $\varepsilon_n$ where $n=2,3,\ldots$ indicates an elliptical shape, triangular shape, etc.) of an initial condition is fairly well-understood at high $\sqrt{s_{NN}}$ as well as its distribution on an event-by-event basis \cite{Gale:2012rq,Renk:2014jja,Moreland:2014oya,Niemi:2015qia,Bernhard:2016tnd,Giacalone:2017uqx}. 
Thus, the large scale of event-by-event initial conditions are fairly well constrained at this time. 
However, uncertainties still remain regarding a number of features of the initial state such as the nucleon width \cite{Giacalone:2017uqx,Giacalone:2021clp,Samanta:2023rbn,Mantysaari:2025ltqNOT}, the number of hotspots \cite{Moreland:2018gsh}, the relevance of out-of-equilibrium effects  in the pre-equilibrium stage  \cite{Schenke:2012wb,Kurkela:2018vqr,Kurkela:2018wud,daSilva:2022xwu,ExTrEMe:2023nhy}, and the contribution of quark/anti-quark pairs \cite{Martinez:2018ygo,Carzon:2019qja,Carzon:2023zfp,Gardim:2024nyz} to BSQ density fluctuations. 
Moreover, large uncertainties still remain at low $\sqrt{s_{NN}}$ for the initial that require 3+1D initial conditions and multiple conserved charges. 

For \ccake{}~2.0 we have tested three different types of initial conditions that all can be run on an event-by-event basis:
\begin{itemize}
    \item \trento{}: MC with nucleonic \cite{Moreland:2014oya} or both nucleonic and sub-nucleonic fluctuations \cite{Moreland:2018gsh}. Parametrized combination of thickness functions to allow maximal flexibility. Mostly in 2+1 but a 3+1 version is also available \cite{Ke:2016jrd}. 
     \item \trento{}+\iccing{}: If one assumes that the $\varepsilon(\vec{r})$ produced in \trento{} consists only of gluons, then \iccing{} \cite{Carzon:2019qja} samples the energy density profiles (in 2+1D only for the moment) and allows for gluons to split into quark/anti-quark pairs. Thus, the energy density profile then also includes BSQ density fields as well. In principle, this could eventually be coupled to kinetic theory as well, as done in \cite{Carzon:2023zfp} but we do not include that here. 
      \item \trento{}+Free-stream: One can also consider a pre-equilibrium of free-streaming \cite{Broniowski:2008qk} where the particles briefly expand outwards without interactions. This expansion period brings the system closer to a point where hydrodynamics is applicable (see e.g. \cite{Plumberg:2021bme}) but has challenges because it is currently only conformally invariant, which does not match to the lattice QCD equation of state \cite{daSilva:2022xwu}.
      \item \ampt{}: It is designed to perform the entire simulation \cite{AMPT,Lin:2001zk}, but it can also be used solely as an initial condition and pre-equilibrium stage by selecting time slices from the system evolution. In the early stages, \ampt{} generates color strings \cite{hijing} and then performs a partonic cascade simulation \cite{Zhang:1997ej}. At a chosen time, one can extract the partons and smear their four-momentum and charges to reconstruct a continuous $3+1$D energy-momentum tensor $T^{\mu\nu}$ and charge currents $N^\mu_q$ \cite{Fu:2020oxj}, 
      which are then decomposed and used as input for hydrodynamics. 
      Calculations are performed in Cartesian coordinates, and we have written a smearing code to convert the initial parton distribution to an initial condition in both hyperbolic and cartesian coordinates as well. 
      \item \smsh{}: Using a Boltzmann approach, \smsh{} \cite{SMASH:2016zqf} simulates hadronic interactions in a relativistic system. While this hadron transport code is often used for the hadronic afterburner, it can also be set-up to simulate the initial state \cite{Goes-Hirayama:2025nls}. When it generates an initial state, it can provide the full $T^{\mu\nu}$ and all charge currents. \smsh{} simulations are performed in Cartesian coordinates but can be converted to hyperbolic coordinates.
      At the moment, the \smsh{} initial state code is not yet public but we have ensured compatibility that we will explore more in a future work. 
\end{itemize}
There are other types of initial conditions that could easily be used as well in \ccake{} such as \ipglasma{} \cite{Schenke:2012wb}, \kompost{} (for pre-equilibrium) \cite{Kurkela:2018wud,Kurkela:2018vqr}, \mcdipper{} \cite{Garcia-Montero:2023gex}, \ekrt{} \cite{Niemi:2015qia}, or the string baryon stopping model \cite{Shen:2017bsr}. 
Here we just picked a handful to begin testing but nothing prevents further initial conditions as long as they were adapted to the \ccake{} format framework. 

Not all initial state frameworks can be run generically across all $\sqrt{s_{NN}}$. 
For instance, \trento{}, \iccing{}, \ekrt{}, \ipglasma{}, and \mcdipper{}, and \kompost{} are all primarily designed for high $\sqrt{s_{NN}}$ whereas \ampt{}, \smsh{}, and the baryon string model can be run at lower $\sqrt{s_{NN}}$ as well.  In the following sections we will discuss some of the further challenges that appear at the lower $\sqrt{s_{NN}}$ because of the dimensionality of the system.

\begin{table}[ht!]
\centering
\begin{tabular}{c|c|c|c|c|c}
\hline
Parameter & \makecell{O+O\\ $5.36$ TeV} & \makecell{Pb+Pb\\ $2.76$ TeV} & \makecell{Pb+Pb\\ $5.02$ TeV} & \makecell{Au+Au\\ $200$ GeV} & \makecell{Xe+Xe\\$5.44$ TeV} \\ \hline
Normalization $\varepsilon$ or $s$ & $\varepsilon$  & s & $\varepsilon$ & $\varepsilon$ & $\varepsilon$ \\
\trento  & $\times125$ & 82 & $\times125$& $\times 70$ & $\times145$\\
$\tau_i$ & $0.6$ fm & $0.6$ fm & $0.6$ fm & $0.6$ fm  &$0.6$ fm   \\
$\eta/s, \zeta/s, \kappa_{qq'}$ & $\{0.12,0,0\}$ & $\{0.1,0,0\}$ & $\{0.12,0,0\}$ & $\{0.16,0,0\}$ &$\{0.12,0,0\}$ \\ 
 \hline
\end{tabular}
\caption{Summary of the normalization constant, initialization times, and transport coefficients that have been tested against experimental data for different systems in \NuclearConfectionery.}
\label{table:normalization}
\end{table}

In Fig.~\ref{table:normalization} we include a summary of the normalization, initial times, and transport coefficients that we use in this paper to study different systems. In heavy-ion collisions, one uncertainty arises because a dimensionful normalization constant is required to reproduce experimental data, which is typically tuned to the multiplicity of particles in central collisions. What is shown in Fig.~\ref{table:normalization} are reasonable estimates of the normalization constants for those transport coefficients, but we do not tune them precisely to data (something we leave for a subsequent work). 

\subsection{Coordinate system}%
\label{Sec:InitialState:CoordinateSystem}
\noindent
The general difference between the high and low beam energy models is that at low $\sqrt{s_{NN}}$ the models are \emph{always} 3+1D and allow baryon stopping. 
We note that some of the high $\sqrt{s_{NN}}$ models also incorporate aspects of this physics. For example, \mcdipper{} includes baryon stopping, \iccing{}, includes BSQ charge fluctuations, and \ipglasma{} and \trento{}, which provide 3+1D initial conditions. However, these models still have other limitations and do not account for all the necessary low-$\sqrt{s_{NN}}$ physics.

Due to the difference in dimensionality at high versus low $\sqrt{s_{NN}}$ different coordinate systems are used. At low $\sqrt{s_{NN}}$, the nuclei are fully 3+1D such that Cartesian coordinates can best describe the system, that is, time with 3-spatial coordinates: $\left\{t,x,y,z\right\}$. 
However, at high $\sqrt{s_{NN}}$, due to the extreme Lorentz contraction of the nuclei, they are essentially flat pancakes that collide. 
Then, after the collision, the system expands at nearly the speed of light along the beam ($z$-axis), such that the fluid is nearly identical copies of itself along the beam axis, which is known as ``boost-invariant".
To take advantage of this feature of boost invariance and significantly simplify computations, we can use hyperbolic coordinates, which include the proper time $\tau$ and the space-time rapidity $\eta$ such that $\left\{\tau,x,y,\eta\right\}$ is the coordinate system. 
Then, the proper time can be related to Cartesian coordinates using: $\tau = \sqrt{t^2 - z^2}$ and the space-time rapidity is $\eta = \tanh^{-1}(z/t)$. 

\subsection{Dynamical initialization}%
\label{Sec:InitialState:DynamicalInitialization}
\noindent
As demonstrated in Eq.~(\ref{eqn:Rbeam}), the radius along the beam $R_0$ is directly connected to $\sqrt{s_{NN}}$ such that low $\sqrt{s_{NN}}$ implies the nuclei are fully 3D objects. 
Then, given the beam energy we can extract the velocity of the objects, i.e., 
\begin{equation}
    v=\sqrt{1-\l(\frac{2 m_p}{\sqrt{s_{NN}}}\r)^2}
\end{equation}
such that we can determine the time required for two nuclei to pass through each other,
\begin{eqnarray}
    t_\mathrm{pass}&=&\frac{2R_{beam}}{v}\\
    &=&\frac{2 R_0}{\sqrt{{\left(\frac{\sqrt{s_{NN}}}{2 m_p}\right)^2-1}}},
\end{eqnarray}
where we find at low $\sqrt{s_{NN}}$ that it can take on order of $\mathcal{O}(1)$ fm for the nuclei to reach complete overlap.

\begin{figure}[ht]
    \centering
    \includegraphics[width=0.8\linewidth]{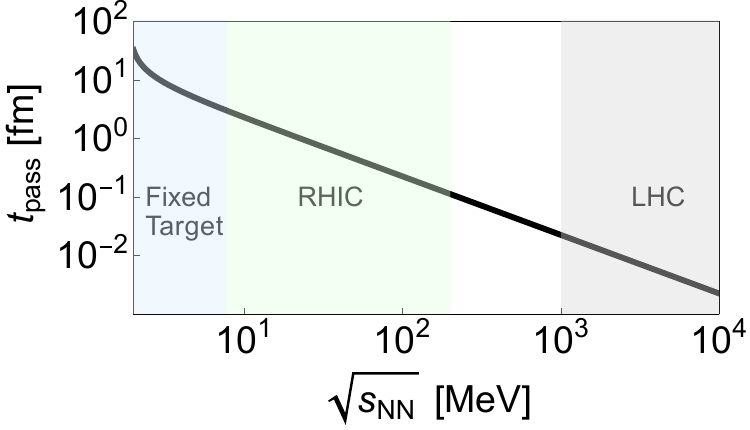}
    \caption{Passing time for two heavy nuclei with an estimated total radius of $R_0=6$ fm across different center-of-mass beam energies $\sqrt{s_{NN}}$. }
    \label{fig:timepass}
\end{figure}

Assuming a heavy nucleus radius of $R_0=6$ fm (which includes both the charge radius and any neutron skin), we can make time-scale estimates of the passing time. In Fig.~\ref{fig:timepass}, we plot the passing time vs. the center-of-mass beam energies for heavy nucleus. Around $\sqrt{s_{NN}}\sim 20$ GeV is where we switch to a passing time of $t_\mathrm{pass}>1$ fm. Lower beam energies, especially as one approaches the fixed-target regime, have a significantly slower passing times. 

Since the hydrodynamic phase lasts only about $10$ fm, waiting until the nuclei have passed through each other fully at low $\sqrt{s_{NN}}$ would leave only a very short window for hydrodynamic evolution. 
To overcome this challenge, groups have begun using dynamical initialization, in which the hydrodynamic phase is initialized before the nuclei have fully passed through each other, so that one can simultaneously solve the initial state and hydrodynamic phase together. 
Such an approach requires source terms within hydrodynamics to initialize the $T^{\mu\nu}$ and $N_q^\mu$ over time, see e.g. \cite{Denicol:2018wdp,Goes-Hirayama:2025nls}.  These sources terms feed into the hydrodynamics as the initial state model reaches a region where hydrodynamics is applicable (for more information, see Sec \ref{Sec:Hydrodynamics:Theory:EquationsOfMotion}).

In order to implement dynamical initialization in \ccake{}, we take advantage of the fact that both \ampt{} and \ccake{} can be run in Cartesian coordinates. In Cartesian coordinates, we start at $t_0=1$ fm, where not all the initial conditions are ready to run in \ccake{}. Thus, any remaining parts of the initial state are implemented as source terms in \ccake{}, which is finalized at $t_\mathrm{pass}=3.6$ fm for this beam energy. 
Then, for hyperbolic coordinates we use $\tau_0=1$ fm where once again not all the initial state is ready to for hydrodynamics such that we also use source term until $t_\mathrm{pass}=3.6$ fm is reached.  
For these runs, we find that we require a different normalization for Cartesian $\mathcal{N}_\mathrm{norm}^\mathrm{Cart.}= 0.42 $ vs. hyperbolic coordinates $\mathcal{N}_\mathrm{norm}^\mathrm{hyperb.}=0.62$. Here we ensure that we reproduce the STAR mid-rapidity data to determine the normalization. For both systems, we take the smeared \ampt{} initial condition at the initial time 
\begin{equation}
    x^0 = \tau_0 = t_0 = 1 \; \text{fm}.
\end{equation}
In this initial condition, all partons generated before the desired time are contributing to the system state. All the partons formed after the initial time are added as source terms during the hydrodynamic evolution. We do this by smearing the parton four momentum $p^\mu_j$ at each SPH particle position using the cubic spline kernel 
\begin{equation}
j^\mu(x_i^0, \vec{r}_i) = 
\frac{1}{\Delta x^0} 
\sum_{j=\mathrm{partons}}^{x^0 - \Delta x^0 < x_j^0 \le x^0}
\frac{\mathcal{N}_\mathrm{norm}}{\sqrt{-g}} \,
p_j^\mu \,
W(\vec{r}_i - \vec{r}_j, h_{\mathrm{source}}),
\end{equation}
where $\Delta x^0$ is the time step, $x^0_j$ is the parton formation time, and $r_j$ is the parton formation position. For the simulations in this section, we take $h_\mathrm{SPH} = 0.3$ fm, initial grid spacing ($0.1$, $0.1$, $0.1$) in fm and $h_\mathrm{source} = 1.4$ fm.

In some extremely rare cases, we obtain nucleons that appear in the initial state that are below the minimum energy condition even in the back-up EoS. In those cases we choose to freeze-out the nucleon instead of including it into our \ccake{} evolution.  However, we note that even for $\sqrt{s_{NN}}=7.7$ GeV we did not find any nucleons that this condition applied to.  

\begin{figure} [ht!]
    \centering
    \includegraphics[width=0.25\linewidth]{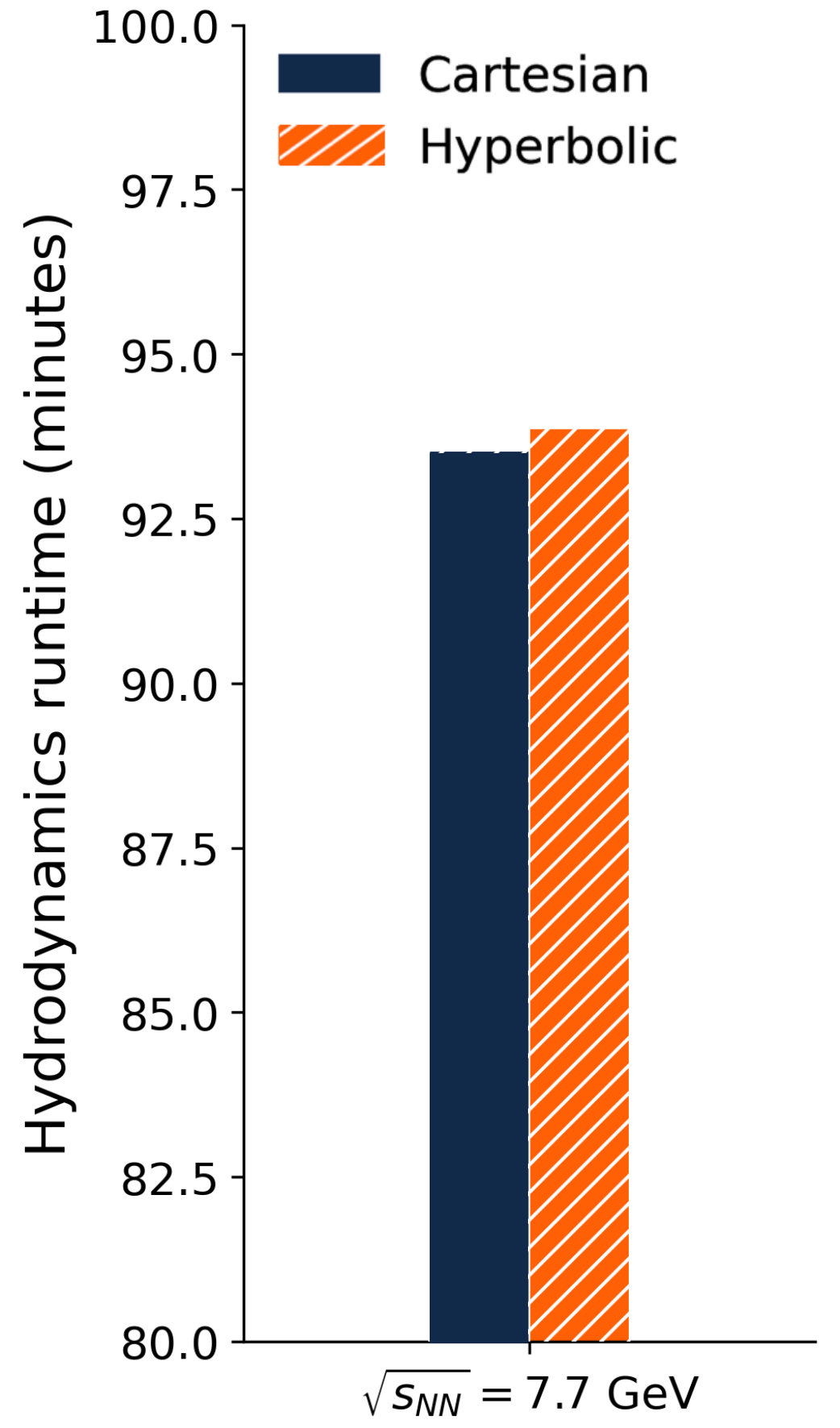}
    \caption{Illustration comparing \ccake{} runtime for the same \ampt{} initial condition. The \ampt{} event is initialized in a similar  way using either a hyperbolic (orange) or a Cartesian coordinate system (blue). }
    \label{fig:timecheckcoordinate}
\end{figure}

In Fig.~\ref{fig:timecheckcoordinate} we first compare the runtime checks for running at $\sqrt{s_{NN}}=7.7$ GeV. 
We compare two different scenarios here: a Cartesian coordinate system, which we expect to be best suited for low beam energies that are fully 3D, and a hyperbolic coordinate system, which we expected will be more appropriate for high beam energies that are more Lorentz contracted.
We find that there is a very slight advantage to run in Cartesian coordinates in terms of runtime, but the differences is very minimal. 

\begin{figure}[ht!]
     \includegraphics[width=0.99\linewidth]{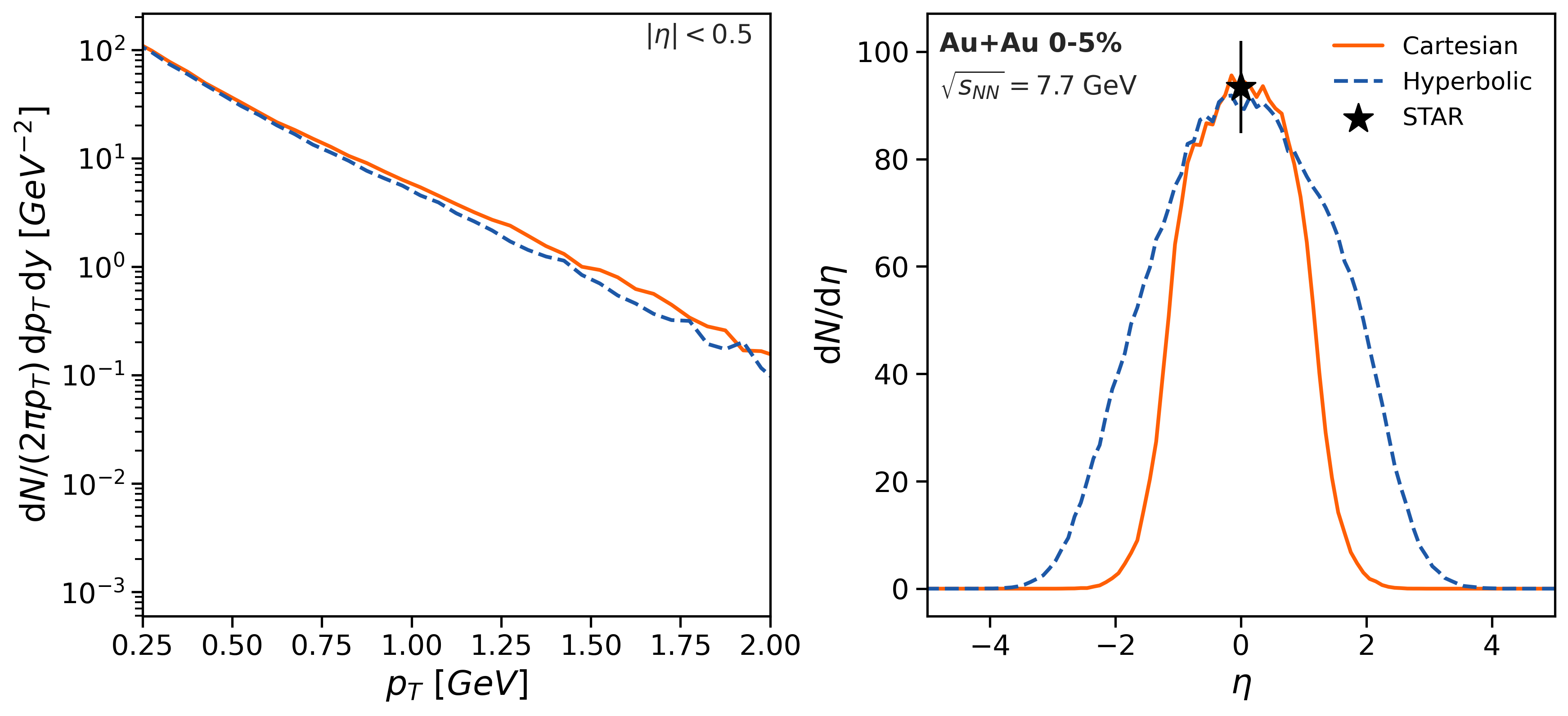}
    \caption{Comparing the observables for the aforementioned \ampt{} event, with the same initialization methods described before. Each initial energy density was normalized to match experimental data from star \cite{STAR:2017sal}.}
    \label{fig:coordinateobservables}
\end{figure}

One might at first assume that it does not matter significantly if one runs in hyperbolic vs. Cartesian coordinates. However, this is based on the assumption that these different choices lead to similar or identical results for experimental observables. 
To test the convergence of these two approaches, we have ran a single event at $\sqrt{s_{NN}}=7.7$ GeV. We then compared  the mid-rapidity spectra (left) and the multiplicity vs. pseudo-rapidity, i.e, $dN/dy$ (right) in Fig.~\ref{fig:coordinateobservables}.
For the multiplicity vs. rapidity results, we also compare our results to STAR data from \cite{STAR:2017sal}.
For the mid-rapidity spectra results, we observe that Cartesian simulation with $t_0=1$ fm vs. hyperbolic with $\tau_0=1$ fm, both followed by dynamical initialization, that the results are nearly identical across $p_T$. 

However, for a wider range of rapidities, differences appear for Cartesian vs. hyperbolic coordinates that can be seen on the right of Fig.~\ref{fig:coordinateobservables}. The Cartesian coordinates lead to a narrower distribution of $dN/dy$ in rapidity whereas the hyperbolic coordinates leads to a wider distribution. However, as expected from the $p_T$-spectra, the two approaches match at mid-rapidity (in some sense this is unsurprising since we tuned the normalization to reproduce the STAR data). 
Given that \ampt{} is calculated in Cartesian coordinates, going from Cartesian into Cartesian coordinates is the more natural approach and we consider those the ``correct'' result.  Thus, we find that hyperbolic coordinates artificially lead to a wider spectra across rapidity.

\begin{figure}[ht!]
    \centering
    \includegraphics[width=0.65\linewidth]{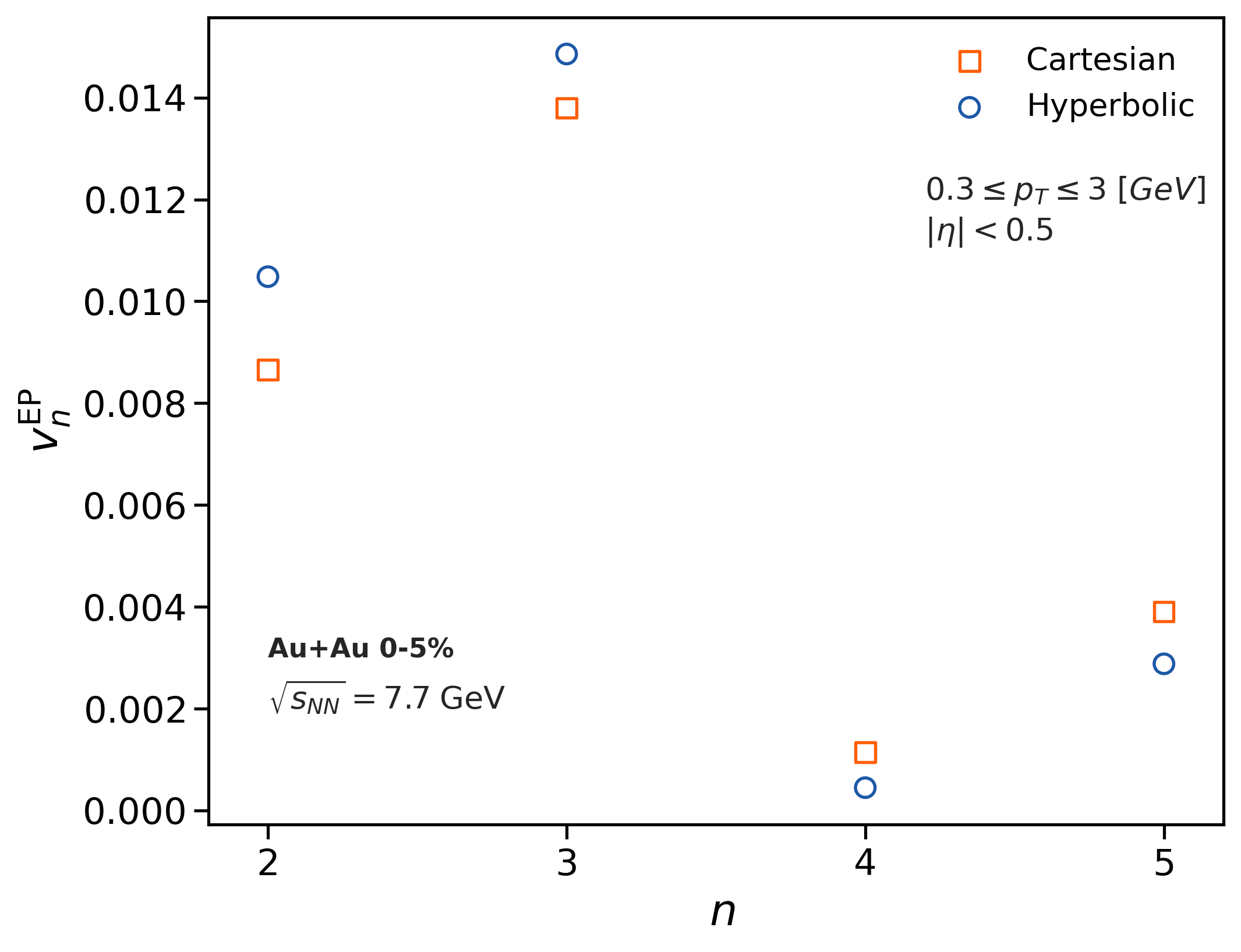}
    \caption{Integrated harmonic flow comparing the different coordinate system }
    \label{fig:coordinatevn} 
\end{figure}

In Fig.~\ref{fig:coordinatevn}  we also check the integrated flow harmonics. 
Here we write $v_n^\mathrm{EP}$ to indicate that this is just the magnitude of the flow harmonic for a single event (for a single event there is no difference between the event plane method and cumulants), where the superscript $\mathrm{EP}$ stands for event plane.
While the flow harmonics are clearly very close to each other (approximately a 20\% effect between our choice in initialization), they are not identical. In this single event, we find that $v_2$ and $v_3$ are enhanced for hyperbolic coordinates vs. $v_4$ and $v_5$ that are suppressed. It is hard to say at this point if this is a systematic effect or something in this single event. 

\section{Hydrodynamics expansion}%
\label{Sec:Hydrodynamics}
\subsection{Equation of state}
\label{Sec:Hydrodynamics:Theory:EquationOfState}
\noindent
The EoS is essential for closing the hydrodynamic equations of motion in heavy-ion collisions. 
The most general EoS spans a four-dimensional phase space defined by $\left\{T,\vec{\mu}\right\}=\left\{T,\mu_B,\mu_S,\mu_Q\right\}$. 
At high energies, the system produces nearly equal numbers of particles and anti-particles, each of which may carry the three conserved charges of baryon number ($B$), strangeness ($S$), and electric charge ($Q$). 
Because quarks and hadrons can carry mixtures of positive and negative charges, the EoS must cover a wide range of $T$ and chemical potentials. 
The temperature should range from 0 (to ensure conservation of energy-momentum) up to $T_\mathrm{max}\sim 700$ MeV for top LHC collisions, while the chemical potentials should ideally cover $\mu_B\sim \pm 1000$ MeV, $\mu_S\sim \pm 1000$ MeV, and $\mu_Q\sim \pm 1000$ MeV to fully account for the phase space reached in heavy-ion collisions (from the highest to lowest beam energies).

At vanishing chemical potentials $\vec{\mu}=0$, the EoS can be calculated directly from lattice QCD in the range $T \sim 130$–$1000$ MeV. 
At finite $\vec{\mu}$, direct calculations are not possible, but one can expand perturbatively around $\vec{\mu}=0$ to obtain pressure susceptibilities:
\begin{equation}
\chi_{ijk}^{BSQ}=\left.\frac{\partial^{i+j+k}(p/T^4)}{\partial(\mu_B/T)^i \, \partial(\mu_S/T)^j \, \partial(\mu_Q/T)^k}\right|_{\mu_B=\mu_S=\mu_Q=0}.
\end{equation}
These susceptibilities can then be used in a Taylor series (or other expansion schemes) to reconstruct the EoS at finite $\vec{\mu}$, e.g.,
\begin{equation}
\frac{p(T,\vec{\mu})}{T^4}=\sum_{i,j,k}\frac{1}{i!j!k!}\,\chi_{ijk}^{BSQ}
\left(\frac{\mu_B}{T}\right)^i\left(\frac{\mu_S}{T}\right)^j\left(\frac{\mu_Q}{T}\right)^k.
\end{equation}
Since these expansions are in powers of $\vec{\mu}/T$, they break down rapidly at large chemical potentials or low temperatures. 
Currently, the only open-source available EoS reconstructed from lattice QCD with baryon, strangeness, and charge chemical potentials are reliable only up to $\vec{\mu}/T \sim 2.5$–$3$ \cite{Noronha-Hostler:2019ayj,Monnai:2024pvy,Abuali:2025tbd}.

Fortunately, the hadron resonance gas EoS can fill in some of the gaps left by the current lattice QCD results.
At sufficiently low $T$ the HRG can be used to calculate the EoS and reproduces lattice QCD well as long as the full list of particles are used (see, e.g., \cite{Alba:2017mqu,SanMartin:2023zhv}). 
However, the challenge is the regime at temperatures just above hadronization where the HRG phase is not applicable, but $T$ is low enough to make the lattice QCD expansions challenging. 
For example, hadronization occurs around $T\sim 130$–$150$ MeV (depending on the values of $\vec{\mu}$) and if one is at $\mu_B=600$ MeV then $\mu_B/T\sim 4$–$4.6$, which is well beyond the regime of validity of our current methods. 
Thus, in \cite{Plumberg:2024leb} backup EoS were developed to ensure thermodynamic stability and continuity if a fluid cell is outside the validity regime of the given EoS. 
The EoS class contains 5 EoS in total: table (lattice QCD), tanh-conformal, conformal, conformal diagonal (twice) wherein the algorithm always searches in the first EoS (lattice QCD) for its location in $\left\{T,\vec{\mu}\right\}$ and then progressively moves on to the next EoS until the point in the QCD phase diagram can be found within the EoS. 
Two conformal diagonal EoS are included because the conformal diagonal EoS has a minimum energy density condition that is dependent on the reach of chemical potentials. The first conformal diagonal does not reach a sufficiently low minimum energy density, so we use a second one to fill that gap (see ~\ref{NewEoS}). 

\subsubsection*{EoS inverter} 
\noindent
A further complication is that the natural hydrodynamic variables differ from those that are most easily used to calculate and tabulate the EoS. 
EoS tables are typically provided on fixed grids of $\left\{T,\vec{\mu}\right\}$ whereas hydrodynamics is written in terms of densities such as the entropy density $s$, and the number densities of the conserved charges, i.e., $\left\{s,\vec{n}\right\}=\left\{s,n_B,n_S,n_Q\right\}$. 
Note that grid codes often use the energy density $\varepsilon$ instead of entropy, but the same challenges of converting from one basis to another remain. 
Thus, there are two options within \ccake{}~2.0 that we provide to the user:
\begin{itemize}
    \item {\bf Online inverter.--} 
        The code takes an EoS table in 4D on a fixed grid of $\left\{T,\vec{\mu}\right\}$ and uses a 4D interpolator and root-finder to obtain all thermodynamic quantities in terms of $\left\{s,\vec{n}\right\}$ while the code is running. If a fluid cell fails to find its location in the EoS; it moves to the backup EoS. This option is the most numerically accurate and allows one to easily change the EoS, but it is slower. 
    \item {\bf Offline inverter.--} 
        A separate code is provided that takes a given EoS on a fixed grid of $\left\{T,\vec{\mu}\right\}$ coupled with the backup EoS and then converts this to a single EoS table on a fixed grid(s) in $\left\{s,\vec{n}\right\}$. This conversion is done once and the new inverted EoS table can be run within \ccake{}~2.0. This option is significantly faster, but it is not quite as numerically accurate and also often requires multiple different grids because $\left\{s,\vec{n}\right\}$ varies significantly more than $\left\{T,\vec{\mu}\right\}$. 
\end{itemize}

It was previously found in \cite{Dore:2022qyz} that a very fine grid around the critical point is needed within the EoS to run relativistic viscous hydrodynamics and properly account for critical scaling of both the EoS and transport coefficients within a simplified hydrodynamic framework. 
Thus, we anticipate that our more realistic simulations will require special care in ensuring that the critical point is accurately included in our numerical calculations, which we leave for a future work. \\

One major performance bottleneck in \ccake{} 1.0 was the 4D root-finding routine of the EoS. Because \ccake{} 2.0 is written with \kokkos{}  which necessitates definitions of variables on the host space and the device space (see Sec. \ref{Sec:Hydrodynamics:Numerics:Kokkos} for details), 4D root-finding significantly slows down runtime on GPUs/running in parallel. 
 With this construction, the online inversion requires data transfer from device space to host space to perform the root-finding, followed by data transfer back to device to evolve the simulation, at each each timestep.

The solution we employ here for \ccake{} 2.0 is to precompute the mapping from (sources) $\{T,\mu_B, \mu_S, \mu_Q\}$ to (targets) $\{s, n_B, n_S, n_Q\}$, and load this conversion table into memory.
The conversion now only requires an interpolation, instead of interpolation and root-finding\footnote{Due to its large complexity, root-finding for non-linear functions is not tractable on GPUs.}; the offline inversion procedure uses the existing root-finder and interpolator.
The downside is that the table can grow very large depending on the resolution of the sources.
Note that in \ccake{}, entropy density is a natural variable whereas in most other (grid based) codes in the field, $\varepsilon$ is used as the natural variable. 
It is straightforward to redefine $\varepsilon$ instead of $s$ as the natural hydrodynamic variable using the same method described above.

One significant challenge in making an efficient 4D inverted table in $\{s, n_B, n_S, n_Q\}$ space is the very non-linear relationship between the sources and targets. 
Thus, a 4D EoS that is on a regular grid in $\{T,\mu_B, \mu_S, \mu_Q\}$ would lead to an oddly shaped region in $\{s, n_B, n_S, n_Q\}$ on an irregular grid. 
However, irregular grids are very challenging for interpolation methods, leading to large numerical error. 
To avoid this problem, our backup EoS makes it possible to obtain a regular grid in $\{s, n_B, n_S, n_Q\}$ even when we are well outside of the regime reached from lattice QCD. 

Another challenge that arises from the non-linear relationship between the sources and targets is that the targets can span many orders of magnitude. 
Thus, if one uses a fixed grid across the entire table, one must choose either an extremely fine grid (to ensure accuracy at low $T$, where thermodynamic quantities vary rapidly with $T$) or a coarse grid (which significantly loses accuracy at low $T$). 
The fine grid option slows down the hydrodynamic code and causes storage issues, due to the enormous size of the table (also a time issue of interpolating over a large table for each run), while the coarse grid introduces errors that are too large to be reasonably used.

To overcome this problem, the conversion tables are constructed to be evenly spaced in logarithmic decades. This uniform spacing ensures that the interpolation is as inexpensive as possible, while the grid arrangement maintains roughly constant relative resolution across scales. The resulting tables are log-scaled overall but linearly spaced locally, keeping interpolation efficient and numerically well-conditioned across the full dynamic range.
In practice, this means that the grid for $s$ might look like $s=\{0.01,0.02,\ldots,0.1,0.2,\ldots,1.0,2.0,\ldots,10\}$.
With this optimization alone, we observe a decrease in runtimes by a factor of eight.

\subsubsection*{EoS convergence tests}
\label{Sec:Hydrodynamics:CodeValidation:EoSConvergence}
\begin{figure}
    \centering
    \includegraphics[width=0.95\linewidth]{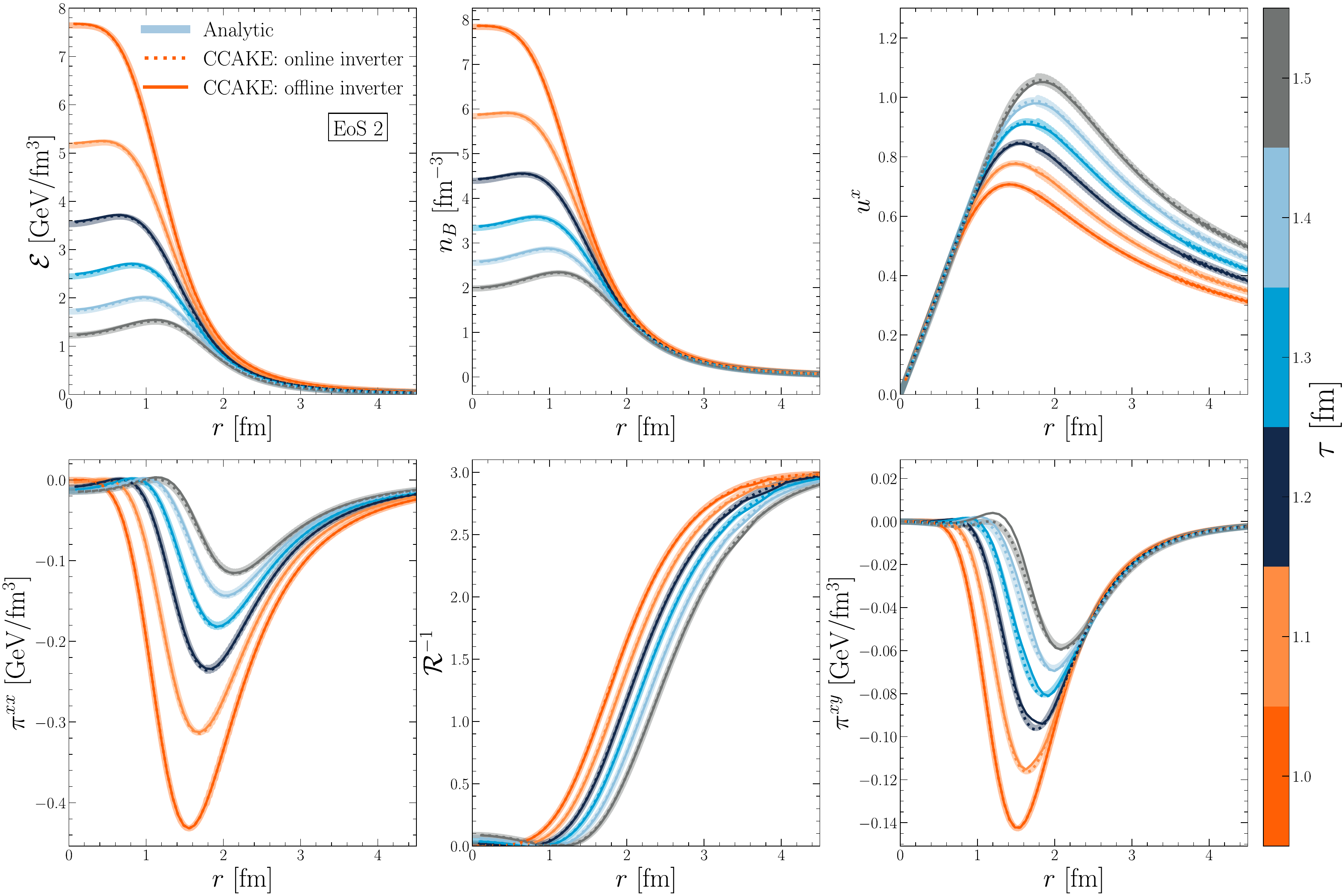}
\caption{Comparison of the BSQ Gubser test with shear viscosity from \cite{Ingles:2025yrv} to \ccake{}~2.0. We compare the EoS with the online vs. offline inverter (see Sec.~\ref{Sec:Hydrodynamics:Theory:EquationOfState} for details).}
    \label{fig:Gubsertest}
\end{figure}

\noindent
We use convergence tests to quantify the effect of the {\bf offline inverter} (this new feature added to \ccake{}~2.0 vs. the {\bf online inverter} that reads in an EoS on a regular grid in $\{T,\mu_B, \mu_S, \mu_Q\}$ (note that lower-order dimensions of the EoS are also allowed in this code) that was included already in \ccake{}~1.0. 
When doing this, we use our grid size that varies logarithmically with $s$, as discussed above.
We find that the offline inverter significantly decreases runtime (and is the only method possible with GPUs), but it sacrifices a slight numerical accuracy in the EoS.  
When comparing to the semi-analytical Gubser flow that includes both shear viscosity and BSQ densities, we find that the offline inverter leads to subtle numerical deviations in the EoS at very late times (see discussion around Fig.~\ref{fig:Gubsertest}).  

Previous work \cite{Ingles:2025yrv} has already confirmed the validity of \ccake{}~1.0 compared to the Gubser check in the presence of conserved charges and shear viscosity. In that paper, the ``online EoS inverter'' was used (wherein the interpolation is done during the hydrodynamic runs across a space of $\left\{T,\mu_B\right\}$ using a multidimensional  table, then a root-finder is used within the hydrodynamic code itself to convert into the natural hydrodynamic variables of $\left\{s,n_B\right\}$. 
While the online EoS inverter is a slower procedure, it is numerically accurate and reproduces the Gubser test with shear viscosity well. 
In principle, since the EoS for the Gubser test is always conformal, one could rather trivially analytically convert it into the natural hydrodynamic variables. However, instead of doing that we use this as an opportunity to test the numerical accuracy of our EoS inverters.

In Fig.~\ref{fig:Gubsertest}, we compare the Gubser solution with a conserved charge and shear viscosity to \ccake{}~2.0 runs that were performed with both the offline EoS inverter and the online EoS inverter. 
Both methods for the EoS inversion are able to capture the semi-analytical solution quite well, especially for the energy density, baryon density, and flow.  
However, we find that at late times that the offline EoS inverter does lead to some slight numerical inaccuracy in the shear stress tensor $\pi^{xy}$. 
It appears that sharp changes in $\pi^{xy}$ vs. the radial component are hard to capture with the offline EoS inverter (these appear around $r\sim 1$--2 fm). 

\begin{figure}[ht!]
    \centering
    \includegraphics[width=0.75\linewidth]{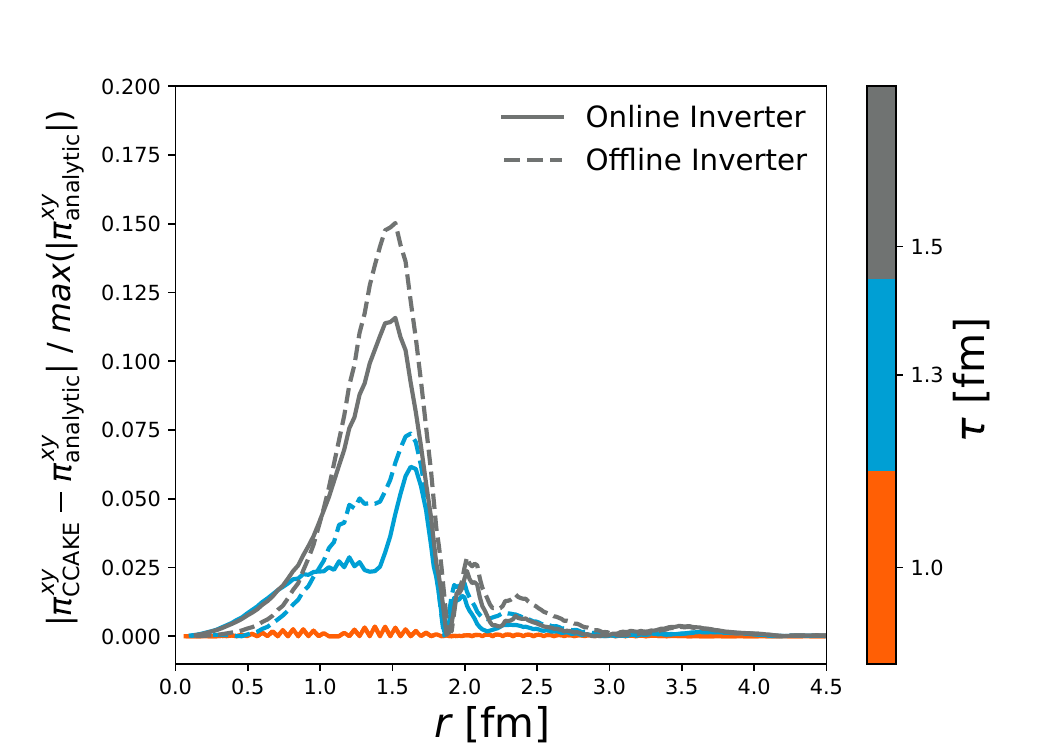}
\caption{Comparison of the discrepancy in shear viscosity between the analytical and \ccake{} result using online vs. offline inverter.}
    \label{fig:pixy_comparison}
\end{figure}

In Fig.~\ref{fig:pixy_comparison} we take a closer look at the shear stress tensor, looking at the difference in the EoS inverters. 
To do so we plot the difference between our \ccake{} results to the analytical results, normalized by the maximum analytical $\pi^{xy}$. We use this observables instead of the percentage difference because many values of $\pi^{xy}$ are very close to 0 (especially at small $r$) such that the percentage difference would look deceptively bad in the regions where $\pi^{xy}\sim 0$ even though the \ccake{} works quite well there (around $r\sim 0$).
Then in Fig.~\ref{fig:pixy_comparison} we find that the differences between the offline and online inverters is smaller at earlier times (and the numerical accuracy is better). Only at  later times we find the differences grow slightly and, indeed, the offline inverter is slightly more numerically inaccurate than the online inverter. However, the differences are concentrated only around $r\sim 1$--$2$ fm, as we saw in the previous plot. 
Additionally, the differences between the offline and online inverters are relatively small such that given the significant speed-up in the code, the offline inverter can have a large advantage to use. 

In future work we plan to quantify this error on experimental observables to better understand if it affects flow or multiplicity.
At the moment, our code has been optimized to handle a smooth EoS without a real phase transition. 
However, previous work \cite{Dore:2022qyz} found that an extremely fine grid is required as one approaches a critical point.

\subsection{Equations of motion}%
\label{Sec:Hydrodynamics:Theory:EquationsOfMotion}
\noindent
The most general form of the conservation equations that govern the hydrodynamic expansion is given by:
\begin{align}
    \begin{aligned}
    \nabla_\mu T^{\mu\nu} & = I^\nu\,,\\
    \nabla_\mu N^{\mu}_q & = I_q\,,
    \label{eq:conservation_source}
    \end{aligned}
\end{align}
where $T^{\mu\nu}$ is the energy-momentum tensor,  $N_q^{\mu}$ is the charge current, $I^\nu$ is a source term of energy and momentum and $I_q$ are the source terms of charge $q$ associated with each current $N_q$.
\\~\\
\noindent
\underline{Hydrodynamic fields:} the tensor decomposition of the energy-momentum tensor and charge current with respect to the time-like four-vector velocity  $u^{\mu}$ can be written as:
\begin{align}
T^{\mu\nu} &= \varepsilon u^{\mu}u^{\nu}- (P+\Pi) \Delta^{\mu\nu} +\pi^{\mu\nu} +2h^{(\mu}u^{\nu)}\, ;\label{Eq:energy-Decom}\\
N^{\mu}_q &= n_q u^{\mu}+J^{\mu}_q\;,  \qquad \qquad q\in\{B,S,Q\}\label{Eq:charge-Decom}
\end{align}
Where $\varepsilon = u_\mu u_\nu T^{\mu\nu}$  is the energy density in the local rest frame of the fluid, $n_q = u_\mu N^{\mu}_q$ is the particle number density in the local rest frame of the fluid. $p=\frac{1}{3} T^{\mu\nu} \Delta_{\mu\nu}$ is the isotropic pressure, $\pi^{\mu\nu}= T^{<\mu\nu>}$, is the shear stress tensor, $h^\mu = T^{<\mu>\nu} u_\nu$  is the energy-momentum diffusion current and finally $J^\mu_q$  is the diffusion current for the conserved charge $q$. By definition, we have the following: 
$u_\mu \pi^{\mu\nu} = 0; \quad \pi^\mu_\mu = 0; \quad u_\mu h^\mu = 0; \quad u_\mu J^\mu_q = 0\,$ where in ideal fluid dynamics, all dissipative currents $\pi^{\mu\nu}=h^\mu=J^\mu=0$ and the quantities $\varepsilon$ and $n_q$ matches the local thermodynamical energy density and charge density of the fluid, respectively. 
\\~\\
\noindent
\underline{Hydrodynamics frame}: In general, for dissipative fluids, one loses the notion of local equilibrium. To define the frame of local equilibrium of the fluid, one possibility is to identify the fluid four-velocity to be identical to the energy flow of the fluid, i.e., 
\begin{equation}
    \varepsilon u^{\mu} = u_{\nu} T^{\mu\nu}, \qquad n_{q} = u_{\mu} N^{\mu}_{q},\label{Eq:landau}
\end{equation}
With this particular choice of ``hydrodynamic frame", heat diffusion is not allowed in the fluid, therefore, we set $h^\mu=0$. This choice is the well-known ``Landau matching condition" also known as the Landau frame.  Another choice of frame is the so called ``Eckart frame", where the fluid four-velocity is chosen to be identical to the particle number flow, i.e., no diffusion is allowed $n^\mu_q=0$. In this work, we will follow the choice of the Landau frame.
The conservation equations are: 
\begin{align}
D\varepsilon + (\varepsilon+p+\Pi)\theta - \pi^{\mu\nu}\nabla_{\langle\mu\rangle} u_\nu &= u_\nu I^\nu\,,\label{eq:e_conservation}\\
(\varepsilon+p)Du^\mu +\Delta^{\mu\nu} \nabla^\lambda\pi_{\nu\lambda} -\pi^{\mu\nu} Du_\nu -\nabla^\mu p  &= \Delta^{\mu}{}_{\nu} I^\nu\,.\label{eq:momentum_conservation}\\
D n_q +n_q \theta + \nabla_\mu J^\mu_q &= I_q\;.
\label{eq:charge_conservation}
\end{align}
\\~\\
\noindent
\underline{Source terms}: in an isolated system, both sources are set to zero $(I^\nu=0 , I_q = 0)$. While the medium in heavy-ion collisions as a whole is an isolated system, there can be overlap in time between different stages of the collision, such that one would not want to switch from one phase to the next at a fixed point in time, but rather one would like to smoothly transition from one model into the next such that a \emph{dynamical initialization} process would be required; see Sec. \ref{Sec:InitialState:DynamicalInitialization}. In addition, those terms will not be zero when the dynamical evolution is coupled to jets where they provide the feedback of the jets to the medium.
\\~\\
\noindent
\underline{Equations of motion}:
In \ccake{}~2.0, we solve the full set of fluid equations of motion in generalized coordinates and for any dimensions for the following transient theories
\\~\\
\noindent
\textbf{Israel--Stewart.--}
Relativistic viscous fluids must remain both causal (velocities less than the speed of light $c$) and thermodynamically stable. From the conservation equations~(\ref{eq:charge_conservation}), one can always construct an equation of motion for a given entropy current, and in the absence of external sources,
\begin{equation}
     \nabla_\mu S^\mu = \beta_\nu \nabla_\mu T^{\mu \nu} - \alpha \nabla_\mu N^\mu_q \, ,
     \label{eq:ideal_entropy}
\end{equation}
where the derivatives of $T^{\mu \nu}$ and  $N^\mu$ are given by the conservation equations. Together with the second law of thermodynamics, this leads to the phenomenological IS equations, which are derived by constraining the entropy production to be positive definite. Taking as a starting point a proposition for the entropy current up to \emph{second order} ($n=2$) in dissipative currents ${\Pi,\pi^{\mu\nu},J^\mu}$ \cite{Israel:1979wp}. Then the equations of motion take the form
\begin{align}
    \tau_\Pi\dot\Pi + \Pi &= -\zeta\theta - \frac{\tau_\Pi\Pi}{2\beta_\Pi}\dot\beta_\Pi - \frac{\tau_\Pi}{2}\Pi\theta+\ldots
    \label{eq:bul_rel}\\
    \tau_\pi\dot\pi^{\mu\nu} + \pi^{\mu\nu} &= 2\eta\sigma^{\mu\nu} - \frac{\tau_\pi\pi^{\mu\nu}}{2\beta_\pi}\dot\beta_\pi - \frac{\tau_\pi}{2}\pi^{\mu\nu}\theta+ \ldots
    \label{eq:she_rel}\\
    \tau_{qq'}\dot J^\mu_{q'} + J^\mu_q &= -\kappa_{qq'}\nabla^\mu\alpha_{q'} + \frac{\tau_{qq'}J^\mu_{q'}}{2\beta}\theta - \frac{\tau_{qq'}}{2\beta}\dot\beta_{q'l}J^\mu_{l}+\ldots
    \label{eq:dif_rel}.
\end{align}
Here, the shear relaxation time $\tau_{\pi}$ provides a time scale to relax back to the Navier--Stokes solution, which preserves causality (at least in the linear regime). 
The $\dots$ appear in Eqs.~(\ref{eq:bul_rel}), (\ref{eq:she_rel}) and (\ref{eq:dif_rel}) refer to coupling and higher-order terms appear. These terms either connect directly to $\pi^{\mu\nu}$ or can couple to other out-of-equilibrium effects like bulk viscosity $\zeta$ of diffusion $\kappa$. A minimal IS approach has been developed in \cite{Marrochio:2013wla} where the equations of motion take the following form: 
\begin{align}
    \tau_\Pi\dot\Pi + \Pi =& -\zeta\theta - {\tau_\Pi}\Pi\theta+...
    \label{eq:min_bul_rel}\\ 
    \tau_{\pi} \Delta_{\mu\nu\alpha\beta}D\pi^{\alpha\beta} +\pi_{\mu\nu}=&2\eta\sigma_{\mu\nu}-\frac{4}{3} \tau_{\pi} \pi_{\mu\nu}\theta
    \label{eq:min_she_rel}\\
    \tau_{qq'}\dot J^\mu_{q'} + J^\mu_q =& -\kappa_{qq'}\nabla^\mu\alpha_{q'} + \frac{\tau_{qq'}J^\mu_{q'}}{2\beta}\theta +...\,,
    \label{eq:min_dif_rel}
\end{align}
and has been tested in \cite{Noronha-Hostler:2014dqa}. \ccake{}~2.0 supports both versions of the IS equations, and some of the results presented here are based on this minimal IS.
\begin{figure*} [t!]
    \centering
    \includegraphics[width=\linewidth]{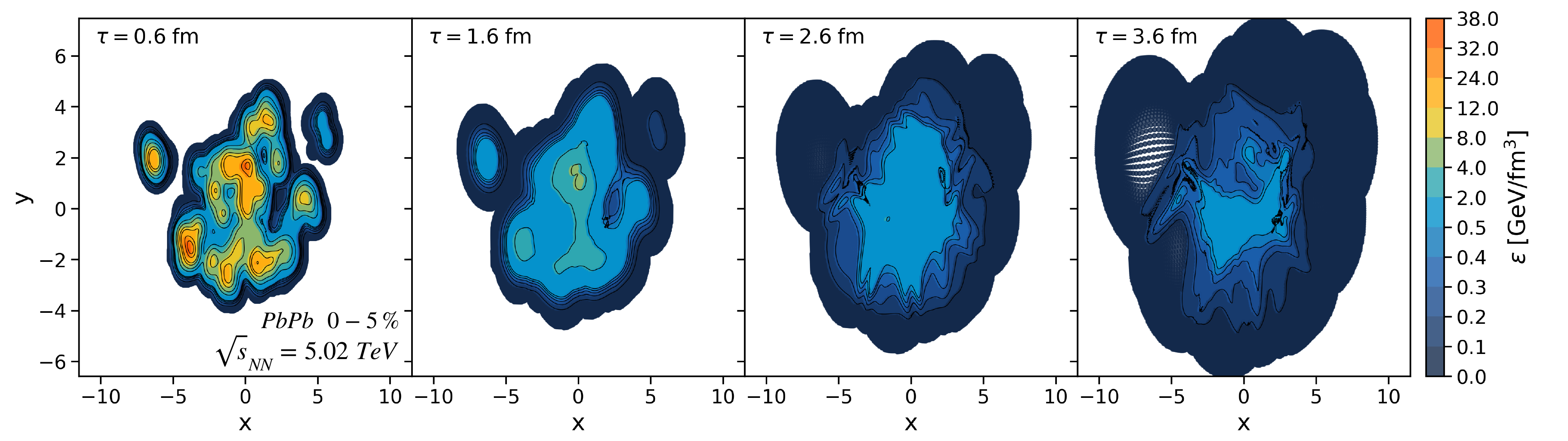}
    \caption{Illustration showing the time evolution of a central Pb+Pb collision at $\sqrt{s_{NN}} = 5.02$ TeV.}
    \label{fig:evolvingcuts}
\end{figure*}

~\\
\noindent
\textbf{DNMR.--}
Unlike the IS theory, DNMR is a microscopic based fluid dynamics theory which is derved directly from the Boltzmann equation via the 14 moment approximation, and with a systematic truncation scheme where the power counting is done in powers of Knudsen and Reynolds numbers Eq.(\ref{eq:knudsen-reynolds}) \cite{Denicol:2012cn}. This is done in the following form: 
\begin{equation}
\begin{aligned}
\tau_{\Pi} \dot{\Pi}+\Pi & =-\zeta \theta+\mathcal{J}+\mathcal{K}+\mathcal{R}\, , \\
\tau_n \dot{J}^{\langle\mu\rangle}+J^\mu & =\kappa_n I^\mu+\mathcal{J}^\mu+\mathcal{K}^\mu+\mathcal{R}^\mu \, ,\\
\tau_\pi \dot{\pi}^{\langle\mu \nu\rangle}+\pi^{\mu \nu} & =2 \eta \sigma^{\mu \nu}+\mathcal{J}^{\mu \nu}+\mathcal{K}^{\mu \nu}+\mathcal{R}^{\mu \nu}\, ,
\end{aligned}
\label{eq:dnmr}
\end{equation}
where the $\mathcal{J}$ terms indicate all terms of first order in Knudsen and inverse Reynolds numbers, the $\mathcal{K}$ terms  are all second order in Knudsen number, and the $\mathcal{R}$ terms contain all terms of second order in inverse Reynolds number. Besides the well-defined expansion parameter, DNMR has the advantage and appealing feature that the calculations of the transport coefficients are done consistently.  

~\\
\textbf{ADNH.--}
Recently, the authors of Ref.~\cite{Almaalol:2022pjc} have developed a new approach to the phenomenological IS theory where new terms were derived in the equations of motion using zero entropy current contributions, the so-called ADNH approach. This technique was used to derive some of the terms from DNMR and can be developed further to derive more terms. In the ADNH approach, the EoM take the form: 
\begin{equation}
\begin{aligned}
    \tau_\Pi \dot\Pi + \Pi &= -(\zeta + \frac{\tau_\Pi}{2}\Pi)\theta - \frac{\tau_\Pi}{2\beta_\Pi}\dot\beta_\Pi\Pi- \frac{\zeta\delta_{J\Pi}^{q} }{\beta}\partial_\mu J^\mu_q + \frac{\zeta \lambda_\Pi^q}{\beta}J_q^\mu\nabla_\mu\delta_{J\Pi}^{q} \\&+  \frac{\zeta \delta_{\Pi \pi}}{2\beta} \pi^{\mu\nu} \sigma_{\mu\nu}\,,\\
    \tau_\pi \dot\pi^{\mu\nu} + \pi^{\mu\nu}& = 2\eta\sigma^{\mu\nu} - \frac{\tau_\pi}{2}\pi^{\mu\nu}\theta - \frac{\tau_\pi\dot\beta_\pi}{2\beta_\pi}\pi^{\mu\nu}
    - \frac{2\eta\delta^{q}_{J\pi}}{\beta}  \nabla^{\langle\mu}J_q^{\nu\rangle} + \frac{2\eta \lambda_\pi^q}{\beta}J_q^{\langle\mu}\nabla^{\nu\rangle}\delta^{q}_{J\pi}\\&-  \frac{\eta \delta_{\Pi \pi}}{\beta} \Pi \sigma_{\mu\nu}+\,\frac{\Lambda_{\pi\pi}}{\beta} \pi_{\lambda\nu}(\sigma_{\mu}^{\lambda}+\frac{1}{3}\Delta_{\mu}^{\lambda}\theta))\,,\\
    \tau_{qq'}\dot J^\mu_{q'} + J^\mu_q & = -\kappa_{qq'}\nabla^\mu\alpha_{q'} + \frac{\tau_{qq'}J^\mu_{q'}}{2\beta}\theta - \frac{\tau_{qq'}}{2\beta}\dot\beta_{q'l}J^\mu_{l}
    - \frac{\kappa_{qq'}\delta_{J\Pi}^{q'}}{\beta} \nabla^\mu\Pi- \frac{\kappa_{qq'}\delta_{J\pi}^{q'}}{\beta} \nabla_\nu\pi^{\mu\nu}\\&- \frac{\kappa_{qq'}\Tilde{\lambda}_\Pi^{q'}}{\beta}\Pi\nabla^\mu\delta_{J\Pi}^{q'}- \frac{\kappa_{qq'} \Tilde{\lambda}_\pi^{q'}}{\beta}\pi^{\mu\nu}\nabla_\nu\delta_{J\pi}^{q'} \, ,
\end{aligned}
\label{Eq:diffusion-evol}
\end{equation}

The public version of \ccake{}~2.0 supports the possibility of running any of the dynamics above by switching on/off any of the available terms. We refer the reader to an example YAML configuration file in \cite{ccakesite} where a set of parameters have been designated to turn on/off each of these terms. In practice, the public version of the code has all the tested terms and excludes the following: 
\begin{itemize}
    \item For Israel--Stewart and ADNH equations of motion, $\dot{\beta}$ terms are required in the equations of motion that require time derivatives of $c_s^2$. There are both numerical challenges to calculate $c_s^2$ derivatives live during the hydrodynamic expansion as well as theoretical challenges (these derivatives are not available in initial state models) such that we leave the inclusion of the $\dot{\beta}$ terms for future work. However, we caution that it has been found in \cite{Dore:2020jye} that these terms were necessary for exploring a critical point in an Israel--Stewart approach.
    \item For the DNMR equations, at the momentum, we only include the shear and bulk viscosity terms. The relevant transport coefficients for these terms are discussed in Sec.  \ref{Sec:Hydrodynamics:Theory:TransportCoefficients}. We exclude $\mathcal{K}$ terms and all couplings with diffusion terms. This exclude the transport coefficients relevant to these terms. (to be included in a future work).
    \item The second order $\mathcal{R}$ terms for shear and bulk viscosities have been included, but their relevant transport coefficients were set to zero, i.e., $\varphi_1=\varphi_3=\varphi_6=0$.
\end{itemize}

\subsection{Transport coefficients }%
\label{Sec:Hydrodynamics:Theory:TransportCoefficients}
\noindent
In relativistic systems, several types of transport coefficients appear. 
Because we adopt Landau matching for our frame of reference, there is no heat flow. 
Thus, the relevant transport coefficients currently used within \ccake{}~2.0 are shear viscosity, bulk viscosity, and diffusion. 
Here, we briefly review the expected behavior of these transport coefficients based on microscopic models and phenomenology, before describing how they are implemented in \ccake{}~2.0. 

\begin{figure}[ht!]
    \centering
    \includegraphics[width=0.7\linewidth]{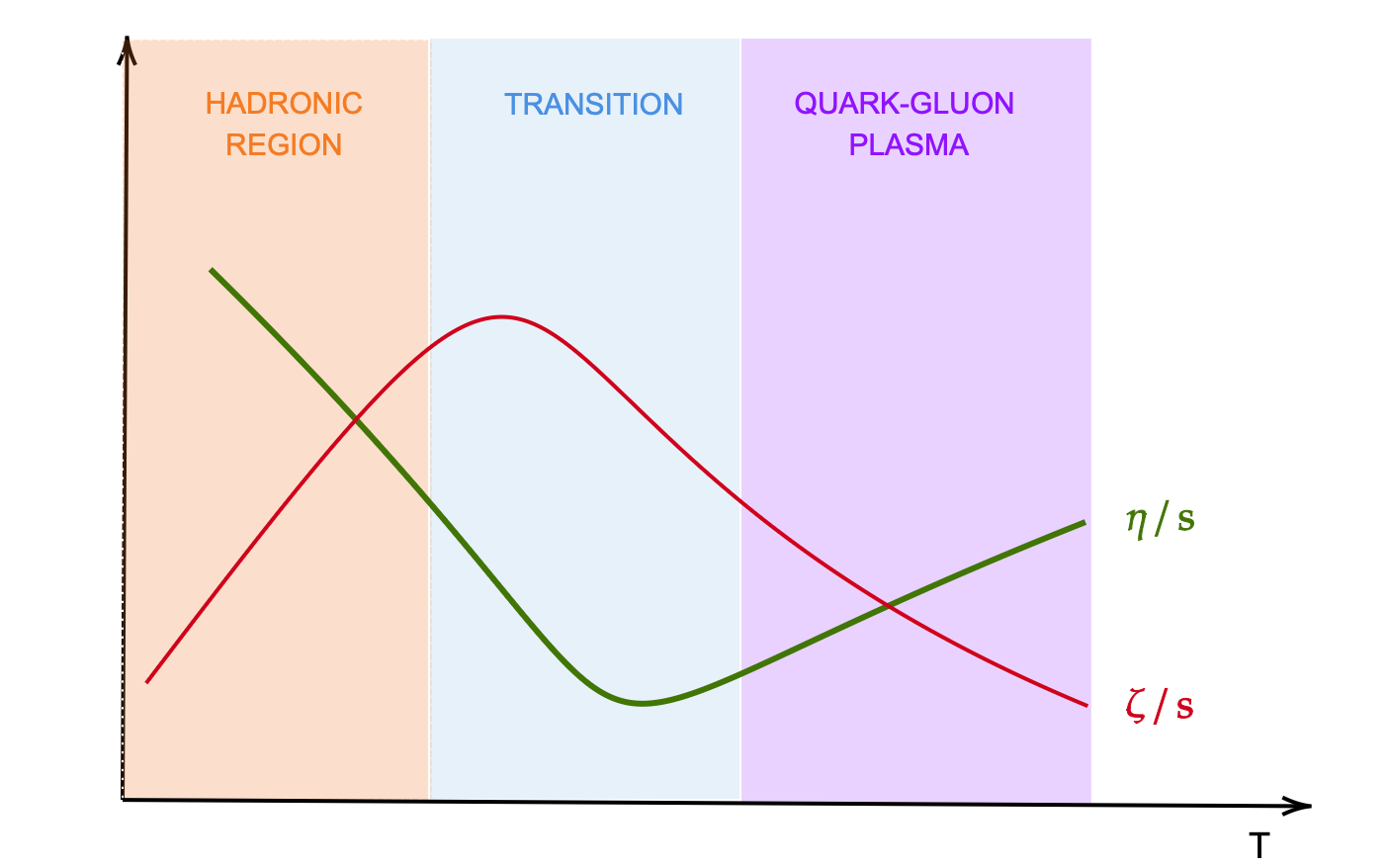}
    \caption{Behavior of transport coefficients at vanishing chemical potential as predicted by Bayesian analysis.
    } 
    \label{fig:transp_coeff}
\end{figure}

~\\
\noindent
\underline{Shear viscosity:}
Shear viscosity is the most important transport coefficient in heavy-ion collisions, as it strongly influences the hydrodynamic evolution of the system. 
Since $\eta$ can be interpreted as a linearization over equilibrium, it is often discussed in its dimensionless form, comparing the relaxation time of the fluid to quantum time scales. 
This is given by the shear-viscosity-to-enthalpy ratio,
\begin{equation}
    \frac{\eta T}{w}=\frac{\eta T}{\varepsilon+p},
\end{equation}
where the enthalpy $w$ is the energy density $\varepsilon$ plus the pressure $p$. 
In the limit of vanishing net densities ($n_B=n_S=n_Q=0$), this reduces to $\eta/s$, because the Gibbs-Duhem relation, given by, $\varepsilon+p=sT+\sum_{q=B,S,Q} n_q\mu_q$, simplifies to $s=(\varepsilon+p)/T$.
Physically, the shear viscosity describes the amount of friction between fluid layers as the system relaxes to equilibrium.  

This dimensionless ratio, $\eta/s(T)$, plays a central role in characterizing the QCD medium, and its temperature dependence has been widely studied. 
As discussed in Sec.~\ref{Sec:FrameworkOverview:PhysicsOverview}, the community generally expects $\eta/s(T)$ to reach a minimum near the QCD phase transition, although Bayesian analyses are still consistent with both a constant and a temperature-dependent $\eta/s$ \cite{Bernhard:2019bmu,JETSCAPE:2020mzn,Auvinen:2020mpc,Nijs:2020roc,Parkkila:2021tqq,Gotz:2025wnv}.  

While $\eta/s(T)$ is expected to have a minimum at vanishing densities, the behavior of $\eta T/w(T,\vec{\mu})$ at finite $\vec{\mu}$ remains largely unknown and is an active area of research. 
Recent pQCD calculations at finite $\mu_B$ and finite $\vec{\mu}$ \cite{Danhoni:2022xmt,Danhoni:2024ewq,Danhoni:2024kgi} suggest a complicated dependence that can increase or decrease depending on the region of the phase diagram. 
Holographic approaches tend to predict a decrease \cite{Rougemont:2017tlu,Grefa:2022sav}, whereas PNJL models show an increase \cite{Soloveva:2020hpr,Soloveva:2021quj} (although they compute $\eta/s$ rather than $\eta T/w$). 
Hadronic models suggest a decrease in $\eta T/w$ at fixed $T$ with increasing $\vec{\mu}$ \cite{Demir:2008tr,Tiwari:2011km,Kadam:2014cua,Kadam:2015xsa,Kadam:2015fza,Rose:2017bjz,Kadam:2018hdo,Mohapatra:2019mcl,Hammelmann:2023fqw}. 
However, since the phase transition line bends to lower $T$ \cite{Borsanyi:2011bn,Bellwied:2015rza,Borsanyi:2020fev,HotQCD:2018pds,Ding:2024sux}, the value of $\eta T/w$ at the transition actually increases significantly (because shear viscosity in hadronic models scales inversely with $T$) \cite{McLaughlin:2021dph}.  
Meanwhile, Bayesian studies at finite density remain scarce, but preliminary results suggest an increase in shear viscosity \cite{Auvinen:2017fjw,Shen:2023awv,Gotz:2025wnv}.  
Thus, the behavior of shear viscosity at large densities remains an open question.  

Within \ccake{}~2.0 we provide three options for shear viscosity:
\begin{itemize}
    \item {\bf Constant:} Fix $\eta/s$ to a constant value across the entire $\left\{T,\mu_B,\mu_S,\mu_Q\right\}$ phase diagram.
    \item {\bf Table:} In the future we will have an option to use $\eta T/w(T,\mu_B,\mu_S,\mu_Q)$ tables from \cite{Danhoni:2024kgi}, which will be included as an additional EoS column. 
    \item {\bf Parameterized:} Use simple parameterizations of $\eta/s$ (with no density dependence) that feature a minimum around the phase transition. 
\end{itemize}

~\\
\noindent
\underline{Bulk viscosity:}
Bulk viscosity $\zeta$ is also relevant in heavy-ion collisions, as it describes the resistance of the fluid to compression and expansion. 
Because $\Pi$ is the trace of the out-of-equilibrium contribution to $T^{\mu\nu}$, bulk viscosity is directly related to deviations from conformality (recall that conformal systems satisfy $\varepsilon=3p$, implying $c_s^2 \rightarrow 1/3$). 
Whenever the trace anomaly $\varepsilon-3p \neq 0$, bulk viscosity is expected to increase. 
Lattice QCD results \cite{Borsanyi:2010cj} show a peak in the trace anomaly near the deconfinement crossover transition, which implies a corresponding peak in bulk viscosity. 
At both very high and very low $T$, however, the system approaches conformality and the bulk viscosity becomes small.  

The exact location and magnitude of the bulk viscosity peak remain uncertain. 
Although generally less important than shear viscosity, bulk viscosity does affect observables such as the average transverse momentum of identified particles $\langle p_T\rangle$ \cite{Ryu:2015vwa}. 
Significant uncertainty also persists in modeling bulk viscosity at particlization \cite{Noronha-Hostler:2013gga,Noronha-Hostler:2014dqa,JETSCAPE:2020mzn}, and known issues arise when converting a conformal pre-equilibrium phase into a non-conformal fluid, leading to artificially large bulk pressure $\Pi$ \cite{daSilva:2022xwu,ExTrEMe:2023nhy}.  

Within \ccake{}~2.0 we provide two options for bulk viscosity:
\begin{itemize}
    \item {\bf Scaling with $c_s^2$:} Since bulk viscosity scales with $c_s^2$ in various theories, we include a form that depends on $c_s^2$. Near a critical point, we also expect critical scaling (see \cite{PhysRevE.55.403,Moore:2008ws,Monnai:2016kud,Dore:2020jye}), so the full form includes contributions from both a non-critical regime (NC) and critical scaling (CS):
    \begin{eqnarray}\label{eqn:bulk}
    \left(\frac{\zeta T}{w}\right)_\mathrm{NC}&=&A \left(\frac{1}{3} - c_s^2\right)^\alpha, \\
    \left(\frac{\zeta T}{w}\right)_\mathrm{CS} &=&\left(\frac{\zeta T}{w}\right)_\mathrm{NC}
         \left[1 +\left(\frac{\xi}{\xi_0}\right)^3\right],
    \end{eqnarray}
    where $A$ is the normalization, $\alpha$ is a power (e.g.\ $\alpha=2$ for weak coupling and $\alpha=1$ for strong coupling \cite{Czajka:2018bod}), $\xi$ is the correlation length, and $\xi_0$ sets the critical scale. 
    Since $c_s^2 \rightarrow 0$ across the first-order phase transition line (in $n_B$), a discontinuity is built in. 
    For EoS with $c_s^2>1/3$, Eq.~(\ref{eqn:bulk}) is not valid if $\alpha$ is odd, as it would give $\zeta T/(\varepsilon+p)<0$.
    \item {\bf Table:} In the future we will have an option to use $\zeta T/w(T,\mu_B,\mu_S,\mu_Q)$ tables, for example from holography \cite{Grefa:2022sav}, which will be included as an additional EoS column. 
    \item {\bf Constant:} For testing purposes, we allow for the possibility of $\zeta/s=const$. However, we discourage users from relying on this limit for realistic simulations. We know already that at very low and high $T$ the EoS approaches the conformal limit wherein $\zeta/s\rightarrow 0$.
\end{itemize}

~\\
\noindent
\underline{BSQ diffusion:}
The last of the transport coefficients covered by \ccake{} is the BSQ diffusion matrix. Diffusion, $\kappa_q$, describes the speed with which a conserved charge $q\in\{B,S,Q\}$ spreads across a fluid (typically thought of in terms of Brownian motion). For an intuitive understanding of this quantity, one can use the simple example of ink being dropped into water. For such a system, diffusion will describe the speed at which the system changes color. For heavy-ion collisions, QCD has three relevant conserved charges: baryon number, strangeness, and electric charge (therefore, BSQ), which means that there are associated diffusion coefficients for each $\kappa_B,\kappa_S,\kappa_Q$.

However, all quarks and hadrons carry all three of these conserved charges and, in fact, have non-trivial correlations between them. For instance, the strange quark carries $B=1/3,S=-1,Q=-1/3$ such that it will contribute to the diffusion process of all three charges, which makes diffusion extremely hard to compute using the same tools mentioned earlier in this section. 
Thus, one obtain a diffusion matrix,
\begin{equation}
    \kappa\equiv \begin{bmatrix}
    \kappa_{BB} &  \kappa_{BS} & \kappa_{BQ} \\
    \kappa_{SB} &  \kappa_{SS} & \kappa_{SQ} \\
    \kappa_{QB} &  \kappa_{QS} & \kappa_{QQ} 
    \end{bmatrix},
\end{equation}
that is symmetric \cite{Greif:2017byw}. Recent years have led to a initial calculations of this matrix in kinetic theory \cite{Greif:2017byw,Fotakis:2019nbq,Fotakis:2021diq,Fotakis:2022usk,Fotakis:2024hmz} or holography \cite{Rougemont:2017tlu}, but a general consensus has not yet been reached in the field on its overall behavior. 
However, it has been found that the off-diagonal terms can lead to more fluctuations of conserved charges \cite{Fotakis:2019nbq}, that net-proton multiplicities are sensitive to baryon diffusion \cite{Denicol:2018wdp}, and that critical scaling of diffusion can affect the passage through the QCD phase diagram \cite{Du:2021zqz}. 
That being said, no full dynamical simulations exist today in 2+1D or 3+1D that incorporate the full diffusion matrix, including off-diagonal terms. 
Furthermore, the conversion from a fluid into hadrons still remains to be resolved with diffusion and the full BSQ matrix. 
Thus, diffusion remains today the least understood transport coefficient. 

Within \ccake{}~2.0 we currently provide the options for diffusion matrix that has constant values of $\kappa_{qq'}/T^2$ (but that each component of the matrix can differ). Future work will explore parameterized dependencies and tables. 

\subsection{BSQ Diffusion and the Beam Energy Scan}
\label{Sec:Hydrodynamics:Theory:BES}
\noindent
Here we have incorporated the necessary changes in \ccake{} in order to explore a wide range of $\sqrt{s_{NN}}$. 
In this section we will explore the new updates to \ccake{}, discuss their effects on the hydrodynamic evolution, and perform new convergence tests across $\sqrt{s_{NN}}$.

It has been previous shown in \cite{Fotakis:2019nbq} for simplified systems (i.e., 1+1D) that the BSQ diffusion matrix can introduce new fluctuations into $n_B,n_S,n_Q$ densities. 
Moreover, in \cite{Plumberg:2024leb,Gardim:2024nyz} the consequences of initializing BSQ fluctuations in the initial state were explored for ideal BSQ currents. 
Now that \ccake{} includes the full BSQ diffusion matrix, we can study the influence of the BSQ diffusion matrix on initial BSQ charge fluctuations. 

\begin{figure}
    \centering
    \includegraphics[width=0.8\linewidth]{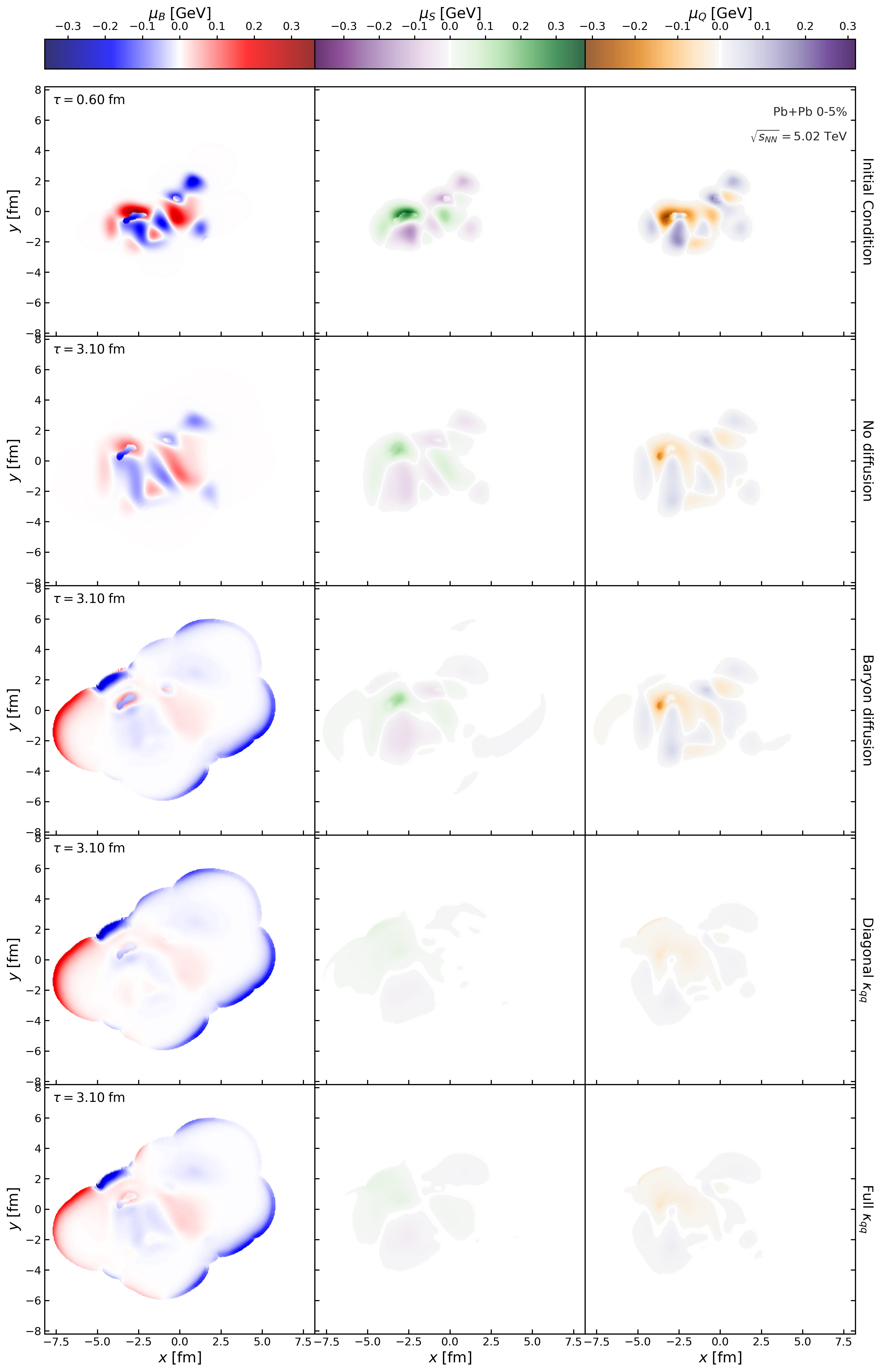}
    \caption{\trento{}+\iccing{}+\ccake{} simulations with the same initial condition.  Distribution of chemical potentials are compared for baryon number (left column), strangeness (middle column), and electric charge (right column). Rows compare the initial condition (top) followed by (starting from the top) simulations  with no diffusion, baryon diffusion only, a diagonal BSQ diffusion matrix, or the full BSQ diffusion matrix.
    }
    \label{fig:bsq_diffusion}
\end{figure}

In Fig.~\ref{fig:bsq_diffusion}, we use a \trento{}+\iccing{} initial condition that initialized BSQ fluctuations ran  with ideal BSQ currents compared different scenarios for the BSQ diffusion matrix: baryon diffusion only, diagonal currents only, and full BSQ matrix.
The full matrix is given by
\begin{equation*}
\frac{\kappa_{qq'}}{T^2} =
\begin{bmatrix}
  0.0482 & -0.0482 & 0.0000 \\
 -0.0482 & 0.1447 & 0.0147 \\
  0.0000 & 0.0147 & 0.0880 \\
\end{bmatrix},
\end{equation*}
with relaxation times $\tau_{qq'} = \frac{1}{T} \, \mathrm{diag}(0.2,\, 0.2,\, 0.2)
$, which were motivated by values tested in \cite{Fotakis:2019nbq}. For this test, we have minimal  IS with shear and bulk viscosity  set to constant values of $\eta/s = 0.16$ and $\zeta/s = 0.08$.
\iccing{} produces initial state fluctuations in  $\left\{\mu_B,\mu_S,\mu_Q\right\}$ from gluons splitting into quark anti-quark pairs, distributed in the transverse plane. 
We then ran the hydrodynamic simulations until $\tau=3.1$ fm to study the influence of our choice of diffusion. 

We find that ideal BSQ charges preserved the most structure from the initial condition into the final state, where the distribution of $\left\{\mu_B,\mu_S,\mu_Q\right\}$ looks similar to what was found in the initial state but it is just spreads out over time. 
Once any diffusion is switched on, much of that structure is smeared out. If we only turn on baryon diffusion, we see some structure in $\mu_S,\mu_Q$ remain because the strangeness and electric charge currents are unaffected. 
Both with the diagonal only matrix as well as with the full BSQ matrix we find that nearly all the structure is eliminated in the strangeness and electric charge channels.

The baryon chemical potential still maintains large fluctuations in $\mu_B$ even to late times and this occurs regardless of the choices made regarding the BSQ diffusion matrix. The nature of these fluctuations appear to be quite different when any baryon diffusion is turned on vs. the ideal case. 
With baryon diffusion the fluctuations tend to spread to the edges of the system where large regions of either anti-matter or matter is formed. It almost appears as if the like charges clustered together in certain regions.
We suspect that there will be experimental consequences for this effect but we leave a full systematic study for a later work where we can construct global observables across many events. 
We believe the reason that baryon fluctuations remain while strangeness and electric charge are washed out are because the value of baryon diffusion is the smallest in the BSQ diffusion matrix compared to strangeness diffusion or electric charge diffusion. 
This effects seems to suppress charge fluctuations for SQ but rather enhance B fluctuations. 
Since we are looking specifically at an event that has no net-BSQ charges it is difficult to say how this picture will change if there is a large net-baryon number and electric charge, which will occur when there is baryon stopping. 

\subsection{Causality and Stability}%
\label{Sec:Hydrodynamics:Theory:CausalityStability}
\noindent
Hydrodynamics is applicable in the limit where the microscopic scale is much smaller than the large scales. However, in a system with multiple different types of dissipative currents there is not a single small or larger scale. 
Quantities that define the applicability of hydrodynamics are Knudsen and inverse Reynolds numbers that look at the ratio of small scale properties to large scale. 
The typical ones for shear and bulk viscosity are
\begin{eqnarray}
\rm{Kn}_{\pi}&=&\tau_\pi \sqrt{\sigma^{\mu\nu}\sigma_{\mu\nu}},\\
\rm{Re}^{-1}_{\pi}&=&\frac{\sqrt{\pi^{\mu\nu}\pi_{\mu\nu}}}{p},\\
\rm{Kn}_{\Pi}&=&\tau_{\Pi}\theta,\\
\rm{Re}^{-1}_{\Pi}&=&\frac{|\Pi|}{p},
\label{eq:knudsen-reynolds}
\end{eqnarray}
where the ratios should be small in the limit that hydrodynamics is applicable. 
In fact, the previously mentioned DNMR equations of motion are derived in order of Knudsen and inverse Reynolds numbers.

One significant challenge is that what constitutes ``small'' and ``large'' is somewhat arbitrary. Thus, it is more useful to have a quantity that acts like an order parameter for hydrodynamic vs. not-hydrodynamic behavior. Causality and stability constraints can be used precisely as such an order parameter. 
While the equations of motion presented in this work guarantee causality for fluids very close to equilibrium, challenges can arise if the system has a $T^{\mu\nu}$ that begins or is perturbed far from equilibrium (e.g., by a jet). 
Fortunately, by studying the characteristic velocities of the equations of motion and the corresponding nonlinear set of partial differential equations, the authors of \cite{Bemfica:2020xym} were able to derive a set of nonlinear causality conditions which could be used to identify how well a fluid described by those equations of motion respects causality. 
The study found both \emph{necessary} conditions  (which are required for causality to be respected) and \emph{sufficient} conditions which, if satisfied, guarantee that the evolution is causal. 

These conditions have been used within the community to determine that the most commonly used in initial conditions often lead to causality violations \cite{Plumberg:2021bme}, that leads to consequences on experimental observables \cite{Chiu:2021muk}. Causality violations can arise because of a mismatch in the bulk pressure when switching from a conformal pre-equilibrium phase into a non-conformal hydrodynamic phase \cite{ExTrEMe:2023nhy}. Additionally, incorporating a penalty into a Bayesian analysis for causality violations  changes the extracted bulk viscosity \cite{Domingues:2024pom}. Recent work has extended the causality constraints to finite $n_B$ \cite{Cordeiro:2025rkt} that we will include in future work. 

In \ccake{}~2.0 there is a class that calculates the necessary and sufficient conditions, labeling each individual fluid cell over time. 
While running the causality conditions is not a default in \ccake{}~2.0, there is a flag that can switch on these conditions and then output the results (including which condition is violated and if it is necessary or sufficient) across the fluid dynamic field at each point in time.  Along with the causality conditions, one can also track Knudsen and inverse Reynolds numbers. 

\begin{figure*}[ht!]
    \centering
    \includegraphics[width=\linewidth,trim={3cm 0 0 0},clip]{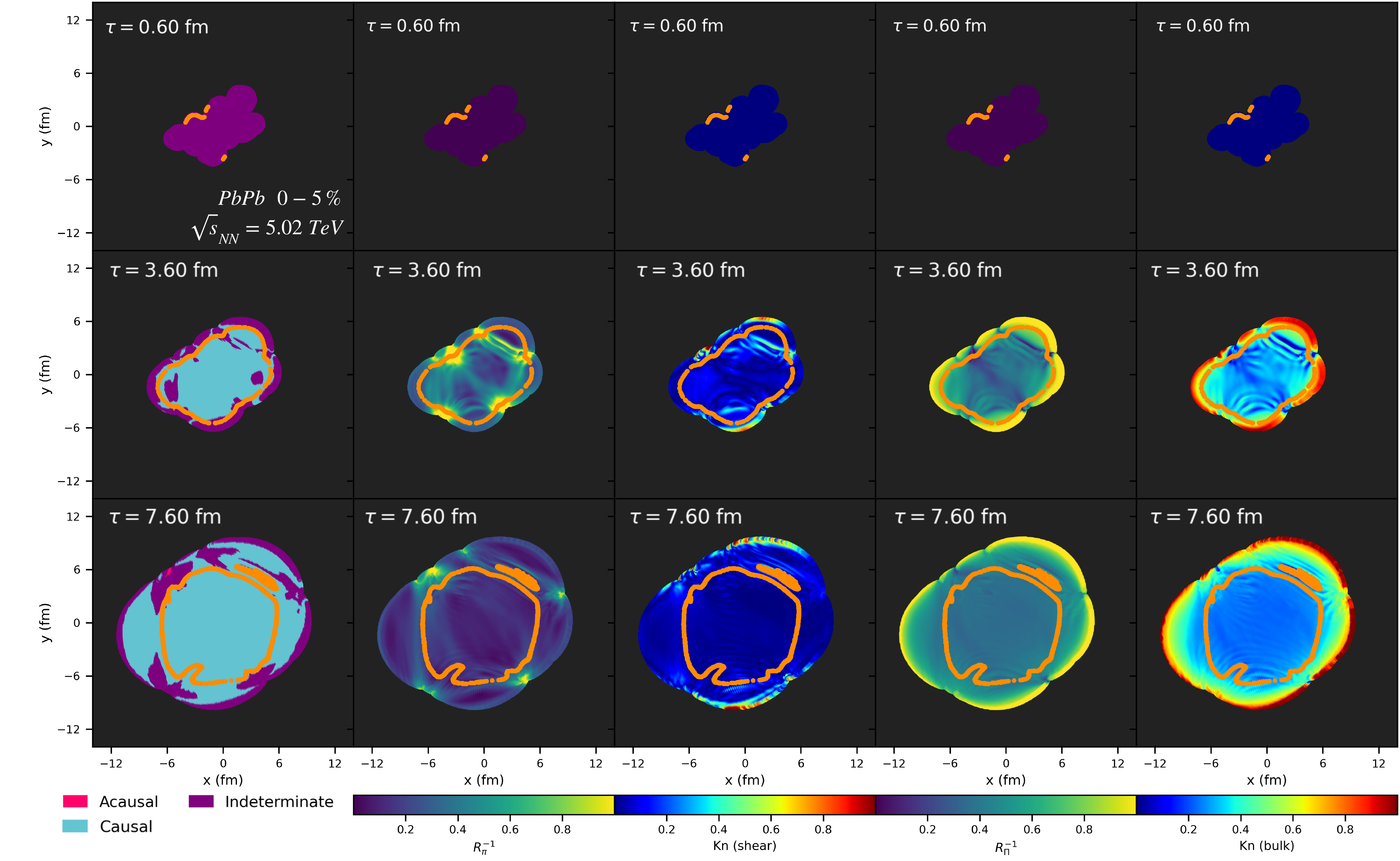}
    \caption{Results for \trento{}+\ccake{} run with minimal Israel--Stewart equations of motion. The columns from left to right show  the causality constraints, the inverse Reynolds number from shear, the Knudsen number from shear, the inverse Reynolds number from bulk, the Knudsen number from bulk. The rows show different times steps starting from early times on top, going to late times on the bottom. The freeze-out surface is plotted in orange. 
    }
    \label{fig:causality_IS}
\end{figure*}

Here we study the interplay between the causality conditions and the Knudsen and inverse Reynolds numbers. 
In Fig.~\ref{fig:causality_IS} we consider a Pb+Pb collision at 5.02 TeV with $\eta/s=0.16$ and $\zeta/s=0.08$ with a freeze-out at $\varepsilon \approx 0.26$ GeV/fm$^3$ that corresponds to $T=150$ MeV at zero chemical potential.  
For these simulations we use a \trento{} initial condition and varying the hydrodynamic equations of motion using  minimal Israel--Stewart  in Fig.~\ref{fig:causality_IS}. 
Here we only consider systems to $\vec{\mu}=0$. 
Unlike previous studies \cite{Plumberg:2021bme,Chiu:2021muk,ExTrEMe:2023nhy,Domingues:2024pom} we do not include a pre-equilibrium phase here but rather initialize only the energy density within \ccake{}.
The rows show the expansion over different times of a single event. The columns demonstrate different quantities that relate to the applicability of hydrodynamics. Starting from the farthest left column, we plot the causality constraints, then the inverse Reynolds number for shear viscosity, the Knudsen number for shear viscosity, the inverse Reynolds number for bulk viscosity, and the Knudsen number for bulk viscosity. For the causality constraints, the shaded regions are in red for acausal, blue for causal, and purple for indeterminate.  We plot the freeze-out surface in orange. 

We find that essentially all fluid cells begin as indeterminate but quickly become causal over time. A ring of indeterminate fluid cells remain around the edges across nearly all time steps, but most of those fluid cells are below the freeze-out surface. 
Only at very late time do we find some acausal fluid cells. 
We can then compare the causality results to the Knudsen and inverse Reynolds numbers (that should be large when fluid dynamics is not applicable and small when it is). 
At early times, these values are all very small since we do not include a pre-equilibrium phase such that it requires time for these values to build up. 

We find distinctive differences for shear vs. bulk viscosity in terms of the applicability of hydrodynamics. The shear viscosity Knudsen and inverse Reynolds numbers are largest in regimes of the largest changes in the gradients (i.e., between valleys between bumps in the energy density field) whereas the bulk viscosity Knudsen and inverse Reynolds numbers are largest at the edges of the system (almost entirely well below the freeze-out hypersurface). 
Comparing these to the causality constraints, we find that both contribute to the causality constraints and that there appears to be a correlation between the highest Knudsen and inverse Reynolds numbers with the indeterminate fluid cells for both shear and bulk viscosity. 

\subsection{Numerical solver: Smoothed particle hydrodynamics (SPH)}%
\label{Sec:Hydrodynamics:Numerics}
\noindent
Due to the high complexity of the relativistic hydrodynamics equations of motion, it is necessary to use numerical simulations to solve the equations.

Most numerical simulations of fluid dynamics for high energy collisions are Eulerian or grid based. In general, the SPH Lagrangian algorithms \cite{Rosswog:2009sr,Rosswog:2020kwm} have been the standard state of the art for astrophysical applications of fluid dynamics. The SPH desirable features are: i) conservation of energy-momentum and particle number is enforced by construction, ii) possibility to smooth over any undesirable degrees of freedom, and iii) robustness against geometry.

\subsubsection{SPH interpretation of a fluid element}%
\label{Sec:Hydrodynamics:Numerics:RelativisticSPH}
\noindent
To define the SPH method, let $A$ be any desired quantity, it can be written as
\begin{equation}
\label{dirac}
    A(\vec{r}) = \int_V  A(\vec{r '})\delta (\vec{r '}-\vec{r})\; \diff V,
\end{equation}
where $V$ is any volume that contains $r$ and $\delta(\vec{r})$ is the three-dimensional Dirac $\delta$ function. In SPH, the $\delta$ function is approximated by $W(\vec{r},h)$, called the kernel function, with specific properties:
\begin{align}
\label{kerprop}
\int_V W(\vec{r},h) \; \diff V &= 1 ,\\
\lim_{h \to 0} W(\vec{r},h) &= \delta(\vec{r}),
\end{align}
wherein as the smoothing length $h$ is decreased, $W$ gets closer to the Dirac $\delta$. Using the kernel, it is possible to calculate the integral Eq.~(\ref{dirac}) as
\begin{equation}
\label{kerint}
    A(\vec{r}) 	\approx \int_V  A(\vec{r '})W(\vec{r '}-\vec{r},h)\;d V.
\end{equation}
 Then, discretizing integral Eq.~(\ref{kerint}) as a sum over $N$ elements,
\begin{equation}
    \label{sph}
    A(\vec{r}) 	\approx \sum_{i=1}^{N} A(\vec{r}_i)W(\vec{r}_i-\vec{r},h)\Delta V_i  .
\end{equation}
The pieces in which the fluid is discretized are known as ``SPH particles", and by smoothing the fluid over the particle volume, we have a continuously differentiable representation of the state of the system at any given time-step.

Each SPH particle then evolves independently according to its own equations of motion, in a way which depends on the locally smoothed densities in its vicinity.
The kernel function, thus, acts to define the neighborhood over which the smoothed physical state of the system influences the subsequent evolution of a particle.

Let us now define $Q$ as a quantity in the fluid rest frame. The density of $Q$ is defined as
\begin{equation}
    q = \frac{\diff Q}{\diff V},
\end{equation}
where in the laboratory frame,
\begin{equation}
\label{refconv}
    q^* = q u^0\sqrt{-g},
\end{equation}
in which the superscript $*$ indicates that the quantity is measured in the laboratory frame. Now, using Eq.~\ref{sph}, it is possible to write
\begin{equation}
    q_a^* =  \sum_{b=1}^{N} q^*(\vec{r}_b)W(\vec{r}_a-\vec{r}_b,h)\Delta V^*_b .
\end{equation}
By defining $M$ as any conserved quantity in the system, it is possible to write
\begin{equation}
\label{quan_sph}
    \Delta V^*_j  = \frac{m_j}{\sigma^*_j},
\end{equation}
where $\sigma^*_j$ is the density of $M$ and $m_j$ is the $M$ fraction of particle $j$. Then, using the notation $q^*(\vec{r}_a) = q^*_a$,
\begin{equation}
    q_a^*=  \sum_{b=1}^{N} m_b \frac{q^*_b}{\sigma^*_b} W(\vec{r}_a-\vec{r}_b,h).
\end{equation}
Then for $\sigma^*$,
\begin{equation}
\label{sigmastar}
    \sigma_a^*=  \sum_{b=1}^{N} m_b  W(\vec{r}_a-\vec{r}_b,h),
\end{equation}
and
\begin{equation}
    M_\mathrm{tot} = \sum_{b=1}^{N} m _b .
\end{equation}
For relativistic hydrodynamics, an issue arises regarding the conserved quantity.
The natural choice for a conserved quantity in hydrodynamics, mass, cannot be used
anymore because it can be converted into energy during the fluid's evolution. The
system may be radiating large amount of energy as well, making total energy also
not a good choice. One could also think that the fluid may be carrying some quantum
number and use it as conserved quantity. However, often these quantum numbers can
be either positive or negative, implying that there are points in the system where
the density is vanishing. The entropy can work when there is no viscosity, but once one introduces dissipative quantities, entropy is not conserved anymore.

If we keep the number of particles fixed in time, then, it is a natural conserved quantity for both the dissipative and the ideal case. By doing this, the total conserved quantity becomes
\begin{equation}
    M_\mathrm{tot} = \sum_{b=1}^{N_\mathrm{SPH}} m _b = N_\mathrm{SPH},
\end{equation}
if we assign the same mass for all the particles, the only possible mass is
\begin{equation}
    m_b = 1 \;\;\; \forall b,
\end{equation}
and then the specific density,
\begin{equation}
    \sigma^*_b = \frac{1}{V^*_b},
\end{equation}
is just the inverse of the particle volume in the laboratory frame.

\subsubsection{Kernel function}
\noindent
There are several types of kernel functions $W(r, h)$ used in SPH. The one used here is the cubic spline kernel, where $r$ is the Euclidean distance between particles and $h$ is the smoothing length.
The kernel function is given by
\begin{equation}
W(r, h) =
\begin{cases}
\sigma_d \left( 1 - \frac{3}{2} q^2 + \frac{3}{4} q^3 \right) & \text{if } 0 \leq q \leq 1, \\
\sigma_d \frac{1}{4} (2 - q)^3 & \text{if } 1 < q \leq 2, \\
0 & \text{if } q > 2,
\end{cases}
\end{equation}
where $q = \frac{r}{h}$, and $\sigma_d$ is the normalization constant for dimension $d$ such that $\sigma_1$ = $\frac{2}{3h}$, $\sigma_2$ = $\frac{10}{7\pi h^2}$ and $\sigma_3$ = $\frac{1}{\pi h^3}$. The distances are defined by
\begin{equation}
    r_{ij} = \sqrt{\sum^D_d (x^d_i - x^d_j)^2}.
\end{equation}
The generalized gradient of the kernel function, required to calculate gradients of physical quantities during the hydrodynamic evolution, is:
\begin{equation}
\nabla W(r, h) = \frac{\sigma_d}{h} \cdot
\begin{cases}
\left( -3q + \frac{9}{4} q^2 \right) \frac{\mathbf{r}_i - \mathbf{r}_j}{r} & \text{if } 0 \leq q \leq 1, \\
-\frac{3}{4} (2 - q)^2 \frac{\mathbf{r}_i - \mathbf{r}_j}{r} & \text{if } 1 < q \leq 2, \\
0 & \text{if } q > 2.
\end{cases}
\end{equation}
The numerical accuracy of the SPH fluid description will depend on the number of the SPH particles and the smoothing length $h$, and convergence checks to see how these parameters affect the observables can be found at Eq.~(\ref{Sec:Observables:Tests}).

\subsubsection{Derivatives in relativistic SPH}%
\label{Sec:Hydrodynamics:Numerics:DerivativesRelativisticSPH}
\noindent
To obtain a partial spatial derivative it is only necessary to take a derivative of the kernel
\begin{equation}
\label{sphderivativerel}
    \partial_k q_a^*=  \sum_{b=1}^{N} m _b \frac{q^*_b}{\sigma^*_b} \partial_k W(\vec{r}_a-\vec{r}_b,h) =  \sum_{b=1}^{N} m_b \frac{q^*_b}{\sigma^*_b} \partial_k W_{ab}.
\end{equation}
There are  various ways to compute the spatial derivatives that will appear below, in addition to Eq.~(\ref{sphderivativerel}) \cite{willian}
by expanding $q^*_a$ as
\begin{equation}
\partial_k q^*_a = \frac{1}{(\sigma^*_a)^n} \partial_k\left[q_a^* (\sigma^*_a)^n \right] - n \frac{q_a^*}{\sigma_a^*} \partial_k \sigma^*_a.
\end{equation}
The first term can be written using the SPH sum as
\begin{equation}
    q^*_a (\sigma^*_a)^n  = \sum_{b=1}^{N} m _b q^*_b(\sigma^*_b)^{n-1} W(\vec{r}_a-\vec{r}_b,h),
\end{equation}
which allows us to obtain
\begin{equation}
\label{generalsphderivative}
    \partial_k q^* = \sum^N_{b=1} m_b \left[q^*_b \frac{(\sigma^*_a)^{n-1}}{(\sigma^*_b)^{n}} - n q^*_a \frac{(\sigma^*_a)^{n-1}}{(\sigma^*_b)^{n}}\right]\partial_k W_{ab}.
\end{equation}
The case where $n=0$ corresponds to the usual SPH derivative, for $n=-1$, one obtains a derivative where the permutation of particle $i$ by particle $j$ changes its sign, which is useful for the calculation of forces. The case of $n=1$ is the opposite, where permuting particles keep the sign of the derivative, which is useful when the derivative is contracted with another vector forming a scalar.

\subsubsection{Equations of motion}%
\label{Sec:Hydrodynamics:Numerics:EquationsOfMotion}
\noindent
To solve the equations of motion defined in Sec.~\ref{Sec:Hydrodynamics:Theory:EquationsOfMotion}. We need to evolve two thermodynamical quantities, the dissipative currents and position and velocity of particles.
In the SPH method, to recover the smooth solution, we can write for any quantity $A$:
\begin{equation}
A(\vec{r},\tau)=\sum_{b=1}^{N_\mathrm{SPH}}m_{b }\frac{1}{\gamma
_{b }\tau }\left( \frac{A }{\sigma }\right) _{b }W[\vec{r}-%
\vec{r}_{b}(\tau);h]\,.
\end{equation}
Therefore, the natural choice of dynamical variables in the SPH method is
\begin{equation}
\phi_a = \left\{ \vec{r},\vec{u},\left( \frac{s}{\sigma }%
\right),\left( \frac{\Pi }{\sigma }\right),\left( \frac{\pi^{\mu\nu} }{\sigma }\right) ,\left( \frac{n_q }{\sigma }\right),\left( \frac{J^\mu_q }{\sigma }\right)\right\}_{a },\label{dynamicalvariables}
\end{equation}
with $a=1,\ldots,N_\mathrm{SPH}$ indicates the SPH particle carrying these quantities. Since $\sigma = 1/V$, for the entropy, we interpret
\begin{equation}
    S_a= \left( \frac{s}{\sigma} \right)_a,
\end{equation}
as the total (extensive) entropy carried by each SPH particle, and a similar understanding can be applied for the other quantities.

Since SPH is a Lagrangian method, we need to calculate the total derivative
\begin{equation}
    \ddt{\phi_a}
     \label{eq:na}
\end{equation}
for each one of the dynamical variables.
When a charge diffusion is non-zero, it couples the charge conservation equations with all the other equations. The strategy is then to write the equations for $s/\sigma$, $q^\mu_q/\sigma$, $\Pi/\sigma$, and $\pi^{\mu \nu}/\sigma$ as
\begin{align}
\label{eq:genform}
    \ddt{} \left( \frac{s}{\sigma}\right)_a&= M^{ S }_{j ; a}\ddt{u^j} +  R^{ S }_{b ; a} \ddt{N_b} + F^S_{a},\\
   \ddt{} \left( \frac{\Pi }{\sigma }\right)_a&= M^{ \Pi }_{j ; a}\ddt{u^j} +  R^{ \Pi }_{b ; a} \ddt{N_b} + F^\Pi_{a},\\
    \ddt{} \left( \frac{J^\mu_q }{\sigma }\right)_{a } &= M^{\mu  ; J }_{j ; a}\ddt{u^j} +  R^{\mu   ; J }_{b ; a} \ddt{N_b} + F^{\nu ; J}_{a},\\
     \ddt{}\left( \frac{\pi^{\mu\nu} }{\sigma }\right)_a &= M^{\mu \nu  ; \pi }_{j ; a}\ddt{u^j} +  R^{\mu \nu  ; \pi }_{b ; a} \ddt{N_b} + F^{\mu \nu; \pi}_{a},
\end{align}
and in a similar way for vectors and scalars, or quantities that do not depend on the charge index. Then, we substitute the charge and flow equations, to write
\begin{align}
\label{eq:nbmatrix}
     R^{a}_{b}\ddt{N_b} &=  M^{a }_{j}\ddt{u^j} + F^{a}\\
     M^{i}_{j}\ddt{u^j} &=  R^{i }_{b}\ddt{N_b} + F^{i}.
\end{align}
 We invert the first equation and substitute it into the second
 \begin{equation}
      M^{i}_{j}\ddt{u^j} = R^i_b \left( (R^{a}_{b})^{-1} M^a_j \ddt{u^j} + (R^{a}_{b})^{-1} F_a\right) +F^i.
 \end{equation}
 Then we can solve for the four acceleration via
\begin{equation}
\label{eq:unugeneral}
\ddt{u^i} =\underbrace{\Big(M^{i}{}_{j}-R^{i}{}_{b}\,(R^{-1})^{b}{}_{a}\,M^{a}{}_{j}\Big)^{-1}}_{\displaystyle (\mathcal{M}^{i}{}_{j})^{-1}}
\;\underbrace{\Big(R^{j}{}_{b}\,(R^{-1})^{b}{}_{a}\,F^{a}+F^{j}\Big)}_{\displaystyle \mathcal{F}^{j}}.
\end{equation}
Once the four-acceleration is known, it can be substituted back in Eq.~(\ref{eq:nbmatrix}). These two quantities can be used to calculate the other equations. Therefore, the objective of the calculations is to determine the matrices $M, R$ and the vectors $F$.

Using both the first law of thermodynamics in Eq.~(\ref{eq:ideal_entropy}) and the conservation equation in Eq.~(\ref{eq:e_conservation}), the entropy production for the dissipative case when sources are present is
\begin{align*}
   T D s + \mu_a^q D n_a^q = u_\mu I^\mu - (\varepsilon+p+\Pi)\theta + \pi^{\mu\nu}\nabla_{\langle\mu\rangle} u_\nu\,.
   \label{eq:NOTideal_entropy}
\end{align*}
In this case, since the derivative $D$ is acting on Lorentz scalars, $D = u^\mu \partial_\mu$,
\begin{align*}
     u^0 T \ddt s + u^0 \mu_a \ddt  n_a = u_\mu I^\mu - (\varepsilon+p+\Pi)\theta + \pi^{\mu\nu}\nabla_{\langle\mu\rangle} u_\nu\,.
\end{align*}
We can use the chain rule to write the term proportional to the expansion rate as
a derivatives of $dS/dx^0$ and $dN_a/dx^0$, where $S = s/\sigma$ and $N_a = n_a/\sigma$.
From the chain rule we obtain additional terms, which exactly cancel the LHS. After rearranging the remaining terms, one obtains
\begin{align}
     u^0 \sigma T \ddt S + u^0 \sigma \mu_a \ddt{N_a} = u_\mu I^\mu - \Pi \theta + \pi^{\mu\nu}\nabla_{\langle\mu\rangle} u_\nu\,.
\end{align}

Since we want to work with derivatives of contravariant components of the
vector and since the metric is a constant from the point of view of
the covariant derivative, we rewrite it as
$\pi^{\mu\nu} \nabla_\mu u_\nu = \pi^\mu_{\,\nu} \nabla_\mu u^ \nu$.
We split this sum as a term which depends on $\nabla_0 u^ \nu$ and another
which depends on $\nabla_i u^\nu$. We write:
\begin{align*}
    \nabla_0 u^\nu = \ddt{u^\nu} -v^k\partial_k u^\nu + \Gamma^\nu_{0\sigma} u^ \sigma,
\end{align*}
and
\begin{align}
\label{eq:unu}
    \ddt{u^ \nu} = \left(\delta^\nu_k - \delta^\nu_0 \frac{u_k}{u_0}\right)\ddt{u^k} - \delta^\nu_0 \frac{u^\alpha u^\beta}{2 u_0}\ddt{g_{\alpha \beta}}.
\end{align}
We also expand the covariant derivative of the spatial components and after some
algebra, we obtain the final result
\begin{align}
    \pi^{\mu\nu} \nabla_{\mu} u_\nu &=
    \left(\pi^0_{\,j} - \pi^0_{\,0} \frac{u_j}{u_0}\right) \ddt{u^j}
    +\left(\pi^k_{\,\nu} - \pi^0_{\,\nu} v^k\right)\partial_k u^\nu
    +\pi^\mu_{\,\nu} \Gamma^\nu_{\mu\sigma} u^\sigma + \pi^0_0 \mathcal{G}_u \nonumber\\ &=  M^{\pi \,u}_j \ddt{u^j} +  F^{\pi \,u},
    \label{eq:shear_viscous_contraction}
\end{align}
with $M^{\pi \,u}_j$ defined as
\begin{align}
     M^{\pi \,u}_j = \left(\pi^0_{\,j} - \pi^0_{\,0} \frac{u_j}{u_0}\right),
\end{align}
and $F^{\pi \,u}$ as
\begin{align}
     F^{\pi \,u} = \left(\pi^k_{\,\nu} - \pi^0_{\,\nu} v^k\right)\partial_k u^\nu
    +\pi^\mu_{\,\nu} \Gamma^\nu_{\mu\sigma} u^\sigma + \pi^0_0 \mathcal{G}_u.
\end{align}

At last, we return to the entropy production expression from Eq.~(\ref{eq:genform}), inserting the above result and obtain
\begin{align}
\label{eq:dsdnrewr}
    \ddt S  = M^{S}_j \ddt{u^j} + F^{S} + R^S_a \ddt{N_a},
\end{align}
where
\begin{align}
    M^S_j & = - \left[\left(\pi^0_{\,0}-\Pi\right) \frac{u_j}{u_0} - \pi^0_{\,j}\right]\frac{1}{\sigma u^0T}, \\
    F^{S} & = \left[ \begin{aligned} &
    u_\mu j^\mu
    + \left(\pi^k_{\,\mu}-v^k\pi^0_{\,\mu}\right)\partial_k u^\mu
    - u^0\Pi \vec \nabla \cdot \vec v
         \\
        &
        + \pi^\sigma_{\,\mu} \Gamma^\mu_{\sigma\nu}u^\nu - \Pi u^0 \ddt{\ln \sqrt{-g}}
        + (\pi^0_{\,0}-\Pi)\mathcal{G}_u
        \end{aligned}\right]\frac{1}{u^0 \sigma T},\\
    R^S_a & = -\mu_a /T.
\end{align}
The same process can be applied to each one of the dynamical equations for variables to obtain the full set of $M,R$, and $F$'s necessary for the fluid evolution. 

\begin{figure} [ht!]
    \centering
    \includegraphics[width=0.75\linewidth]{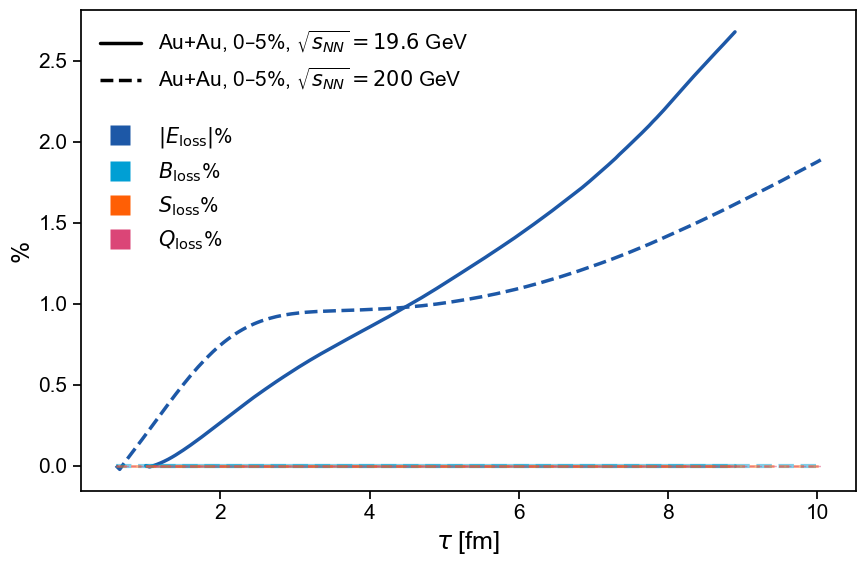}
    \caption{Conservation plots of energy and BSQ conserved charges for Au+Au $0$--$5\%$ centrality for $\sqrt{s_{NN}}$ = 19.6 and 200 GeV. }
    \label{fig:central_conservation}
\end{figure}

\subsubsection{Exact conservation}%
\label{Sec:Hydrodynamics:Numerics:ExactConservation}
\noindent
Even with the analytical checks performed in this section, we cannot test every single term in our equations of motion in 3+1 because some terms disappear in these highly symmetry systems. 
For instance, all the analytical checks are performed for conformal EoS such that $\varepsilon=3p$. 
In the limit of the conformal EoS, the trace anomaly has no trace, which implies that bulk viscosity is exactly zero. 
Thus, we do not have the possibility of performing analytical checks for any terms related to the bulk viscosity.

One method for testing even these complicated equations of motion is to check energy conservation. 
While energy non-conservation does not allow one to easily pin-point exactly where  a problem is, it can still indicate that there is a problem if energy conservation is strongly violated (note that it is typically conserved at the percent level or even significantly better than that).
Thus, in this section we check energy conservation. 

In addition to energy conservation, one might ask how well \ccake{} conserves BSQ charges as well. 
Due to its Lagrangian nature, a feature of the SPH method is that in the absence of dissipative currents, the entropy and charges are conserved exactly and naturally. This can be easily seen in the equations of motion, the equations are
\begin{align}
    \ddt{S}_i &= \ddt{s}_i + s_i \theta_i = 0,\\
    \ddt{N_X}_i &= \ddt{n_X}_i + (n_X)_i \theta_i=0,
\end{align}
where it is clear that the quantities are conserved exactly for each SPH particle.
The energy, on the other hand, is not exactly conserved, due to numerical errors in the integration method, but still, the energy conservation violation is generally expected to be very small if one solves SPH close to the continuum limit.

In Fig.~\ref{fig:central_conservation} the energy conservation for and BSQ charge conservation over time are shown with 3+1D simulations  for both $\sqrt{s_{NN}}=19.6$ GeV (top) and $\sqrt{s_{NN}}=200$ GeV (bottom).
We find (as expected) that \ccake{} has exact conservation of BSQ charges over time. We do find a small amount of energy loss over time at the order of the percent level. Overall, the lower beam energy has a slightly large energy loss but it not a significant difference. 
We note that 2+1D simulations have about an order of magnitude smaller energy loss (and sometimes even less). 

\subsection{Code validation: (semi)analytical checks}%
\label{Sec:Hydrodynamics:CodeValidation}
\noindent
Given the significant number of changes made from \ccake{}~1.0 to \ccake{}~2.0 between the incorporation of \cabana{}, the offline inverter of the EoS, the generalized coordinate system, and the extension from 2+1 into 3+1 dimensions, it was not guaranteed a prior that the equations of motion would accurately pass (semi) analytical checks. 
Thus, we include a variety of (semi) analytical checks that can determine if \ccake{}~2.0 is accurately solving the equations of motion. 

Here we use the following checks to ensure that \ccake{} is accurately solving the equations of motion, with the correct convergence in terms of both the grid size and the smooth scale $h$:
\begin{itemize}
    \item The Landau--Khalatnikov solution
\cite{Landau:1953gs,Landau:1955tlt,Belenkij:1955pgn,khalatnikov1954} tests the accuracy of \ccake{} when solved using Cartesian coordinates
\item The 2+1D BSQ Gubser solution  from \cite{Ingles:2025yrv} tests the validity of the offline inverter for the EoS
\item The 3+1D checks in hyperbolic coordinates from \cite{Bradley:2024jxq} tests the accuracy of our solutions in hyperbolic coordinates. 
\end{itemize}

Let us begin with the  Landau--Khalatnikov solution, which is a well-known model for the hydrodynamical evolution and can be used to test the longitudinal expansion in cartesian coordinates.
We consider a uniform distribution of matter of size $2l$ in the z axis with zero velocity. The solution for early times is the relativistic Riemann Simple Wave, a rarefaction wave that propagates with a speed of $c_s$. Once the simple wave reaches the center of the fluid, the solution must be matched to the Landau--Khalatnikov solution  \cite{Wong:2014mxa}.

Defining \begin{equation}
\label{psi2}
    \psi =\ln{\left[\left(\frac{\varepsilon}{\varepsilon_0}\right)^{\frac{c_s^2}{1+c_s^2}}\right]},
\end{equation} and assuming $\varepsilon =c_s^2 P$ we can write the relativistic simple wave solution as
\begin{align}
\label{simplewave:xt}
    \frac{x}{t} &= \frac{\tanh(-\psi / c_s) - c_s}{1 - \tanh(-\psi / c_s) c_s}, \\
\label{simplewave:v}
     v &= \tanh{\left(-\frac{\psi}{c_s}\right)},
\end{align}
where $x = z-l$.

Then for each point $z$, for a given time $t$, we invert the equation to find $\varepsilon(z,t)$ and use it to calculate the velocity $v(z,t)$. This is valid for all the fluid until the wave reaches the center of the fluid at $t_l = l/c_s$. After this, the Landau--Khalatnikov solution is given by
\begin{equation}
\label{landaukhalatnikov}
    \begin{cases}
        x = e^{-\psi}\left(\frac{\partial \chi}{\partial \psi} \sinh{y}+\frac{\partial \chi}{\partial y}\cosh{y}\right)\\
    t = e^{-\psi}\left(\frac{\partial \chi}{\partial \psi} \cosh{y}+\frac{\partial \chi}{\partial y}\sinh{y}\right)
    \end{cases},
\end{equation}    
where $y=\arctanh v$ and 
\begin{equation}
    \label{chi}
    \chi(\psi,y) = -l\sqrt{3}e^\psi \int_{y/\sqrt{3}}^{-\psi} e^{2\psi'}I_0\left[\sqrt{\psi'-\frac{y^2}{3}}\right]d\psi',
\end{equation}
with $I_0$ being the modified Bessel function of the first kind. Then, this system is solved numerically following the method described in \cite{Wong:2014mxa}. 

The Landau--Khalatnikov solution is only valid for the more central portions of the fluid, and the simple wave must be used at the edges. Therefore, the full solution is obtained by combining both.
For $t \leq l/c_s$, the solution (\ref{simplewave:xt},\ref{simplewave:v}) describes the whole fluid evolution, for $t > l/c_s$, we apply (\ref{landaukhalatnikov}) where $y < -\psi/c_s$ , and the simple wave (\ref{simplewave:xt},\ref{simplewave:v}) where $y \geq -\psi/c_s $.

\begin{figure*}[ht!]
    \centering
    \includegraphics[width=\linewidth]{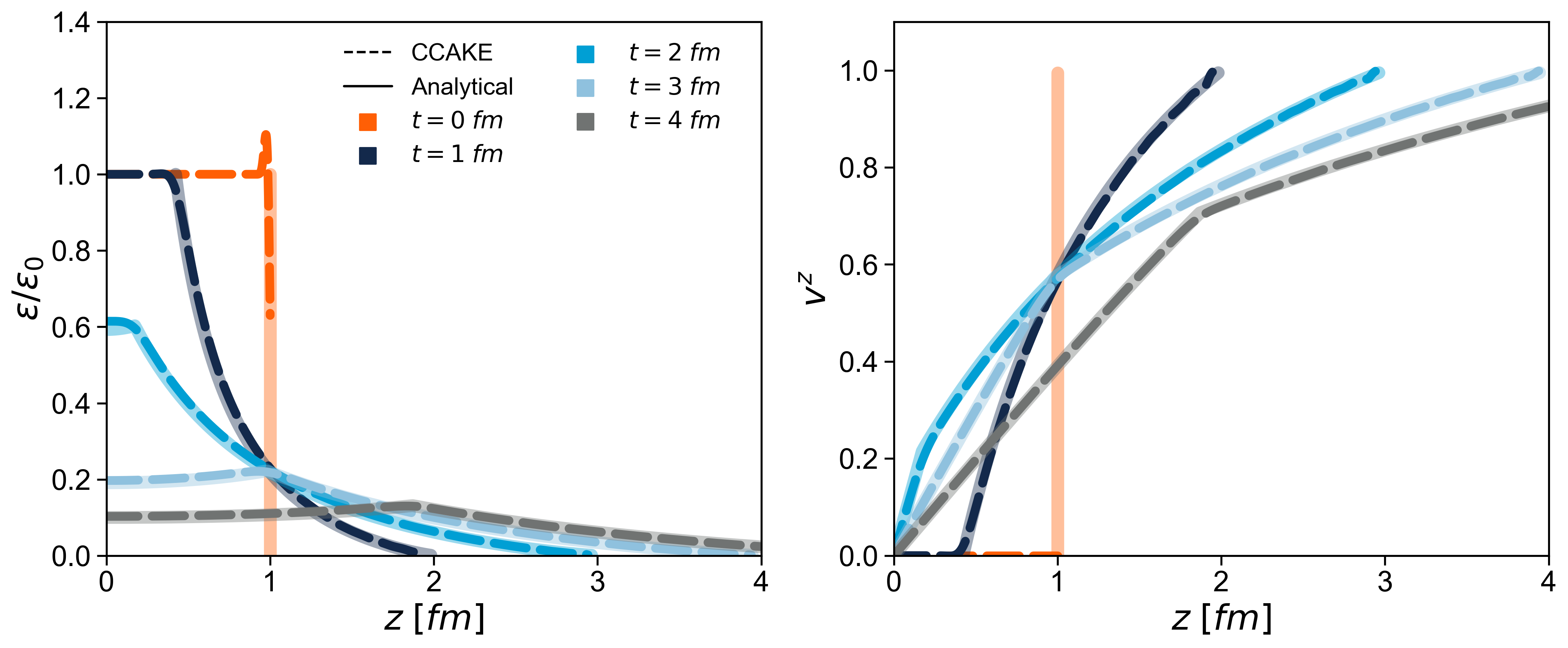}
    \caption{Simulated longitudinal cartesian hydrodynamics compared to Landau--Khalatnikov solution, here we have $20,010$ SPH particles and $h=0.0012$ fm, $l=1$ fm. The \ccake{} results are represented by the dashed curves and the analytical ones by the solid curves.}
    \label{fig:landau}
\end{figure*}

\begin{figure*}
    \centering
    \includegraphics[width=\linewidth]{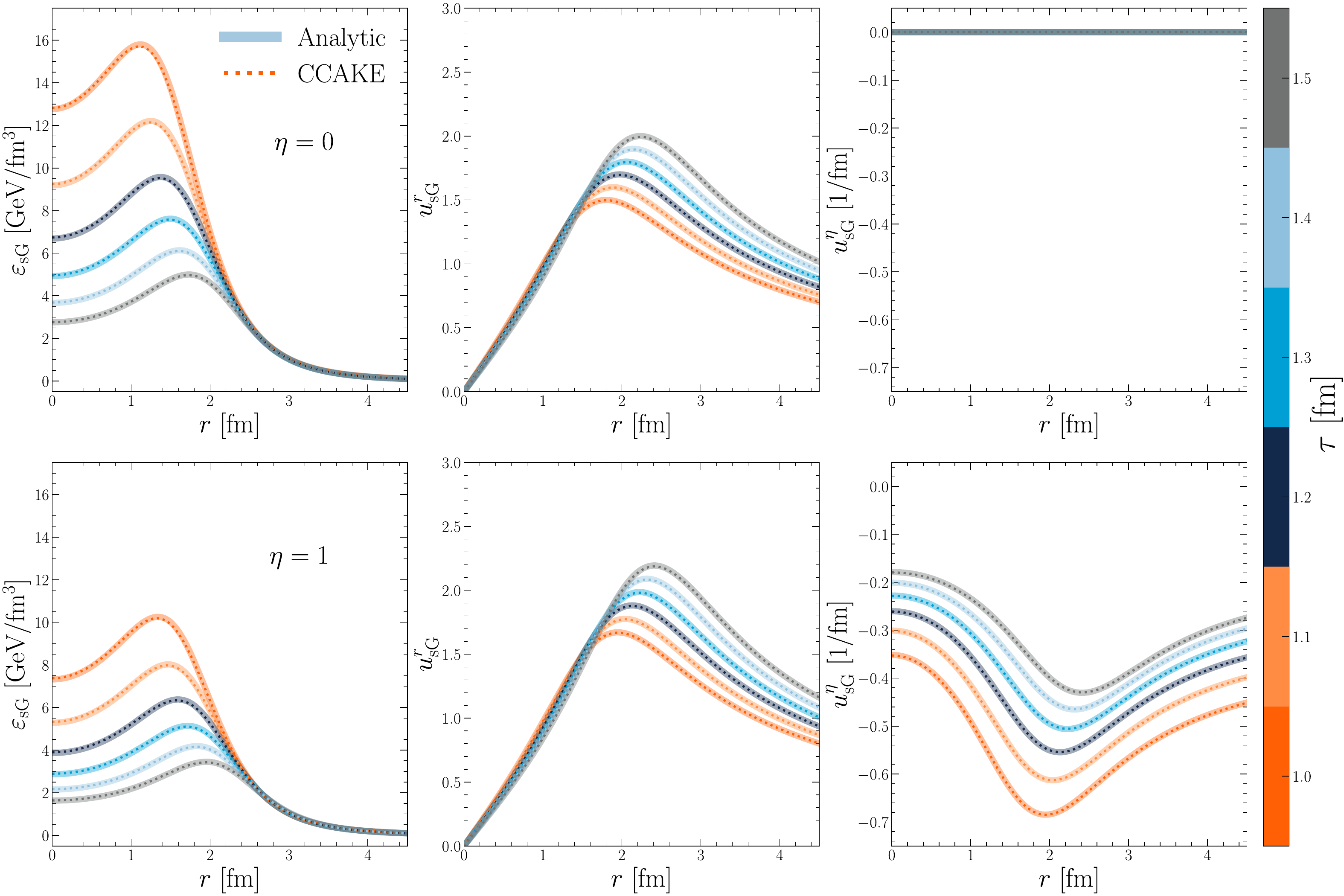}
\caption{Comparison of the ideal ``shifted Gubser" solution \cite{Bradley:2024jxq} to \ccake{}~2.0.  From left to right, we show the energy density $\varepsilon_\text{sG}$, the radial flow velocity $u^r_\text{sG}$, and the rapidity component of the velocity $u^\eta_\text{sG}$, for $\eta = 0$ (top) and $\eta = 1$ (bottom).}
    \label{Figure-shiftedGubserSolution}
\end{figure*}
The results of the simulation can be found in Fig.~\ref{fig:landau}. It is possible to see that SPH code has an excellent agreement with the Landau--Khalatnikov solution, but it has some difficulty reproducing the matching region where there is a peak. The initial time step also have a mismatch caused by the fact that the simple wave solution is simply describing only the ``edge'' of the fluid, before the wave start to propagate.

Analytical and semi-analytical solutions exist in 2+1D for the highly symmetrized Gubser flow. The original Gubser flow can be found in \cite{Gubser:2010ze,Gubser:2010ui} where they were extended to relativistic viscous hydrodynamics solutions and benchmarked against hydrodynamics for the first time in \cite{Marrochio:2013wla}. Then Gubser flow was extended to finite baryon densities (but as an ideal fluid) in \cite{Denicol:2018wdp} or as a fluid with shear viscosity but without evolution of the chemical potentials \cite{Du:2019obx}. 
Only in \cite{Ingles:2025yrv} was a relativistic viscous hydrodynamic solution fully coupled to BSQ conserved charges considered, which is what we will use here to benchmark \ccake{}~2.0.

In Gubser flow, it is possible to use different types of conformal EoS. Here we use what was defined as \textsc{EoS2} in \cite{Ingles:2025yrv} that is the same as our conformal EoS defined in Sec.~\ref{Sec:Hydrodynamics:Theory:EquationOfState}.
Then, the following equations of motion are obtained for the temperature profile, chemical potentials, and shear stress tensor:
\begin{subequations}
\label{eq:EoS2-evol}
\begin{align}
    \frac{1}{\hat T} \frac{d \hat T}{d \rho}
    &=
    \left[
        \frac{1}{3}
        \left(
            \hat{\bar \pi} - 2
        \right)
        +
        \hat{\bar \pi}
        \sum_Y
        \left(\frac{T_\ast}{\mu_\ast}\frac{\hat\mu_Y}{\hat T}\right)^2
    \right]
    \tanh\rho\,,
    \\
    \frac{1}{\hat \mu_Y}
    \frac{d \hat\mu_Y}{d\rho}
    &=
    -\frac{2}{3}
    (1 + \hat{\bar \pi})\tanh \rho,
    \label{eq:EoS1-evol-mu}\\
     \frac{d\hat{\bar \pi}}{d\rho}
        &=
        \frac{4}{3\mathcal C}\tanh{\rho}
        -
        \frac{\hat{\bar\pi}}{\tau_R}
        -
         \frac{4}{3}\hat{\bar \pi}^2\tanh{\rho},
\end{align}
\end{subequations}
where  $\hat \pi \equiv \hat \pi^\eta_\eta$, $\hat{\bar \pi} \equiv \hat \pi / (\hat{\mathcal E} + \hat{\mathcal P})$, and $\hat \theta = -3\hat\sigma^\eta_\eta = 2\tanh\rho$. Here $T_*$ and $\mu_*$ are just constants, which we choose to be $T_*=\mu_*=1$ GeV.

As recently pointed out in \cite{Shi:2022iyb}, any solution of the hydrodynamic equations of motion can used to generate an infinite number of new solutions by applying to it a transformation consistent with those equations of motion.  This insight can be used to convert the ideal 2+1D Gubser solution (which depends only on $\tau$ and $r$) into a 3+1D solution with non-trivial dependence on $\tau$, $r$, and $\eta$ \cite{Bradley:2024jxq}.  This is specifically done by implementing a ``temporal shift," i.e., a constant translation of the Gubser solution in Minkowski time $t$: $t \to t' = t + t_0$.  In Milne coordinates, the transformation is given by
\begin{align}
    \tau \to \tau'\!\l( \tau, \eta, r \r) &= \sqrt{\tau^2 + 2 t_0 \tau \cosh \eta + t_0^2}, \label{tau_prime} \\
    \eta \to \eta'\!\l( \tau, \eta, r \r) &= \arctanh\l(\frac{\tau \sinh \eta}{\tau \cosh \eta + t_0}\r), \label{eta_prime} \\
    r \to r'\!\l( \tau, \eta, r \r) &= r. \label{r_prime}
\end{align}

The original Gubser viscous solution is most conveniently written in terms of the temperature $T_\text{G}$, where the subscript `G' denotes `Gubser', $T_\text{G}$ is related to the energy density by $\varepsilon_\text{G} \equiv f_* T_\text{G}^4 $ (where $f_* \equiv 11$) and is defined by \cite{Gubser:2010ze, Gubser:2010ui} 
\begin{align}
    T_\text{G}(\tau, r) &= \frac{\hbar c}{f_*^{1/4} \tau} \hat{T}\!\l(\rho(\tau, r)\r), \label{vGubser_T} \\
    \hat{T}(\rho) &= \frac{\hat{T}_0}{\l(\cosh \rho\r)^{2/3}}\l( 1 + \frac{H_0}{9 \hat{T}_0} \l( \sinh\rho \r)^3 {_2F_1}\l( \frac{3}{2}, \frac{7}{6}, \frac{5}{2}; -\sinh^2\rho \r) \r), \label{vGubser_That}
\end{align}
the variable $\rho$ is defined by
\begin{equation}
\sinh \rho = \frac{q^2\l( \tau^2 - r^2 \r) - 1}{2 q \tau}, \label{rho_definition}
\end{equation}
and $H_0 \propto \eta/s$.

The flow $u_\text{G}^\mu$ is defined by
\begin{alignat}{3}
    u_\text{G}^\tau &= \frac{1 + q^2\l( \tau^2 + r^2 \r)}{\sqrt{1 + 2q^2\l( \tau^2 + r^2 \r) + q^4\l( \tau^2 - r^2 \r)^2}} \label{Gubser_utau}, \\
    u_\text{G}^r &= \frac{2 q^2\ r \tau}{\sqrt{1 + 2q^2\l( \tau^2 + r^2 \r) + q^4\l( \tau^2 - r^2 \r)^2}}. \label{Gubser_ur}
\end{alignat}
Gubser's complete viscous solution for $T_\text{G}$ and $u_\text{G}^\mu$ is thus given by Eqs.~\eqref{vGubser_T}, \eqref{Gubser_utau}, and \eqref{Gubser_ur}.

The shifted version of Gubser's solution (denoted by the subscript `sG') is then obtained by applying the transformation Eqs.~\eqref{tau_prime}-\eqref{r_prime} to the solution Eqs.~\eqref{vGubser_T}-\eqref{Gubser_ur}.  The result is \cite{Bradley:2024jxq}
\begin{align}
    T_\text{sG}(\tau, r, \eta) &= T_\text{G}(\tau', r), \label{shifted_viscous_Gubser_T} \\
    \varepsilon_\text{sG}(\tau, r, \eta) &= \varepsilon_\text{G}(\tau', r), \label{shifted_viscous_Gubser_e}
\end{align}
where $\tau'$ is defined by \eqref{tau_prime}, and the corresponding flow profile is
\begin{align}
    u_\text{sG}^r(\tau, r, \eta) &= u_\text{G}^{r'}\l( \tau', r \r), \label{shifted_viscous_Gubser_ur} \\
    u_\text{sG}^\tau(\tau, r, \eta) &= u_\text{G}^{\tau'}\l( \tau', r \r) \l(\frac{\tau + t_0 \cosh \eta}{\sqrt{\tau^2 + 2 t_0 \tau \cosh \eta + t_0^2}} \r), \label{shifted_viscous_Gubser_utau} \\
    u_\text{sG}^\eta(\tau, r, \eta) &= -u_\text{G}^{\tau'}\l( \tau', r \r) \l(\frac{t_0 \sinh \eta}{\tau \sqrt{\tau^2 + 2 t_0 \tau \cosh \eta + t_0^2}} \r), \label{shifted_viscous_Gubser_ueta}
\end{align}
with such a method we obtain an analytical solution in 3+1D in hyperbolic coordinates that will allow us to benchmark \ccake{}~2.0. 

In Fig.~\ref{Figure-shiftedGubserSolution} we present the exact shifted Gubser analytical solution in 3+1D compared to \ccake{}~2.0. 
In the figure we used the temporal shift parameter $t_0 = 0.5$ fm, an initial time of $\tau_0$ = 1 fm, and $q = 1$ fm$^{-1}$.
We show results for ideal ($H_0 = 0$) evolution only, with $\hat{T}_0$ adjusted so that the energy density $\varepsilon_\text{sG} = 80$ GeV$/\text{fm}^3$ at the point $(\tau,r,\eta)=(0.5$ fm, 0, 0).
Our initial grid spans $[-5\text{ fm},+5\text{ fm}]$ in the $x$ and $y$ directions (with grid spacings $\Delta x_{x} = \Delta x_{y} = 0.05$ fm), and $[-2,+2]$ in the $\eta$ direction (with $\Delta x_{\eta} = 0.025$).
Additionally, we have chosen a timestep of $d\tau = 0.001$ fm and a smoothing scale $h = 0.1$ fm, which corresponds to roughly $6\times 10^6$ SPH particles in the simulation.
Overall, we find a nearly perfect agreement between the analytical solution and \ccake{}~2.0, showing that we accurately solve the 3+1D equations of motion. 

\subsection{Convergence Tests}%
\label{Sec:Observables:Tests}
\noindent
With SPH, the smoothing scale $h$ sets the lowest, resolvable scale of the fluid dynamic simulations. 
However, within that scale $h$, one must \emph{oversample} the fluid with a large enough number of SPH particle in order to accurately determine the hydrodynamics variables. 
Thus, studies of convergence must take into account the total number of SPH particles $N_\mathrm{SPH}$ as well as the number of nearest neighbors $N_{nb}$. 
These quantities are  interconnected because a finer initial grid size increases both $N_\mathrm{SPH}$ as well as $N_{nb}$. Furthermore, a large $h$  increases $N_{nb}$ because it increases the range at which the code finds more nearest neighbors. Astrophysical studies often determine $h$ dynamically based on a fixed $N_{nb}$, but such an approach (known as adaptive-SPH \cite{Monaghan:1992rr,Shapiro:1995we,Owen:1995qv,Liebendoerfer:2001gu,Rosswog:2009sr}) that requires new terms in the gradient of the kernel function that we have not yet derived for our coordinate systems such that we leave this for a future work. 
In \cite{Zhu:2014qka} it was shown that a large $N_{nb}\gtrsim 1000$ for 3+1D astrophysical simulations is necessary for convergence to continuum solutions (although modern astrophysical codes often use closer to $N_{nb}\gtrsim 300$ \cite{Rosswog:2020kwm}).  However, their systems are significantly different (larger total system that live significantly longer) than heavy-ion collisions such that we must test what is the necessary $N_{nb}$ for convergence in heavy-ion observables.

We always begin our initial condition on a fixed grid.
We calculate the total volume of our system for the initial condition:
\begin{equation}
    V_\mathrm{sys}=\prod_{i=x,y,\dots}^D\left(X_{i,\,\mathrm{max}}-X_{i,\,\mathrm{min}}\right)
\end{equation}
where $X_{i,\,\mathrm{max}}-X_{i,\,\mathrm{min}}$ is the range of the initial grid size in a given dimension (for heavy-ion collisions $X_{i,\,\mathrm{max}}\sim 10$ fm and $-X_{i,\,\mathrm{min}}\sim -10$ fm).
Then we can estimate our total number of SPH particles $N_\mathrm{SPH}$ as the $V_\mathrm{sys}$ times the number density of SPH particle in each direction $n_\mathrm{SPH}$.  Since the initial condition is a grid, $n_\mathrm{SPH}$ is just inverse of the step size of each dimension in the grid multiplied together, i.e.,
\begin{equation}
N_\mathrm{SPH}^\mathrm{est}=n_\mathrm{SPH} V_\mathrm{sys}=\prod_{i=x,y,\dots}^D\frac{X_{i,\,\mathrm{max}}-X_{i,\,\mathrm{min}}}{\Delta x_i}
\end{equation}
where $\Delta x_i$ is the step size in the $i$th dimension. 
To give an example, for 2+1D initial conditions that start at a fixed $\tau_0$ with a grid of $\Delta x_x=\Delta x_y=0.06$ fm and a range of $X_{x,\,\mathrm{max}}-X_{x,\,\mathrm{min}}=X_{y,\,\mathrm{max}}-X_{y,\,\mathrm{min}}=24$ fm, we obtain $N_\mathrm{SPH}^\mathrm{est}=40,000$. 
Now, the caveat here is that within our hydrodynamic simulations, we only run SPH particles that have more energy density than some very small threshold $\varepsilon_\mathrm{min}$ such that our actual number of $N_\mathrm{SPH}< N_\mathrm{SPH}^\mathrm{est}$ because our initial conditions are elliptical (not a square) so there are always points in our initial conditions that have vanishing or nearly vanishing energy densities. 

Next we can determine the number density of SPH ``cells.''  Each SPH cell has a radius of $r=2h$ such that their entire length in a given dimension is $w=4h$. 
Then in $D$ dimensions this leads to a \emph{cubic} volume per SPH cell of  $V_h^\mathrm{cub}\equiv \left(4h\right)^D$ where the number density of SPH cells is just $n_h=1/V_h^\mathrm{cub}$ since there is one cell in each volume. 
Our number of SPH cells within a cubic volume is:
\begin{eqnarray}
    N_h^\mathrm{cub}&=&n_h V_\mathrm{sys}\\
    &=&\frac{\prod_{i=x,y,\dots}^D\left(X_{i,\,\mathrm{max}}-X_{i,\,\mathrm{min}}\right)}{\left(4h\right)^D}.
\end{eqnarray}
Now we can estimate our number of nearest neighbors $N_{nb}^\mathrm{cub}$ is our number of SPH particles within a cubic volume of $V_h^\mathrm{cub}$, i.e., 
\begin{eqnarray}
    N_{nb}^{\mathrm{cub},\,\mathrm{est}}&=&\frac{N_\mathrm{SPH}}{N_h^\mathrm{cub}}\\
    &=&\frac{\left(4h\right)^D}{\prod_{i=x,y,\dots}^D \Delta x_i}.
\end{eqnarray}
However, SPH fluids cells are \emph{Not} cubes but rather spheres. Thus, we now need to calculate our number density of nearest neighbors within a cubic volume $n_{nb}^\mathrm{cub}=N_{nb}^{\mathrm{cub},\,\mathrm{est}}/V_h^\mathrm{cub}$ and then multiple it by the actual volume of an SPH fluid cell in 3+1D is $V_\mathrm{SPH}=4/3\, \pi (2h)^3$ or in 2+1D it is an area $A_\mathrm{SPH}= \pi (2h)^2$, i.e.,
\begin{eqnarray}
    N_{nb}^\mathrm{est}&\equiv &n_{nb}^\mathrm{cub}V_\mathrm{SPH}\\
    &=&\frac{4}{3}\frac{\pi (2h)^3}{ \Delta x_x \Delta x_y \Delta x_z},
\end{eqnarray}
where we work it out for the case of 3+1D.

That provides us with an estimated $N_{nb}^\mathrm{est}$ for a given initial condition.
After the initial time step, the SPH particles will move with the system such that these estimates no longer hold. Thus, at late times one must calculate the actual number of nearest neighbors that surrounds some point $\vec{r}$, i.e., 
\begin{equation}
    N_{nb}(\vec{r})=\sum_{\alpha}^{N_\mathrm{SPH}}\Theta(2h-|\vec{r}_\alpha-\vec{r}|),
\end{equation}
which will vary with $\vec{r}$.

For 2+1D simulations, different types of convergence tests have been performed for the \vusphydro{} code. First in \cite{Noronha-Hostler:2013gga} $h$ was held constant while $N_\mathrm{SPH}$ was systematically increased (thus $N_{nb}$ was also increased in this process). It is a bit challenging to compare $N_{nb}$ in 2+1D to 3+1D simulations since one picks up a factor of 
\begin{equation}
    N_{nb}^{2D}=\frac{4h}{\Delta z}N_{nb}^{3D}
\end{equation} 
where for a typical heavy-ion simulation $\frac{4h}{\Delta z}= 20$ when $h=0.3$ fm and $\Delta z=0.06$ fm. 
With this caveat in mind, \cite{Noronha-Hostler:2013gga} found for 2+1D simulations that $N_{nb}^{2D}\gtrsim 400$ where if we assume the same grid size and $h$ for the $z$ direction as well (i.e., $h=0.3$ fm and $\Delta z=0.06$ fm), then we obtain $N_{nb}^{3D}\gtrsim 8000$. 
However, given the astrophysical results that $N_{nb}^{3D}\gtrsim 1000$ seems reasonable, it may be that we do not need quite as fine of a grid for 3+1D simulations, which must be checked now. 

Before arriving at our convergence tests here, we also note that in \cite{Noronha-Hostler:2015coa} another 2+1D study was performed where the grid size was kept fixed while $h$ was varied. 
However, it was pointed out in that work that a larger $h$ smooths out small scale structure in the initial condition up to the scale of $h$ such that too large of an $h$ may loose important information from the initial condition.
Thus, it was recommended to use as small of an $h$ as possible while increasing the grid size instead. 
Furthermore, another challenge of a large $h$ is that it significantly slows down the hydrodynamic code as well.

In this work, we perform our convergence tests with a focus on $N_{nb}^\mathrm{est}$ while holding $h$ fixed at either $h=\left\{0.3,0.4\right\}$ fm. One challenge with heavy-ion collisions is that the dynamics in the transverse plan may require a different choice in the grid size than in the rapidity or longitudinal directions (depending on the choice of coordinates).
Here we will always fix $\Delta x_x=\Delta x_y$  but will allow $\Delta x_z$ or $\Delta x_\eta$ to vary in our convergence tests.
For simplicity's sake then our nearest neighbor estimate becomes:
\begin{eqnarray}
    N_{nb}^\mathrm{est}&=&\frac{\left(4h\right)^3}{\left(\Delta x_x\right)^2 \Delta x_z},
\end{eqnarray}
where $\Delta x_z$ may instead be $\Delta x_\eta$, depending on the coordinate system. 
Here we also find it useful to define our 2D and 3D nearest neighbors, i.e.,
\begin{eqnarray}
    N_{nb}^{2D}&\equiv &\frac{\left(4h\right)^2}{\left(\Delta x_x\right)^2 }, \\
     N_{nb}^{3D}&\equiv &\frac{4h}{\Delta x_z}  N_{nb}^{2D},\\
     \frac{N_{nb}^{3D}}{N_{nb}^{2D}} &= & \frac{4h}{\Delta x_z}, 
\end{eqnarray}
since we have already performed convergence tests on the transverse plane and have a decent understanding of the nearest neighbors required there. 
One can imagine different choices of $\Delta x_z$ vs. $\Delta x_x$ that could lead to degenerate solutions of $N_{nb}^{3D}$. 
However, they may not lead to equivalent results for the convergence test because generally the transverse plane has more structure than the rapidity direction (at least for large $\sqrt{s_{NN}}$. 
Thus, we expect that our results from the convergence tests will find that $\Delta x_z\geq \Delta x_x$ (and not the opposite). 
\begin{table}[ht!]
\centering
\begin{tabular}{c c c c c}
\toprule
\multicolumn{5}{c}{$\sqrt{s_{NN}}=200~\mathrm{GeV}$} \\
\midrule
$N_{nb}^{\mathrm{est}}$ & $h$ [fm] & $\Delta x_x$ [fm] & $\Delta x_z$ [fm] & Converged? \\
1728 & 0.3 & 0.1 & 0.1 & yes \\
\midrule[1.2pt]
\multicolumn{5}{c}{$\sqrt{s_{NN}}=19.6~\mathrm{GeV}$} \\
\midrule
$N_{nb}^{\mathrm{est}}$ & $h$ [fm] & $\Delta x_x$ [fm] & $\Delta x_z$ [fm] & Converged? \\
1152 & 0.3 & 0.1 & 0.15 & yes \\
576 & 0.3 & 0.1 & 0.3 & for $v_2,v_3$ \\
\bottomrule
\end{tabular}
\caption{Estimated number of nearest neighbors for a given initial condition. We note that at lower beam energies due to long run time, one may decide to run a coarser grid that provides reasonable results for $v_2$ and $v_3$ but not for higher harmonics.}
\label{tab:neighbours}
\end{table}

In Tab.\ \ref{tab:neighbours} we run through the different combinations of $h,\Delta x_x,\Delta x_z$ that we have checked that led to different values of $N_{nb}^\mathrm{est}$ at two different beam energies of $\sqrt{s_{NN}}=200 $ GeV and $\sqrt{s_{NN}}=19.6$ GeV. In this paper, we focus both on internal convergence tests like energy conservation and internal eccentricity calculations as well as final state flow observables. 
The internal eccentricities that we calculate are:
\begin{eqnarray}\label{eqn:eccentricities}
    \varepsilon_{2,x}|_{\eta} &\equiv& 
    \sqrt{
        \frac{
            \left\langle x^2 - y^2 \right\rangle^2_{\varepsilon \gamma} 
            + \left\langle 2xy \right\rangle^2_{\varepsilon \gamma}
        }{
            \left\langle x^2 + y^2 \right\rangle^2_{\varepsilon \gamma}
        }
    }, \\
    \varepsilon_{2,p}|_{\eta} &\equiv& 
    \sqrt{
        \frac{
            \left\langle T^{xx} - T^{yy} \right\rangle^2_1 
            + \left\langle 2T^{xy} \right\rangle^2_1
        }{
            \left\langle T^{xx} + T^{yy} \right\rangle^2_1
        }
    }
\end{eqnarray}
where $\varepsilon_{2,x}$ and $\varepsilon_{2,k}$ are the spatial and momentum eccentricities, respectively. Here, the Lorentz factor is
\begin{equation}
    \gamma = \sqrt{1 - u_x^2 - u_y^2 - \frac{u_\eta^2}{\tau^2}}
\end{equation}
and 
\begin{equation}
    \left\langle f(x,y) \right\rangle_w = \frac{\int dx\,dy\,w(x,y)f(x,y)}{\int dx\,dy\,w(x,y)}
\end{equation}
and we hold $\eta=const.$ within a small rapidity cut.
We calculate these eccentricities across the transverse plane for different rapidity cuts. 

\subsubsection{Convergence of SPH particle number / grid spacing}%
\label{Sec:Observables:Tests:SPHParticleGrid}
\noindent
At top RHIC and LHC energies, i.e.,  $\sqrt{s_{NN}}=200$ GeV to $5.02$ TeV, previous convergence tests for the transverse plane found that $\Delta x_{x}=\Delta x_{y}=0.06$ fm provided a reasonable grid size to reproduce $v_1$--$v_6$ up to the percent level accuracy (higher harmonics are generally more sensitive to the grid size) in the original \vusphydro{} \cite{Noronha-Hostler:2013gga}. 
These tests were performed for a variety of 2+1 initial conditions \trento, \mcglauber{}, \mckln{}, \ipglasma{}. 
However, we do not yet have intuition on the correct grid size for $\Delta x_{\eta}$ and also even what the transverse plane grid size should be for 3+1D simulations and as $\sqrt{s_{NN}}$ is varied to low beam energies. 
Furthermore, with the improvements of the code from \vusphydro{} to \ccake{}~2.0, other sources of numerical error have been improved and we no include higher-order terms in the equation of motion such that such a fine grid size may no longer be needed. 
Thus, we test independently varying both $\Delta x_{x}$ (we always assume $\Delta x_{x}=\Delta x_{y}$ so we do not mention $\Delta x_{y}$ further) and $\Delta x_{\eta}$, while keeping $h$ fixed.

For these these test we require a 3+1D initial state code that can be run across a wide range of beam energies. Thus, we use \ampt{} which is open-source, in 3+1D, has all three conserved charges, and can be run at a wide-range of beam energies to perform a convergence test. The added advantage of \ampt{} is that it is originally in Cartesian coordinates such that we will also use it to compare Cartesian vs. hyperbolic coordinates. 
Here we will test two different beam energies $\sqrt{s_{NN}}=200$ GeV and $\sqrt{s_{NN}}=19.6$ GeV for Au+Au collisions. At $\sqrt{s_{NN}}=200$ GeV we use only the energy density profile, but at $\sqrt{s_{NN}}=19.6$ GeV we incorporate the conserved charges as well. When running we use  $\eta/s = 0.08$ and in 3+1D we must use MC sampling of the freeze-out surface but since we do not what that to influence our results we include an enormous sample size of $N_{\left\{s\right\}}$ = 200,000.

We begin with testing the effects of the transverse plane (since we already have reasonable intuition of a good $\Delta x_{x}$ size from 2+1D simulations) and then once we have determined $\Delta x_{x}$ we will vary $\Delta x_{\eta}$. 
We will also use our ratio of $\frac{N_{nb}^{3D}}{N_{nb}^{2D}}$ estimate as a guideline for reasonable guesses for $\Delta x_{x}$ and $\Delta x_{\eta}$. 
Here we only show the results for all charged particles and we have not focused on obtaining the precise normalization or best fit transport coefficients for the given $\sqrt{s_{NN}}$ and centrality since this is only for a single event and also only for testing purposes. 
We do not expect any significant changes in our results if one varies the normalization or transport coefficients. 
We purposefully used a mid-central event because it allows for a quicker run time but also because it has less matter such that our convergence tests should be reasonable across a wide range of centrality class. If one performs a convergence test only on the most-central events, there is a risk of the numerical convergence breaking down for peripheral collisions since the system size is significantly smaller. 
Similarly, for very small systems like $p$+Pb or $pp$, one should redo the convergence tests described here but we leave this challenge for another work since the value of $h$ should likely also change (and is decrease with system size). 

On the bottom of each plot in the sections below, we always show the absolute value of the percentage difference for a given observable $\mathcal{O}$, i.e.,
\begin{equation}
    \Delta \left[\%\right] \equiv \frac{|\mathcal{O}_{\Delta x_i>\mathrm{cont}}(m)-\mathcal{O}_\mathrm{cont}(m)|}{\langle \mathcal{O}\rangle}
\end{equation}
where $m$ is whatever variable is along the $x$-axis, we always choose $\mathcal{O}_\mathrm{cont}$ to be the finest grid-size possible, and we normalize by the average $\langle \mathcal{O}\rangle =\sum_m \mathcal{O}(m)/N_m$, i.e., for the differential spectra $m=p_T$ such that $\langle \mathcal{O}\rangle=dN/dy$.  The reason we do not normalize by $\mathcal{O}_\mathrm{cont}(m)$ is because at low $p_T$ or large $|\eta|$ the spectra/flow becomes very small, which makes the error appear artificially large. 

\begin{figure}[ht!]
    \centering
    \includegraphics[width=1\linewidth]{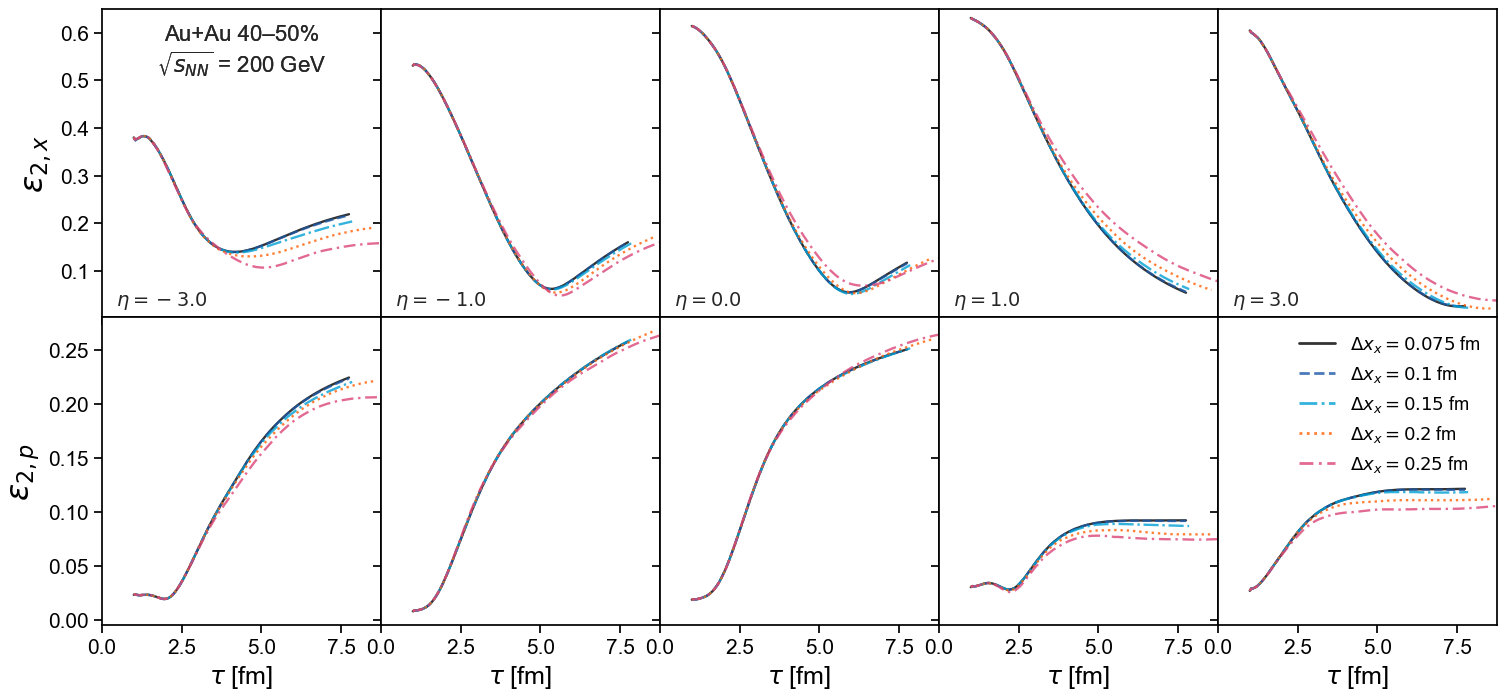}
    \caption{Eccentricity results across rapidity slices for Au+Au $\sqrt{s_{NN}}$ = 200 GeV, $\eta/s$ = 0.08 and vanishing charge densities for fixed longitudinal step-size $\Delta x_{\eta}$ = 0.1.}
    \label{fig:200_x}
\end{figure}
~\\
\noindent
{\bf Varying the transverse plane grid size.--}
Keeping $\Delta x_{\eta}=0.1$ fm fixed, we now vary $\Delta x_{x}$. For these simulations we have turned off BSQ conserved charges since there is very little baryon stopping at $\sqrt{s_{NN}}=200$ GeV.
From very high energies, we expect that integrated flow should be sensitive to the transverse grid size. Previously, at top LHC energies we have used $\Delta x_{x}=0.06$ fm (or sometimes even smaller) for 2+1D simulations. Here we generally find that we do not need quite as small of a grid size for the 3+1D simulations. However, $\Delta x_{x}$ is still quite small. 

To test the required $\Delta x_{x}$ in 3+1D simulations at the Beam energy scan we start with Au+Au collisions of 200 GeV and look at the internal eccentricities in Fig. \ref{fig:200_x}. There we see convergence starting around $\Delta x_x=0.15$ fm. We note that $\Delta x_x=0.2$ fm is quite close but shows divergences at late time.  We also note that the coordinate space eccentricities seem to be more sensitive to the choice in $\Delta x_x$ than the momentum space eccentricities.

\begin{figure*}[ht!]
\centering
\includegraphics[keepaspectratio, width=0.9\linewidth]{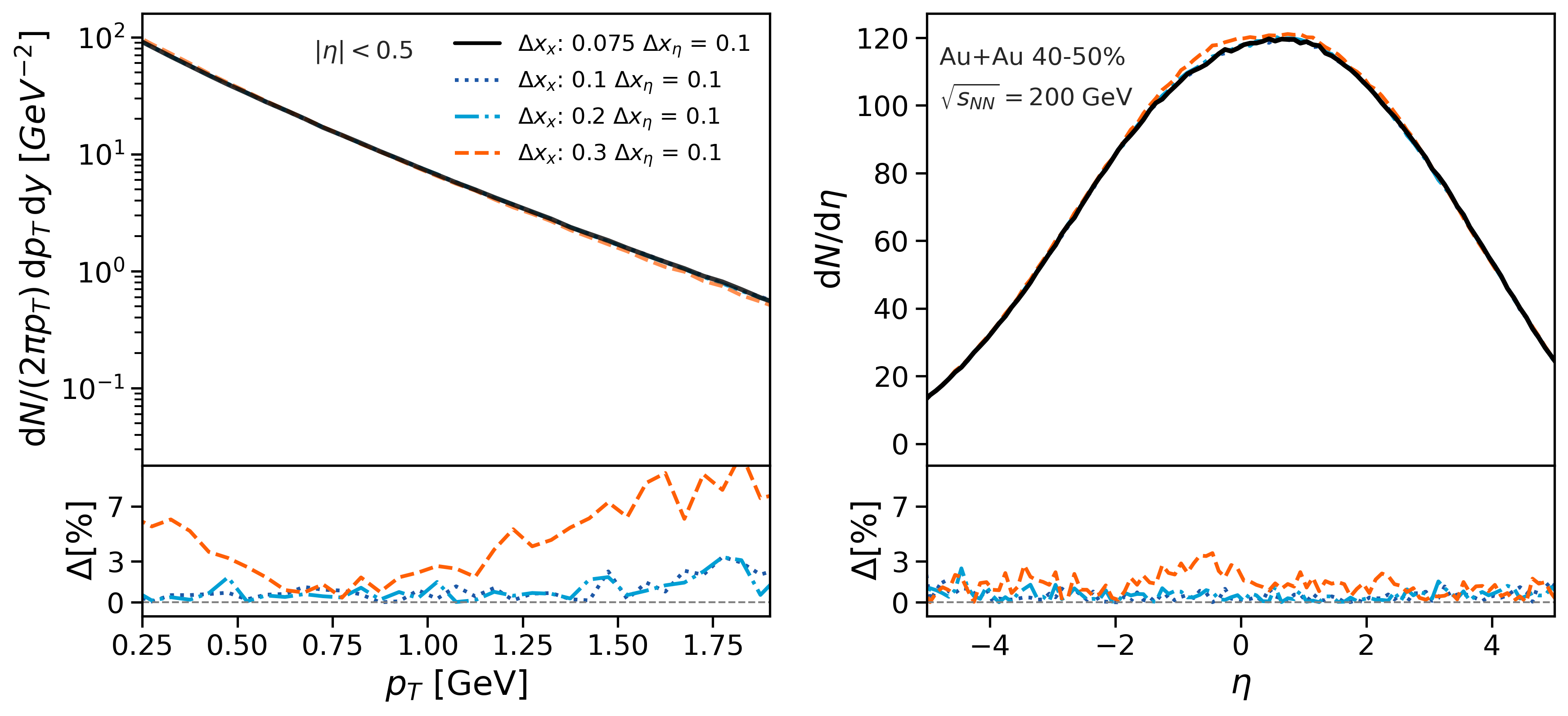}
    \caption{Grid size convergence test for $\Delta x_x \;[fm]$ while holding $\Delta x_{\eta}=0.1$ fixed. The differential spectra in $p_T$ is shown on the left and the rapidity dependent multiplicity is shown on the right. 
    }
    \label{fig:ETAmultiplicity200}
\end{figure*}

In Fig.~\ref{fig:ETAmultiplicity200} we now check the influence of the transverse plane's initial grid size on the differential spectra at mid-rapidity (left) and $dN/d\eta$ (right). 
We check a mid-central event because less central events are more sensitive to initial grid size effects (they have few nearest neighbors due to their small system sizes). 
We note that the differential spectra is shown on a log-scale such that it is difficult to see small differences visually. 
However, we do find that one requires an initial grid size of $\Delta x_x\lesssim 0.2$ fm in order to obtain an error at the percentage level. 
For the $dN/d\eta$ distribution we find that nearly all transverse plane grid sizes are reasonable.

\begin{figure*}[ht!]
\centering
\includegraphics[keepaspectratio, width=0.5\linewidth]{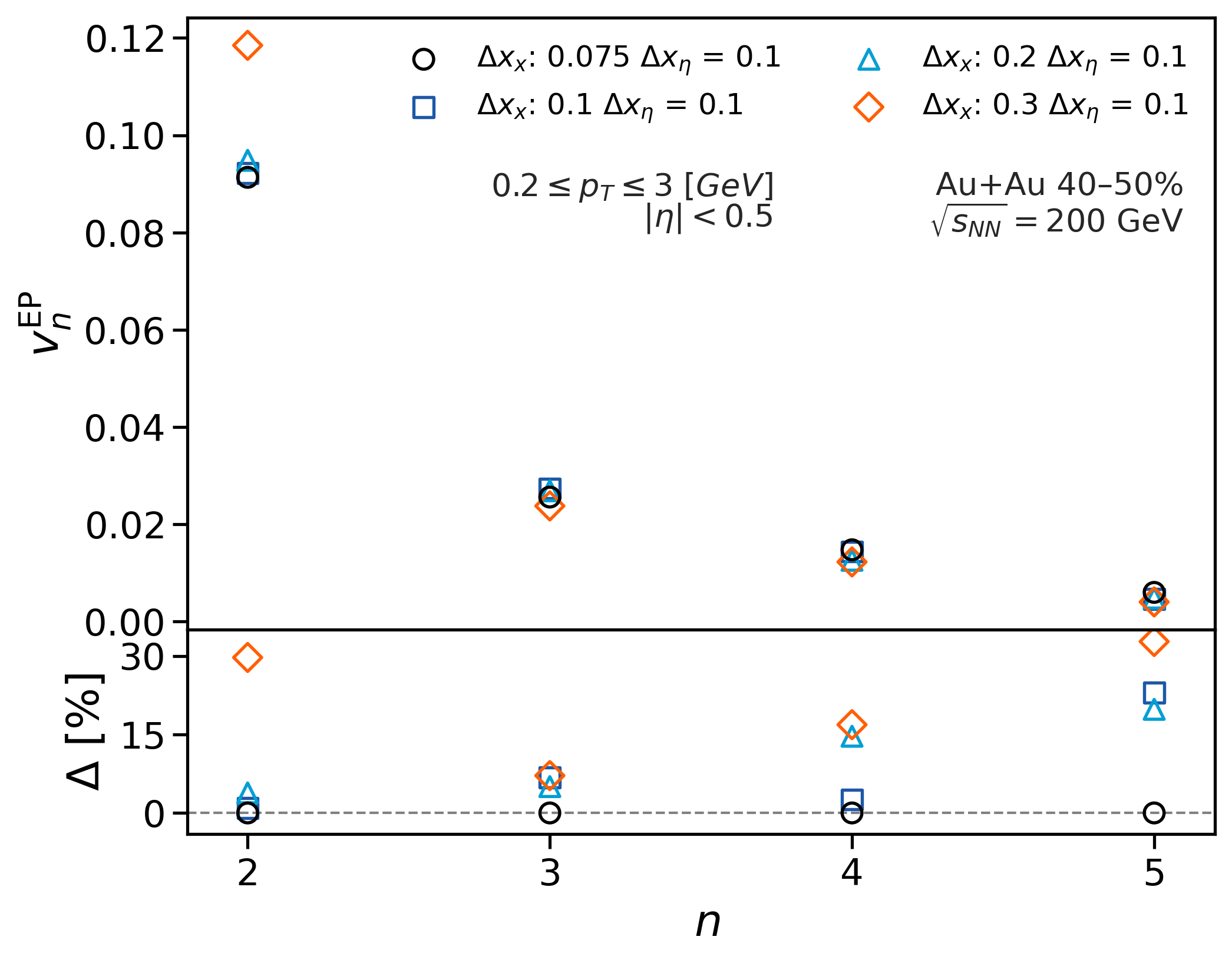}
    \caption{Grid size convergence test for $\Delta x_{x} \;[fm]$ while holding $\Delta x_{\eta}=0.1$ fixed  for the integrated flow harmonics are shown for $n=2$--$5$, calculated at mid-rapidity.}
    \label{fig:xint_vn200}
\end{figure*}

Next in Fig.~\ref{fig:xint_vn200} we compare the influence of the initial grid size on the integrated flow calculate at mid-rapidity for $v_2$--$v_5$. 
We find that integrated flow is, indeed, more sensitive to the grid size in the transverse plane. For instance, $\Delta x_x=0.3$ fm significantly over predicts $v_2$ (on the order of $30\%$). However, if one only consider $v_2$ and $v_3$, then using $\Delta x_x\leq 0.2$ fm appear to be entirely reasonable. A word of caution, though, is that $\Delta x_x\leq 0.2$ fm leads to error in $v_4$ and $v_5$ on the order of $\sim 15\%$.  If one wishes to obtain results for higher-order flow harmonics, then one should include a grid size as small as $\Delta x_x=0.1$ fm for $v_4$ but may want to consider even smaller to obtain $v_5$.  
Given these results, we general consider $\Delta x_x=0.1$ fm to be our fiducial values for $\sqrt{s_{NN}}=200$ GeV, but we also provide times estimates for $\Delta x_x\leq 0.2$ fm since it is a reasonable choice if one is not interested in higher harmonics. 

\begin{figure*}[ht!]
\centering
\includegraphics[keepaspectratio, width=0.9\linewidth]{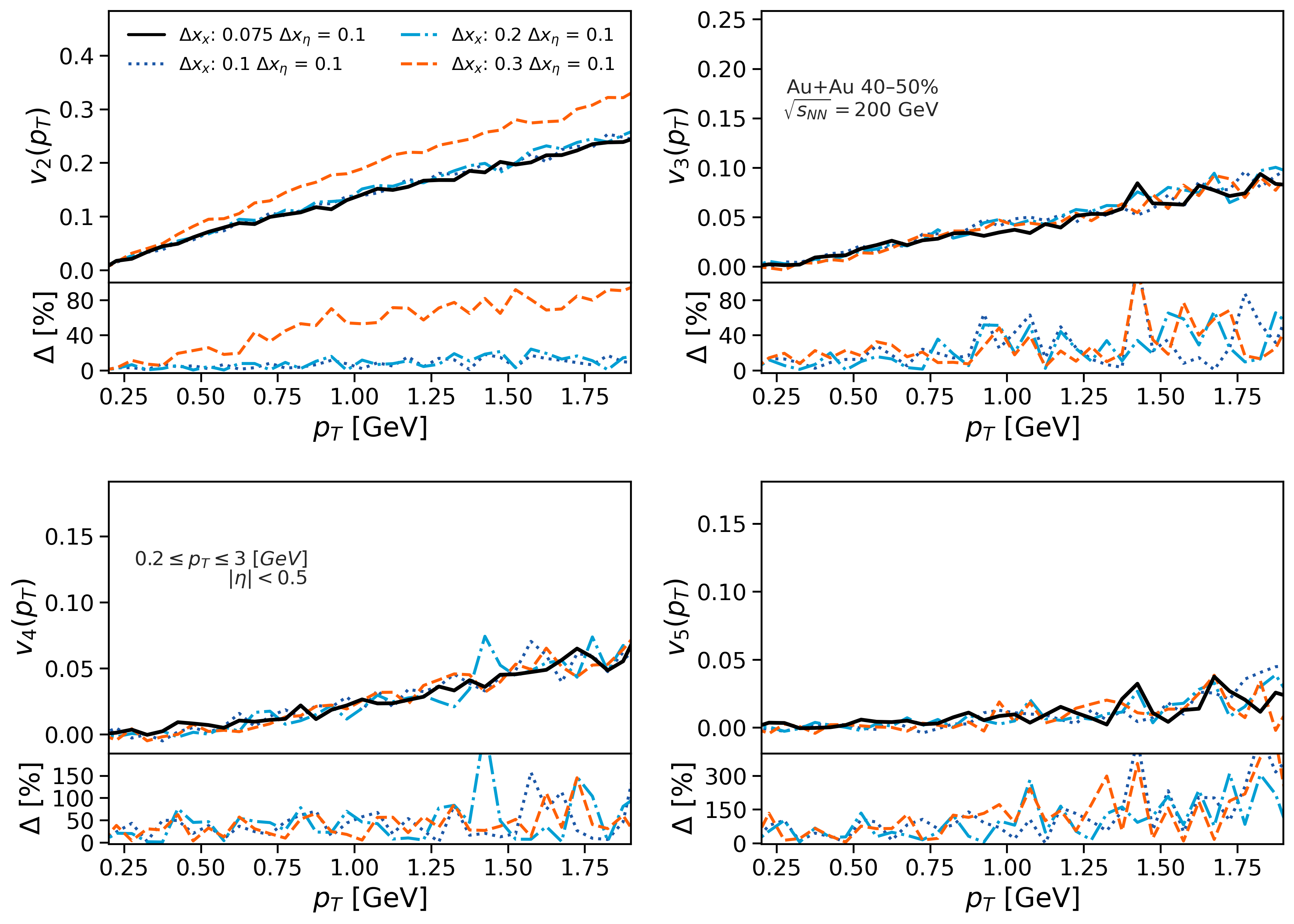}
    \caption{Grid size convergence test for $\Delta x_{x}\;[fm]$ while holding $\Delta x_{\eta}=0.1$ fixed for the differential flow harmonics are shown as a function of $p_T$ for $n=2$--$5$, calculate at mid-rapidity. Here we include $N_{\left\{s\right\}}=10,000$ samples of the freeze-out surface.}
    \label{fig:xvn_pt_conv}
\end{figure*}

In Fig.\ \ref{fig:xvn_pt_conv} we show the differential flow vs. $p_T$ when varying the initial grid size in the transverse plane. Here we note that we have used $N_{\left\{s\right\}}=10,000$ samples of the freeze-out surface for these calculations. 
Generally, we find that our $v_n(p_T)$ flow curves look fairly similar to each other (with the exception of $\Delta x_{x}=0.3$ fm, but that is to be expected given our results for the integrated flow).  
However, we do obtain large fluctuations at different points in $p_T$ that we suspect has more to do with our number of samples rather than the grid size itself.  
Given that this is for a single event, the jaggedness would like disappear when averaging over many events.

\begin{figure*}[ht!]
\centering
\includegraphics[keepaspectratio, width=0.9\linewidth]{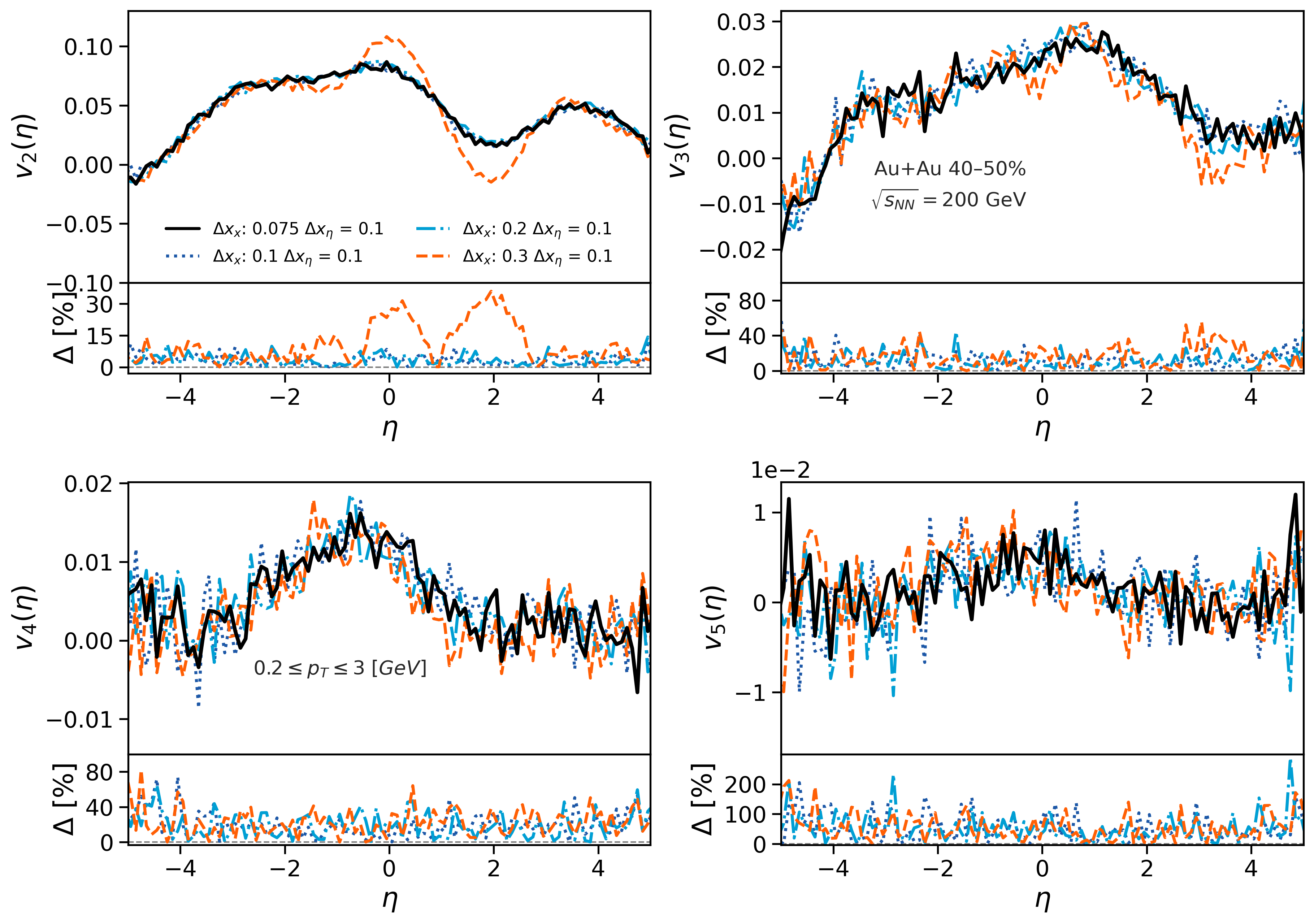}
    \caption{Grid size convergence test for $\Delta x_{x} \;[fm]$ while holding $\Delta x_{\eta}=0.1$  fixed for the differential flow harmonics are shown as a function of $\eta$ for $n=2$--$5$. }
    \label{fig:xconv_vn_eta}
\end{figure*}

In Fig.\ \ref{fig:xconv_vn_eta} we show the differential flow vs. $\eta$ when varying the initial grid size in the transverse plane. Here we note that we have used $N_{\left\{s\right\}}=10,000$ samples of the freeze-out surface for these calculations. 
We find similar results to the differential flow in that one can see the same qualitative trends in the flow (especially for $v_2(\eta)$ and $v_3(\eta)$) but there is quite a bit of noise as one compares neighboring points in $\eta$.  For higher-order harmonics the noise is extremely large and we do not trust current $v_5(\eta)$ results. Significant more work would be required to determine the convergence needed in $\Delta x_x$, $\Delta x_{\eta}$, and  $N_{\left\{s\right\}}$ to obtain smooth results. 

~\\
\noindent
{\bf Varying the rapidity direction grid size.--} Keeping $\Delta x_{x}=0.1$ fm fixed, we now vary $\Delta x_{\eta}$. We have chosen a range of $\Delta x_{\eta}$ from 0.025-0.4 where the lower bound is primarily limited by the capabilities of \ampt{}.  The upper bound is constrained by the size of $h$, such that $\Delta x_{\eta}<2h$ otherwise there may be no particles within a certain SPH fluid cell\footnote{Note a dimensionful factor of $1$ fm$^{-1}$ is used to deal with the dimensions of $h$ vs. $\eta$, which is dimensionless.}.   
Our ``continuum limit'' here is $\Delta x_x=0.1$ fm and $\Delta x_{\eta}=0.025$.

\begin{figure}[ht!]
    \centering
    \includegraphics[width=1\linewidth]{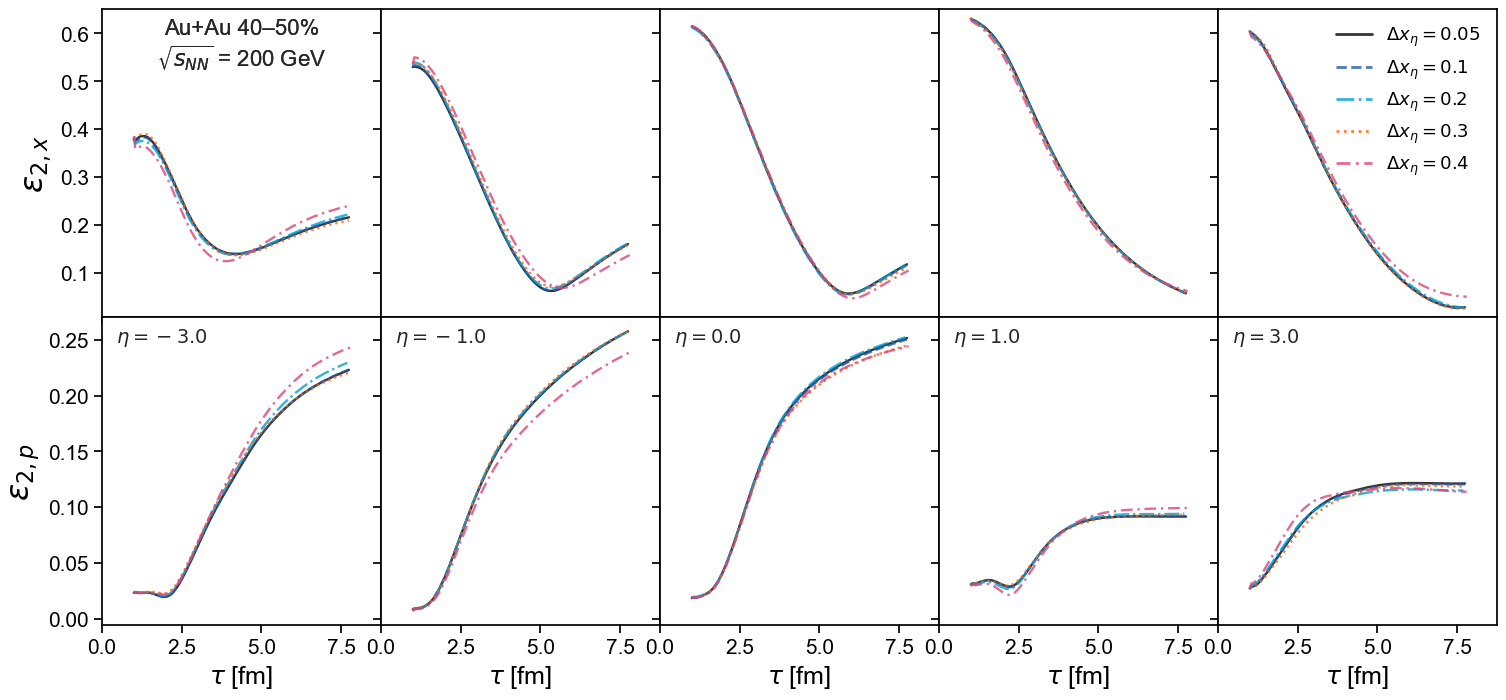}
    \caption{Eccentricity results across rapidity slices for Au+Au $\sqrt{s_{NN}}$ = 200 GeV, $\eta /s$ = 0.08 and vanishing charge densities for fixed transverse step-size $\Delta x_x=0.1$ fm.}
    \label{fig:200_eta}
\end{figure}

In Fig.~\ref{fig:200_eta} we first check the internal eccentricities for the case of a fixed $\Delta x_x=0.1$ fm and varying $\Delta x_\eta$. 
We find that at $\sqrt{s_{NN}}=200$ GeV that there is almost no dependent on the $\Delta x_\eta$ grid size, except for the coarse grid of $\Delta x_\eta=0.4$ fm that just begins to show divergence. Thus, it seems reasonable to say that at least for the internal eccentricities that $\Delta x_\eta\leq 0.3$ fm has converged. 

\begin{figure*}[ht!]
\centering
\includegraphics[keepaspectratio, width=0.9\linewidth]{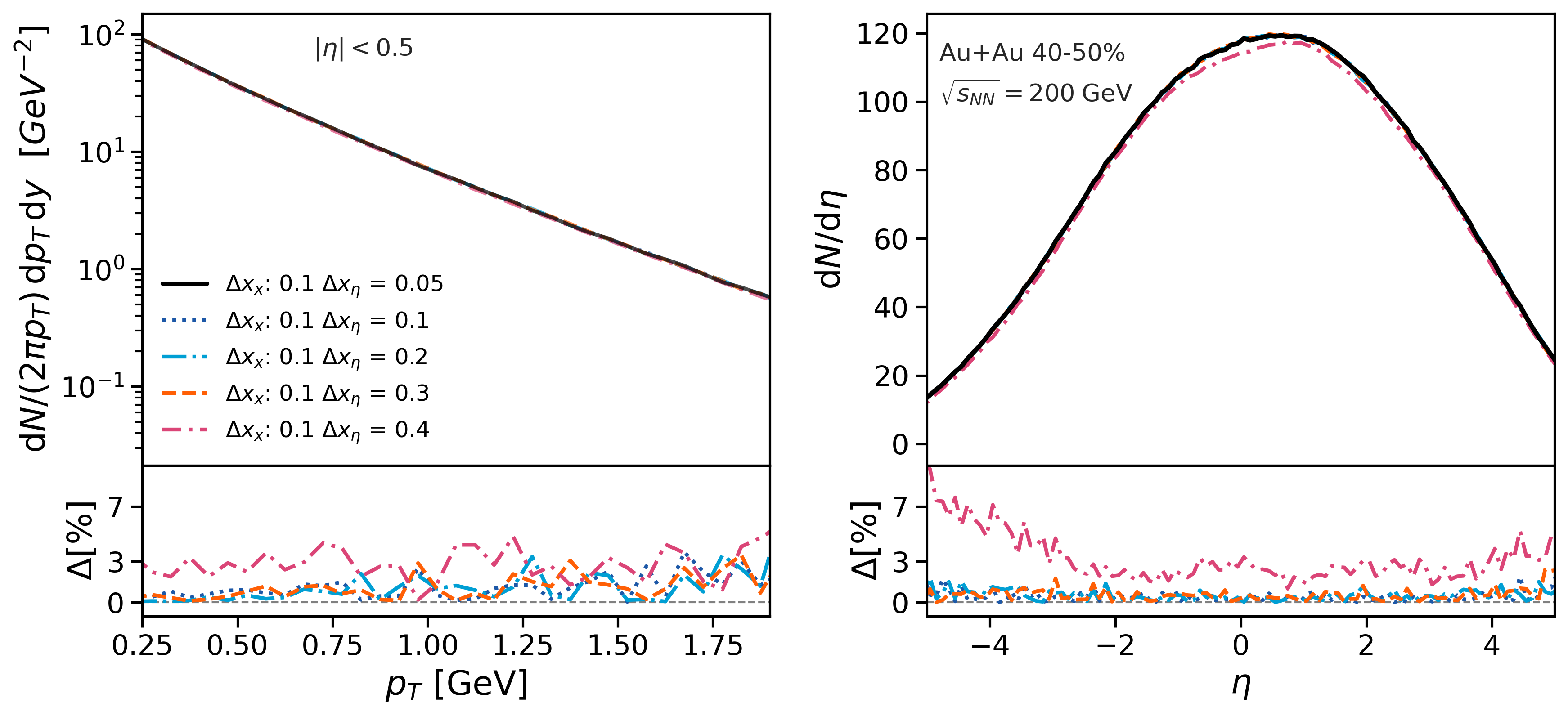}
    \caption{Grid size convergence test for $\Delta x_\eta$ while holding $\Delta x_{x}=0.1$ fm fixed. The differential spectra in $p_T$ is shown on the left and the rapidity dependent multiplicity is shown on the right.
    }
    \label{fig:xmultiplicity200}
\end{figure*}

In Fig.~\ref{fig:xmultiplicity200} we show the differential $p_T$-spectra on the left and the $dN/d\eta$ spectra on the right for our single event.
For the differential $p_T$-spectra we enforce a standard rapidity cut from STAR of $|\eta|<0.5$. 
We find that the differential spectra is nearly independent of our choice in the $\Delta x_{\eta}$ grid, which is to be expected since it is a $p_T$ dependent quantity. 
We do find, though, that for the largest $\Delta x_{\eta}=0.4$ that we obtain an error of a few percentages. 
For the $dN/d\eta$ results (right) we integrate over the entire $p_T$ range. Once again our $dN/d\eta $ multiplicity converges well for most choices of the grid size but $\Delta x_{\eta}=0.4$ has the most significant error up to $7\%$ for forward/backward rapidity. While the $ \Delta \left[\%\right]$ is smaller at mid-rapidity, one can visually see a difference in the values of $dN/d\eta $ such that it is clear that $\Delta x_{\eta}=0.4$ is too coarse of a grid.

\begin{figure*}[ht!]
\centering
\includegraphics[keepaspectratio, width=0.5\linewidth]{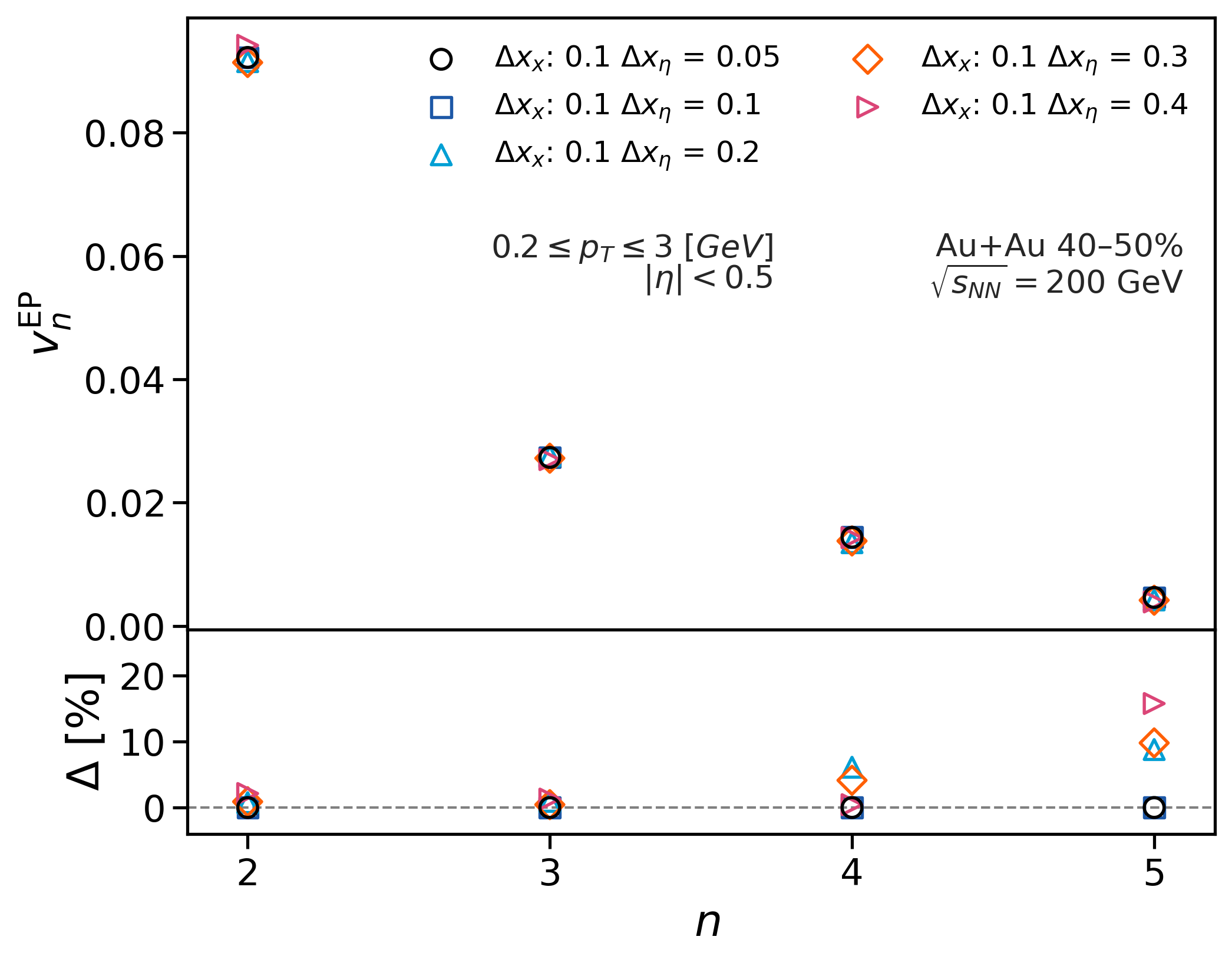}
    \caption{Grid size convergence test for $\Delta x_{\eta}$ while holding $\Delta x_{x}=0.1$ fm fixed  for the integrated flow harmonics are shown for $n=2$--$5$, calculated at mid-rapidity.}
    \label{fig:ETAint_vn}
\end{figure*}

In Fig.~\ref{fig:ETAint_vn} we plot the integrated flow for $v_n$ with $n=2$--$5$ for a single event.  Generally, spectra is one of the easiest observables to obtain (i.e., does not require a fine grid for the initial state) and integrated flow may require a somewhat finer spacing (followed by differential flow that we will discuss later). Thus, we anticipate that we will see larger deviations in the $ \Delta \left[\%\right]$ for the integrated flow than what we saw previously in Fig.~\ref{fig:ETAmultiplicity200}. 

We find that all  initial grid sizes manage to calculate the integrated flow observables up to $v_5$ within a $10\%$ error---even for the coarsest grid.
In fact, for elliptical flow that is the most robust to grid size changes, we see nearly identical results across all initial grid sizes with the one exception being $\Delta x_{\eta}=0.4$ that has an error around $5\%$. 
Similarly, $v_3$ has nearly identical results for for all grid size and only has up to a maximum error of $3\%$ even for $\Delta x_{\eta}=0.4$. Note that normally there is a hierarchy of higher-order harmonics being more sensitive to grid-size effects, but since this is only a single event some small differences may appear. 
Since some studies only focus on $dN/dy$ and integrated $v_2$ and $v_3$, then one may be able to use a coarse grid of $\Delta x_x=0.1$ fm and $\Delta x_\eta=0.2$ for the sake of speed. 

We now compare $v_4$ and $v_5$ and see larger deviations for $\Delta x_{\eta}=0.4$ (up to $7\%$) and find a rather significant deviation for $v_5$ with $\Delta x_{\eta}=0.2$ to be nearly $10\%$. Thus, for higher-harmonics for intergrated flow we suggest using a grid size of $\Delta x_{\eta}=0.1$.

\begin{figure*}[ht!]
\centering
\includegraphics[keepaspectratio, width=0.9\linewidth]{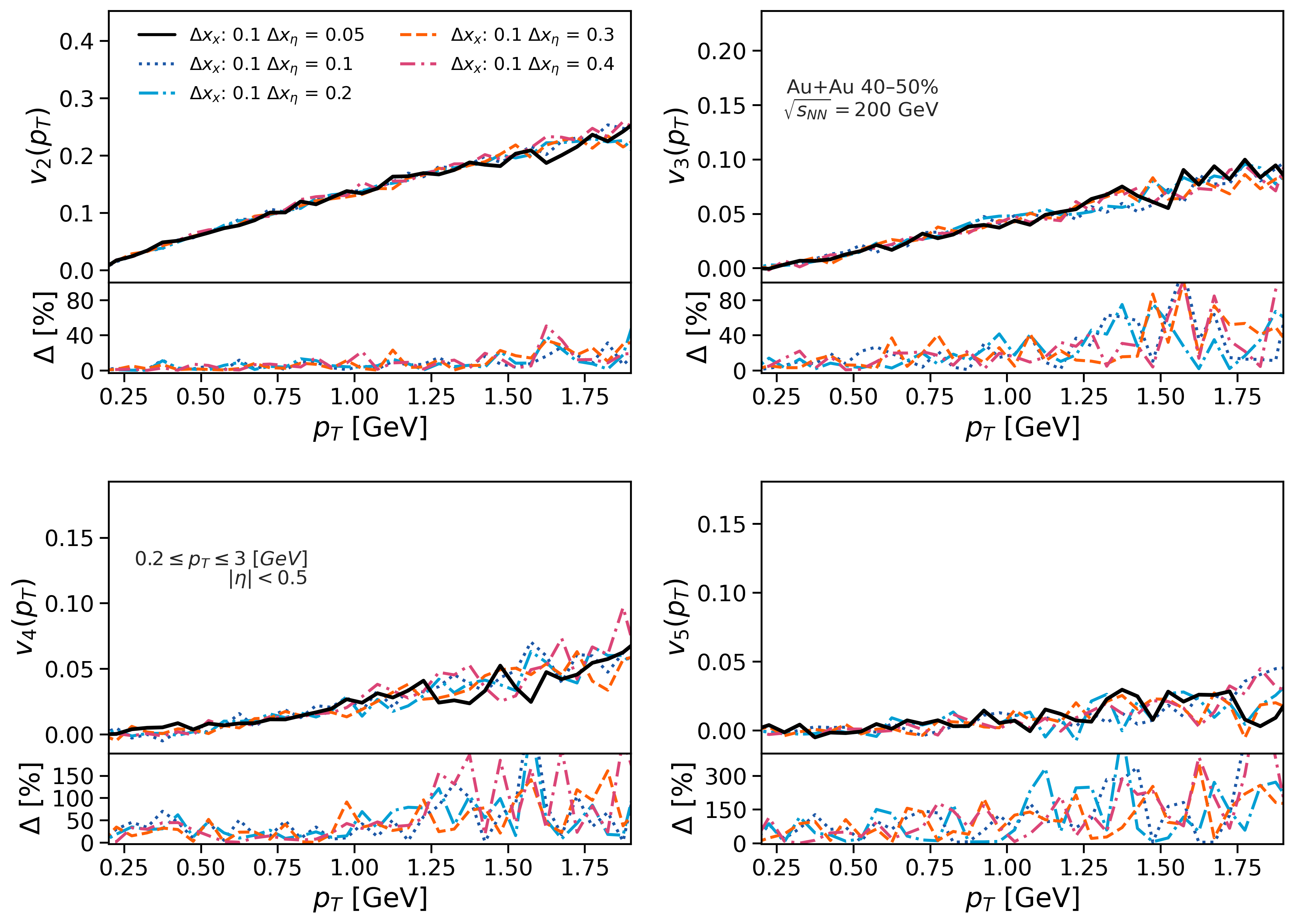}
    \caption{Grid size convergence test for $\Delta x_{\eta}$ while holding $\Delta x_{x}=0.1$ fm fixed for the differential flow harmonics are shown as a function of $p_T$ for $n=2$--$5$, calculate at mid-rapidity. Here we include $N_{\left\{s\right\}}=10,000$ samples of the freeze-out surface. }
    \label{fig:ETAvn_pt_conv}
\end{figure*}

In Fig.~\ref{fig:ETAvn_pt_conv} we compare the different in $p_T$ convergence tests when varying $\Delta x_{\eta}$ for $v_2$--$v_5$ for a single event. 
Since we are keeping our grid in the transverse plane fixed and look only at mid-rapidity we do not expect any significant differences in our differential flow harmonics. Visually, it is nearly impossible to see any deviations from the plot but because we are dealing with small numbers here even time changes end up causing larger percent differences (especially for higher-harmonics).  
At low $p_T$ we do not find any significant error, but we do find that the error increases with increasing $p_T$. 
However, that may be also from the finite number of samples to obtain particles that we require for 3+1D simulations.  Given that it is such a tiny difference we do not explore this point further.

\begin{figure*}[ht!]
\centering
\includegraphics[keepaspectratio, width=0.9\linewidth]{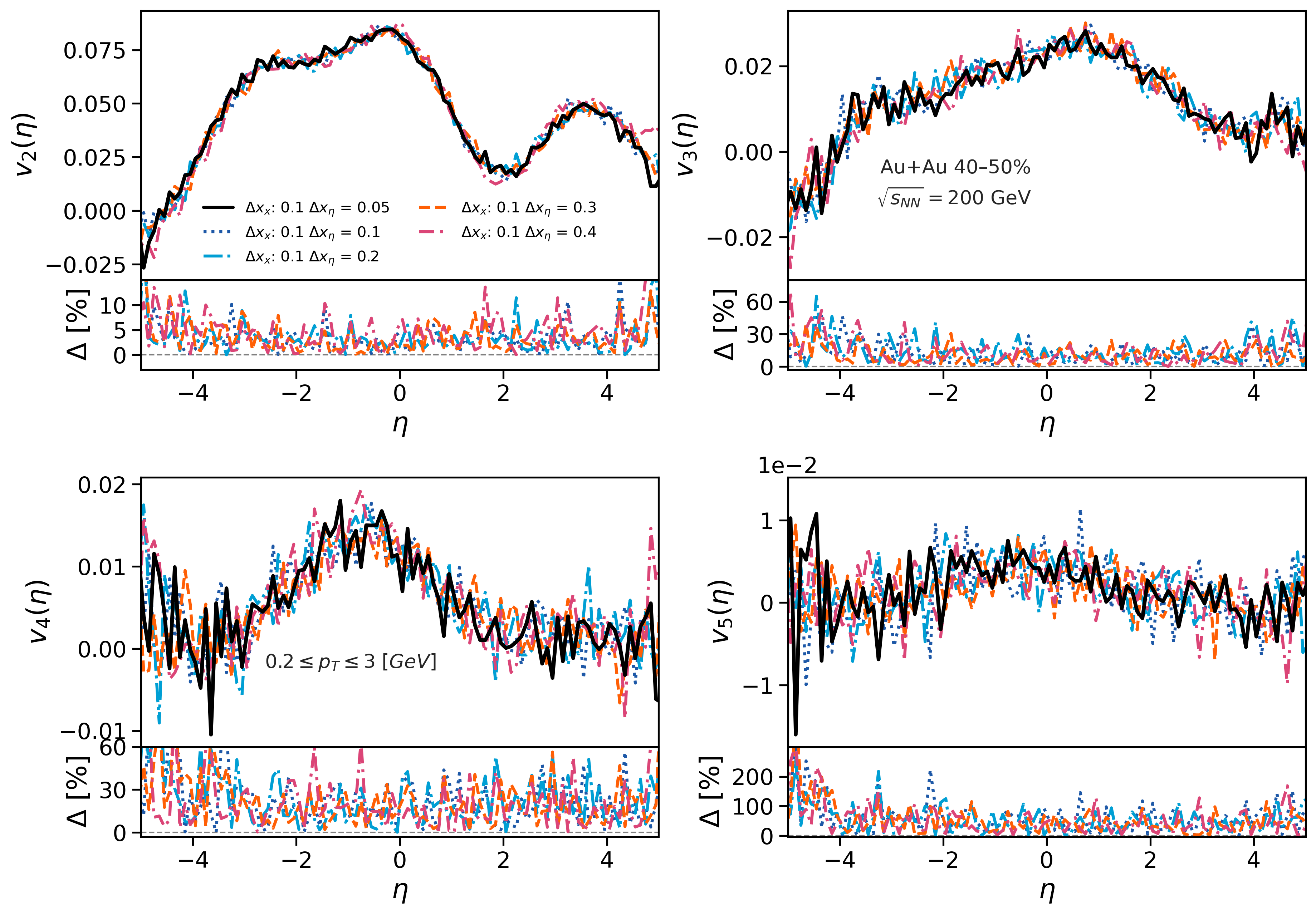}
    \caption{Grid size convergence test for $\Delta x_{\eta}$ while holding $\Delta x_{x}=0.1$ fm fixed for the differential flow harmonics are shown as a function of $\eta$ for $n=2$--$5$. }
    \label{fig:ETAconv_vn_eta}
\end{figure*}

Finally, we come to Fig.~\ref{fig:ETAconv_vn_eta} where we compare the differential flow in $\eta$ that we expect to be very sensitive to our choices in the initial grid size. 
Looking at the elliptical flow, we see  that $\Delta x_{\eta}=0.3$ leads to a significant error in $v_2(\eta)$ such that the qualitative behavior vs. $\eta$ even changes. For this particular event $v_3$ has small fluctuations across $\eta$ but is not quite as dramatic as elliptical flow.  
Thus, there is no clear ``winner''. 
Then for $v_4(\eta)$ and $v_5(\eta)$ we see that we actually obtain a rather bumpy distribution even for our ``continuum limit'' (especially for $v_5(\eta)$). Thus, we argue that if one requires very sensitive $v_5(\eta)$ results, it is probably best to use an even finer grid or increase $N_{\left\{s\right\}}$ further. 
However, most studies do now compare to such observables so we do not strive to a better numerical accuracy beyond what was achieved here. 

To summarize, after performing these convergence tests, we suggest a grid size of $\Delta x_{x}=0.1$ fm and $\Delta x_{\eta}=0.1$ that balances performance and accuracy.  
If one is only interested in multiplicity and integrated elliptical and triangular flow than $\Delta x_{x}=0.1$ fm and $\Delta x_{\eta}=0.2$ would be sufficient as well. However, calculations focused on higher-order harmonics (especially vs. $\eta$) should be very cautious and do a more in-depth study on tracking down the source of numerical error. Would would suggest varying $\Delta x_x$, $\Delta x_\eta$, and $N_{\left\{s\right\}}$ until a clear convergence was achieved.

\clearpage

\subsubsection{Convergence tests at \texorpdfstring{$\sqrt{s_{NN}}=19.6$}{√s=19.6} GeV}%
\label{Sec:Observables:Tests:ConvergenceSqrts}
\noindent
Now we move on to a lower $\sqrt{s_{NN}}=19.6$ GeV where the system is significantly less Lorentz contracted. 
Once again, we use a single event at this beam energy to test convergence. 
Here we keep the initial condition fixed and just increase/decrease the grid size that is fed into \ccake{}.
At this lower $\sqrt{s_{NN}}$ we have less intuition both about the required step steps in the grid as well as how any internal eccentricities relate to final flow harmonics. Thus, we repeat the same process that we did before at $\sqrt{s_{NN}}=200$ GeV.

In this section we will test not only $\Delta x_x$, $\Delta x_\eta$ but also our underlying smoothing scale $h$.  
Lower $\sqrt{s_{NN}}$ implies a smaller amount of momentum transferred between collisions of nucleons in the initial state. When less momentum is transferred the resulting degrees-of-freedom are less gluon and sea quark dominated, and instead are influenced by the valence quarks. The valence quarks lead to large scale internal correlations within the systems (compared to gluons and sea quarks, which are short scale fluctuations). Thus, we anticipate that low $\sqrt{s_{NN}}$ may have a larger smoothing scale $h$ as compared to high $\sqrt{s_{NN}}$.

\begin{figure}[ht!]
    \centering
    \includegraphics[width=1\linewidth]{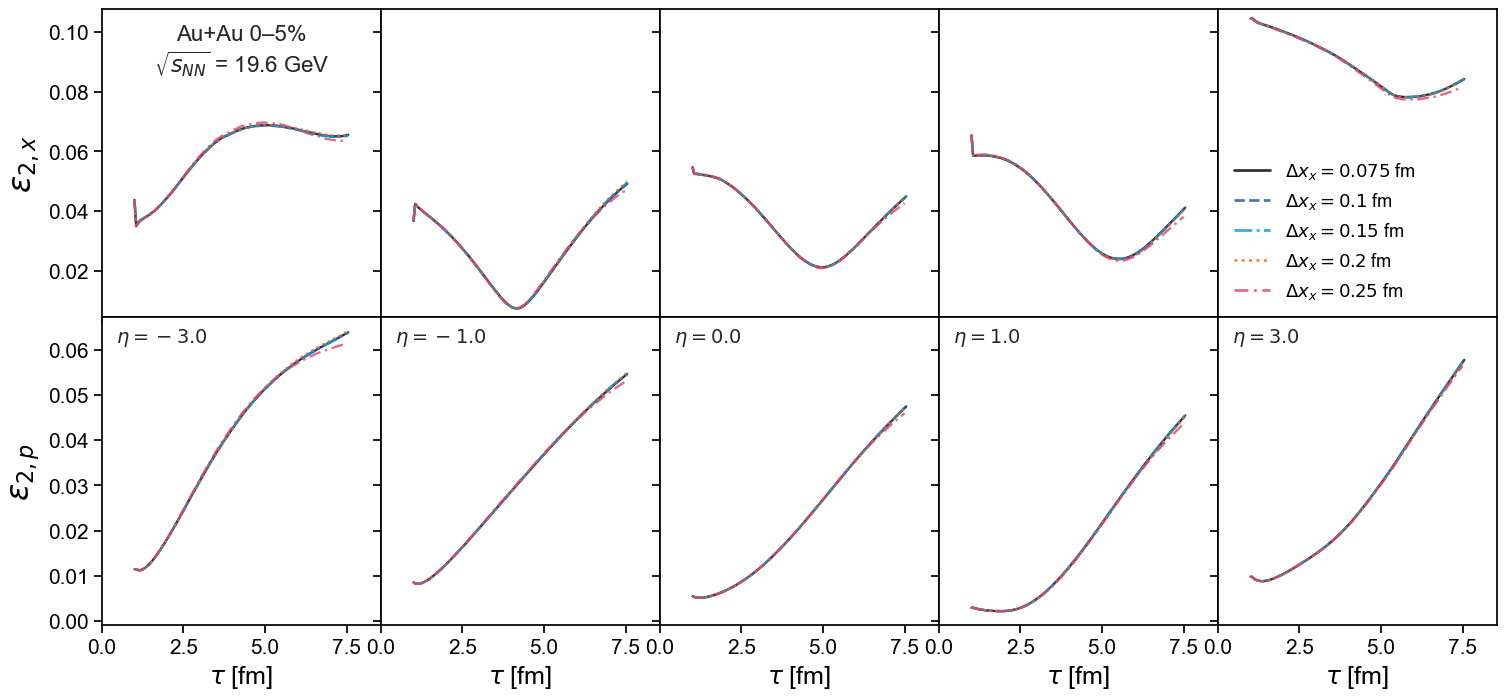}
    \caption{Eccentricity results across rapidity slices for Au+Au $\sqrt{s_{NN}}$ = 19.6 GeV, $\eta /s$ = 0.08 and BSQ charges for fixed longitudinal step-size $\Delta x_\eta$ = 0.1}
    \label{fig:19p6_x}
\end{figure}
\begin{figure}[ht!]
    \centering
    \includegraphics[width=1\linewidth]{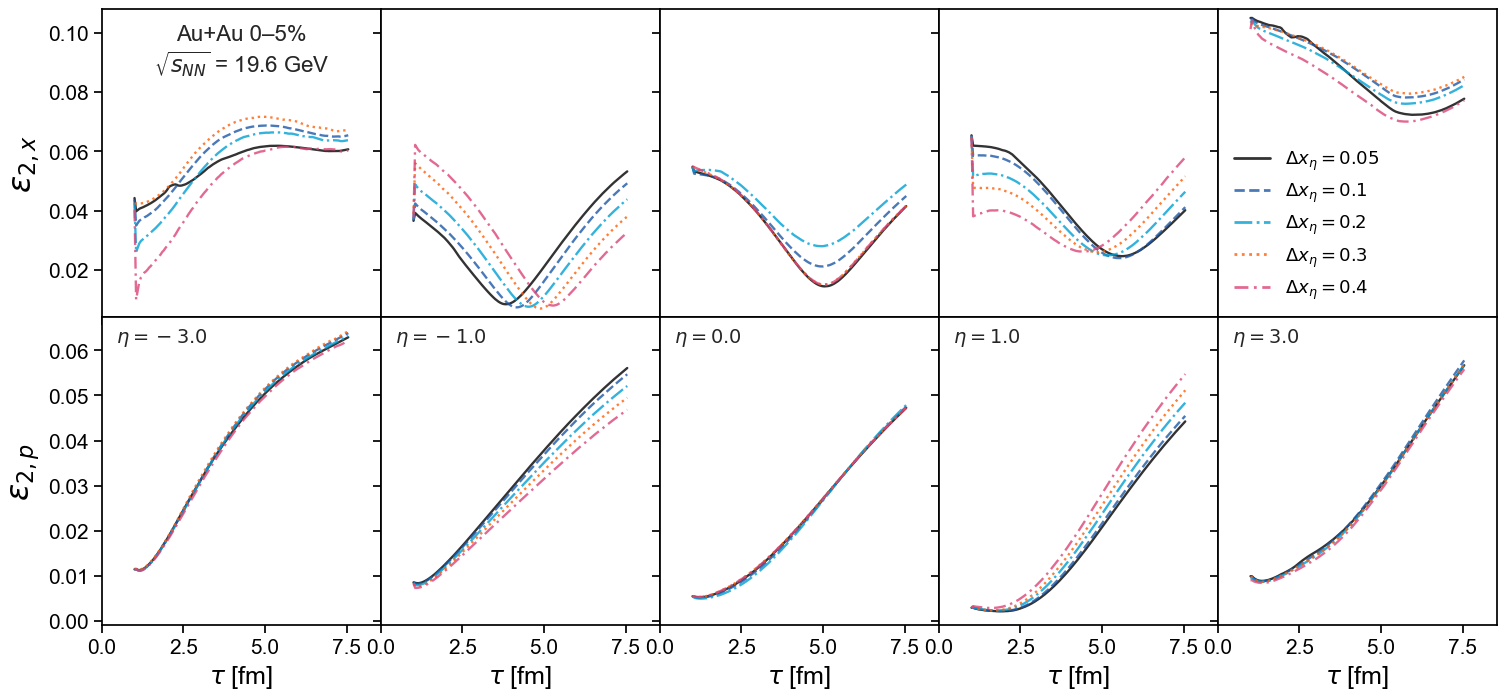}
    \caption{Eccentricity results across rapidity slices for Au+Au $\sqrt{s_{NN}}$ = 19.6 GeV, $\eta /s$ = 0.08 and and BSQ charges for fixed transverse step-size $\Delta x_x= 0.1$ fm.}
    \label{fig:19p6_eta}
\end{figure}

\noindent
{\bf Internal eccentricities.--}
We begin with the internal eccentricities from Eq.~(\ref{eqn:eccentricities}) in Fig.~\ref{fig:19p6_x} for $\Delta x_\eta=0.1$ while varying $\Delta x_x$ and  in Fig.~\ref{fig:19p6_eta} for $\Delta x_x=0.1$ fm while varying $\Delta x_\eta$.
Here we keep $h=0.3$ fm fixed, but will explore varying $h$ when it comes to the experimental observables. 
In Fig.\ \ref{fig:19p6_x}, when we keep the rapidity grid fixed and vary only the transverse plan, we see a clear convergence in both $\varepsilon_{e,x}$ (top row) and  $\varepsilon_{e,p}$ (bottom row) across all rapidity slices.  Generally, any choice of $\Delta x_x\leq 0.2$ fm shows convergence. The results for the transverse plan seem to show approximately the same degree of convergence regardless of rapidity slice. 
These results imply that for the internal eccentricity the beam energy of $\sqrt{s_{NN}}=19.6$ GeV does not show a strong dependence on the initial transverse plane grid size choice.  
These results are consistent with our discussion above that it appears that small scale fluctuations are less relevant at this low beam energies. 

Next we move to Fig.~\ref{fig:19p6_eta} where we keep the transverse plane fixed with converged values of $\Delta x_x=0.1$ fm while varying $\Delta x_{\eta}$. We find that  momentum space eccentricities (bottom row) shows convergence between $\Delta x_\eta=0.1$ and $\Delta x_\eta=0.05$ but $\Delta x_\eta=0.2$ already shows a good deal of divergence at $\eta=\pm 1$. 
In contrast, the spatial eccentricities (top row) do not demonstrate clear convergence even comparing $\Delta x_\eta=0.1$ and $\Delta x_\eta=0.05$, which is pretty surprising. 
Given that the experimental observables are calculated in momentum space, it may be that convergence in the spatial eccentricities is not required. 
However, one clear conclusions from the internal eccentricities is that the initial rapidity grid size appears to be more important than the initial transverse plane grid size. This conclusions is the opposite of what we found at high $\sqrt{s_{NN}}= 200$ GeV. 
Next we will check if this conclusion holds once looking at the experimental observables. 

\begin{figure}[ht!]
    \centering
       \includegraphics[width=\linewidth]{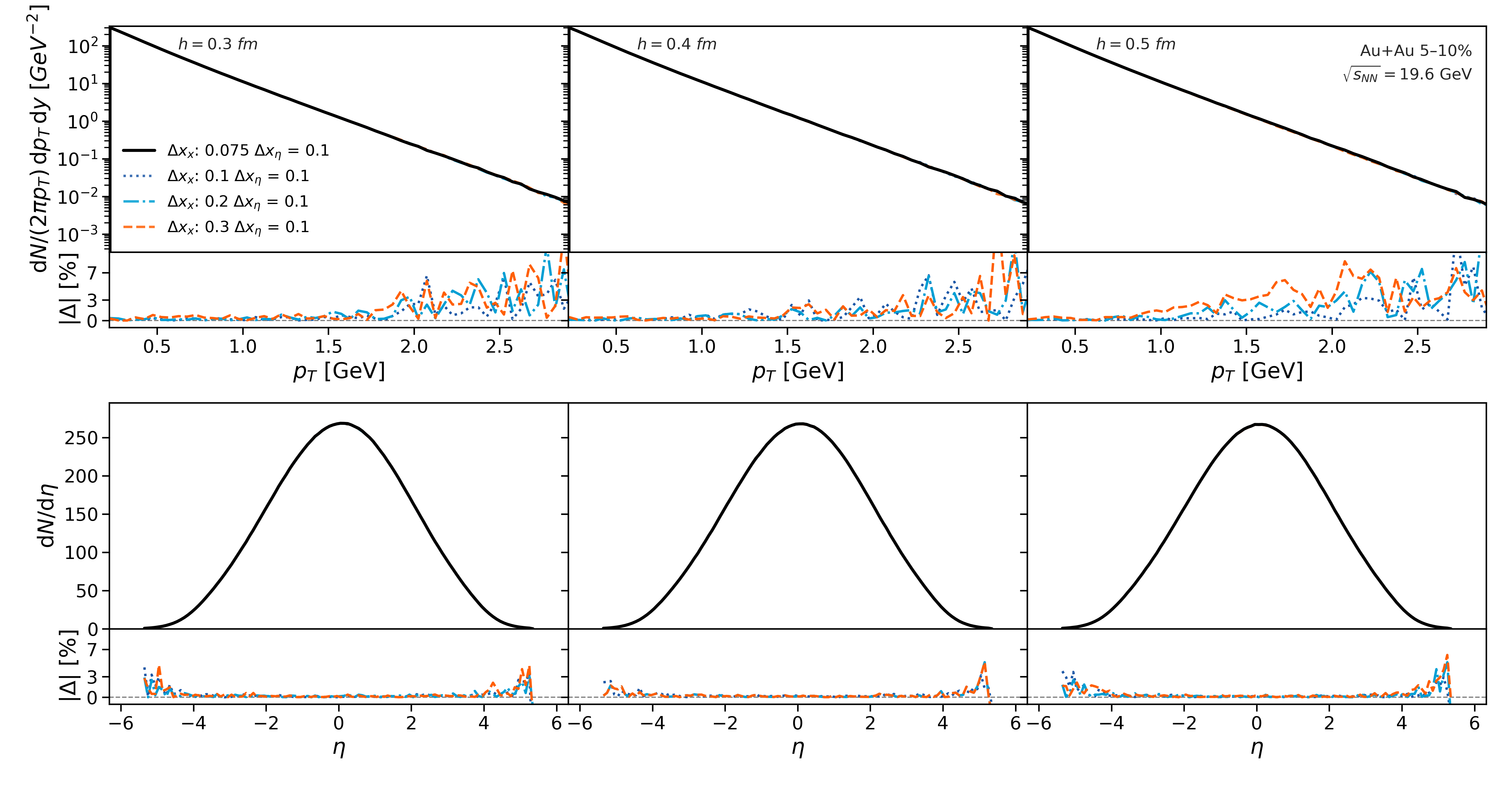}
    \caption{Grid size convergence test for keeping $\Delta x_{\eta}=0.1$ fixed, while varying $\Delta x_x \;[fm]$ . The differential $p_T$-spectra is shown in the top row and the rapidity dependence $dN/d\eta$ is shown on the bottom row. The columns different choices of the smoothing scale $h$ varied from 0.3--0.5 fm. 
    }\label{fig:19p6spec_x}
\end{figure}
\begin{figure}[ht!]
    \centering
        \includegraphics[width=\linewidth]{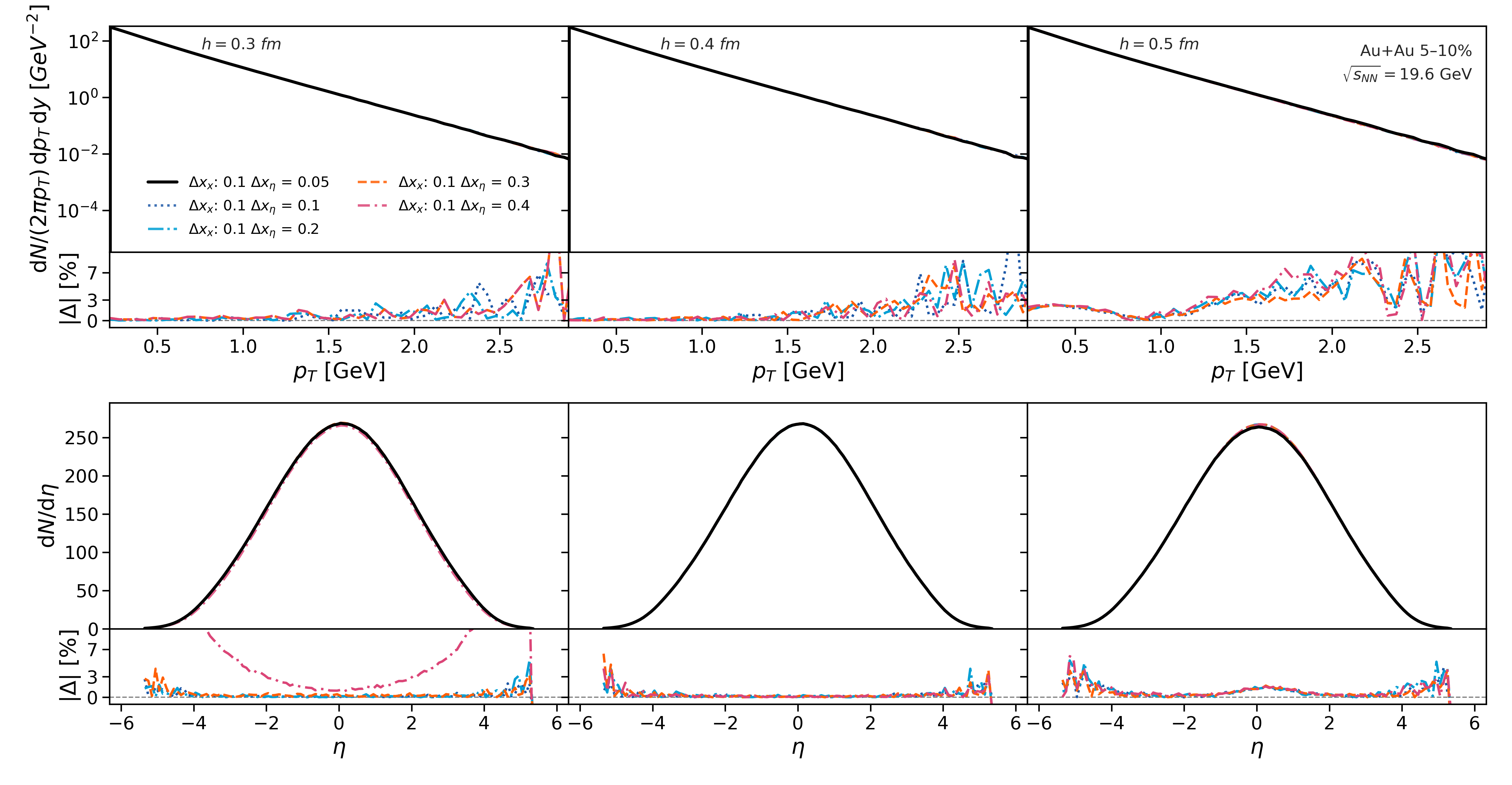}
    \caption{Grid size convergence test for keeping $\Delta x_x=0.1$ fm fixed, while varying $\Delta x_\eta$. The differential $p_T$-spectra is shown in the top row and the rapidity dependence $dN/d\eta$ is shown on the bottom row. The columns different choices of the smoothing scale $h$ varied from 0.3--0.5 fm. 
    }\label{fig:19p6spec_eta}
\end{figure}

~\\
\noindent
{\bf Spectra and $dN/\eta$.--}
Now we come to differential spectra in $p_T$ and $\eta$ observables, which can be measured experimentally. At high $\sqrt{s_{NN}}$ we found these observables fairly robust to grid size choices (with the one exception being an extremely coarse $\Delta x_x$ choice). 
In Fig.~\ref{fig:19p6spec_x}-\ref{fig:19p6spec_eta} we explore the influence of the grid size and smoothing $h$ scale on the differential $p_T$-spectra and the rapidity dependent multiplicity distribution $dN/\eta$. 
In Fig.~\ref{fig:19p6spec_x} we study the effect  of varying  $\Delta x_x$ while fixing $\Delta x_\eta=0.1$ and in Fig.\ \ref{fig:19p6spec_eta} we show the effect of varying  $\Delta x_\eta$ while fixing $\Delta x_x=0.1$ fm. We also explore the consequence of the smoothing scale $h$ (different columns). 

Let us first look at the differential $p_T$-spectra (top row) where we find almost no difference depending on the underlying choice of the initial grid size $\Delta x_x$ in Fig.~\ref{fig:19p6spec_x}, although at high $p_T$ we find larger error mostly because this is calculated as a percentage difference and is normalized by a very small value. Similarly, we find almost no dependence on the $h$ scale. 
We compare also the $dN/d\eta$ multiplicity (bottom row) and find nearly perfect convergence across rapidity, with the exeception of very forward or backward rapidity i.e. $|\eta|>4$, which is for similar reasons as the high $p_T$ spectra because there the overall $dN/d\eta$ is extremely small. 
These results are consistent with our results from the internal eccentricities.

In Fig.\ \ref{fig:19p6spec_eta} we look at the spectra and $dN/dy$ for varying $\Delta x_\eta$, while fixing $\Delta x_x=0.1$ fm. 
Here we find a slightly stronger dependence on the choice of $\Delta x_\eta$. For the differential spectra in $p_T$ (top row), we find at low $p_T$ there is almost no difference in the spectra depending on the choice of $\Delta x_\eta$. However, at high $p_T$ there is a stronger dependence, although once again we point out that the normalization there is very small which enhances these effects.
However, one key difference here when studying the initial grid size in the rapidity-directions vs the transverse plane, we note that there is more of an $h$ dependence in the differential spectra. We see that for $h=0.5$ there is even a few percentage level change  based on the grid size. 
Thus, we can conclude that at low $\sqrt{s_{NN}}$ the observables are more sensitive to our choice in the initial rapidity grid size and smoothing scale. 

Next also in Fig.\ \ref{fig:19p6spec_eta} (bottom row) we look at $dN/dy$.  For $h=0.3$ fm we see that too coarse of a grid in rapidity can be problematic, specifically $\Delta x_\eta=0.4$ can lead to errors above the $10\%$ level.  However, if one chooses a larger $h$ than that difference disappears. 
This result goes back to the idea of having enough nearest neighbors, if $h$ is too small and $\Delta x_\eta$ is too coarse then too few $N_{nb}$ leads to numerical error. Similarly, if $h$ is too large it can over-smooth the system, leaving out important small scale structure \cite{Noronha-Hostler:2015coa} (this appears to be starting to happen somewhat for $h=0.5$ fm). 
That being said we otherwise find a fairly robust regime of $h$, $\Delta x_x$, $\Delta x_\eta$ that produce equivalent results.

\begin{figure}[ht!]
\includegraphics[width=\linewidth]{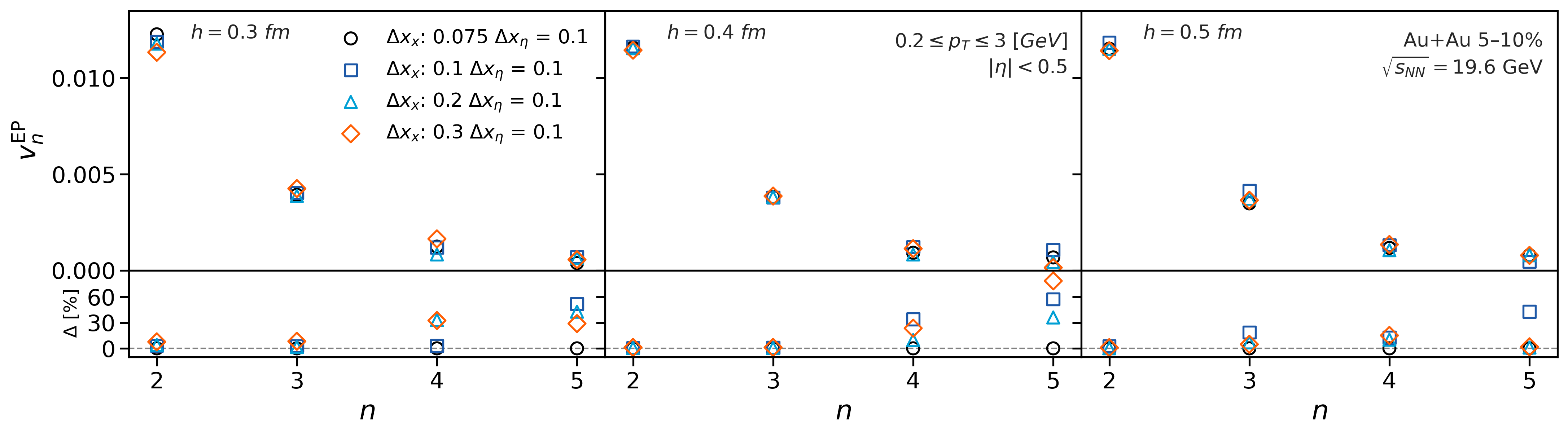}\\
 \includegraphics[width=\linewidth]{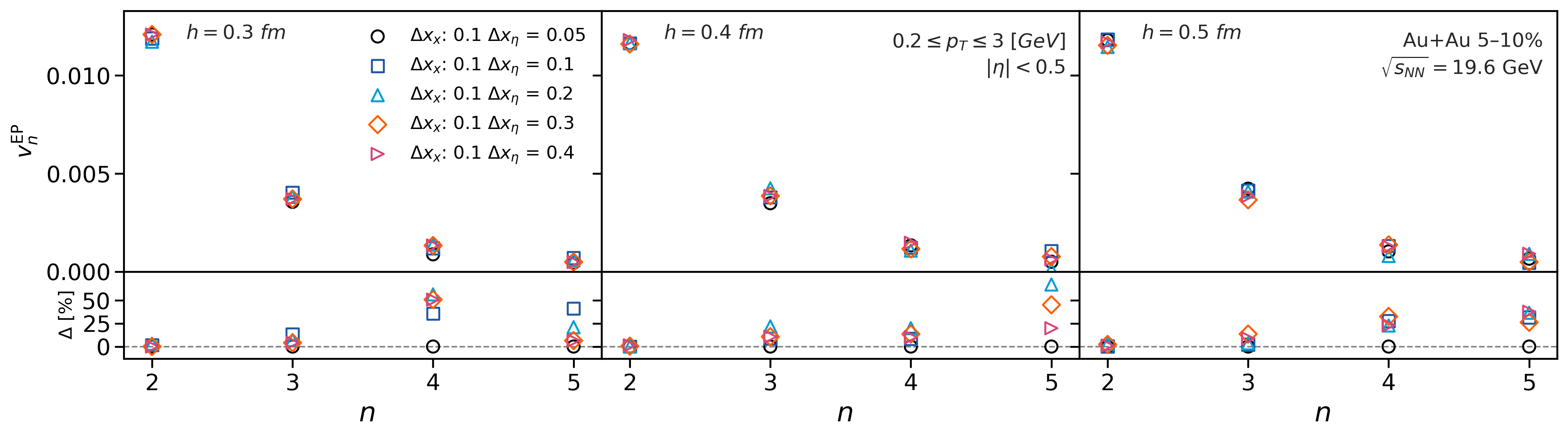}
    \caption{Grid size convergence test for $\Delta x_{\eta}$ while holding $\Delta_x=0.1$ fm fixed (top) and for $\Delta x_{\eta}=0.1$ while varying $\Delta_x \;[fm]$ (bottom). The integrated flow harmonics are shown for $n=2$--$5$. The columns different choices of the smoothing scale $h$ varied from 0.3--0.5 fm. }\label{fig:19p6vn}
\end{figure}

\noindent
{\bf Collective flow.--}
In Fig.~\ref{fig:19p6vn} we show our convergence tests on integrated flow. The top row compares the initial grid variation in the transverse plane, while fixing  $\Delta x_{\eta}=0.1$  and the bottom row compares the initial grid size variation in the rapidity direction, while fixing the $\Delta x_{x}=0.1$ fm. 
Once again, the columns vary the smoothing scale $h$. 

We first look at fixing $\Delta x_{\eta}=0.1$ and varying the transverse plane. 
Let us first start with $h=0.3$ fm, where we find some small dependence on $\Delta x_x$ for $v_2$ but it is less than a $10\%$ effect.  The triangular flow is largely independent of $\Delta x_x$, but $v_4$ and $v_5$ show significantly more sensitivity where on requires $\Delta x_x\lesssim 0.1$ fm to obtain $v_4$ within a $10\%$ error and $v_5$ requires even a smaller grid. 
If we next look at $h=0.4$ we find similar results for $v_2$ and $v_3$, but $v_4$ and $v_5$ are still quite sensitive to the initial transverse plane grid size ($v_5$ shows even less convergence). Finally, for $h=0.5$ we find that even $v_2$ is relatively insensitive to the grid size but $v_3$ begins to show more numerical error. Interestingly enough, $v_4$ and $v_5$ show less sensitivity to the transverse plane grid size, which is likely because $h=0.5$ fm is smoothing out important gradients relevant to $v_4$ and $v_5$. 
Thus, we would suggest running $h=0.3-0.4$ fm with $\Delta x_x\lesssim 0.2$ fm if one is only interested in $v_2$ and $v_3$, but then a smaller grid size if one also wants $v_4$ and $v_5$.

Finally, we come to the bottom row in Fig.\ \ref{fig:19p6vn} where we study the the influence of $\Delta x_{\eta}$ on the integrated flow. We find that the integrated flow harmonics are not as sensitive to our choice in $\Delta x_{\eta}$ (not the percentage difference scale has a smaller range in the bottom row of Fig.\ \ref{fig:19p6vn} compared to the top row). 
Generally, we find that $v_2$ and $v_3$ are fairly insensitive to our grid size choice in $\Delta x_{\eta}$ (at least below the $20\%$ error level).  However, we find that $v_4$ for $h=0.3$ fm is very sensitive to our choice of  $\Delta x_{\eta}$, which is likely because the of potential issues with too small of $h$ with too coarse of grid there when one studies a very sensitive observable like $v_4$.  A slightly larger $h=0.4$ fm appears to resolve this problem, although $v_5$ appears to be worse for $h=0.4$.  Looking at $h=0.5$ fm, the overall effects seems to be relatively consistent, although we actually find the very tiny grid size of $\Delta x_{\eta}=0.075$ appears to differ from nearly all the other results. 
We briefly comment, though, that one should not be too concerned by the large percentage error on $v_4$ and $v_5$ because we are once again normalizing by very tiny numbers (on the order of $\sim 0.001$) such that small differences lead to large percentage errors. We note that looking at $v_4$ data from HADES \cite{HADES:2022osk}, their errors are generally larger than what we are seeing here. 
Thus, given that our numerical error for $v_4$ and $v_5$ is small compared to the experimental data error, we also have a reasonable range of $\Delta x_\eta$ and $h$ that we can choose to run in \ccake{}.

\subsection{Optimization}
\label{Sec:optimization}
\noindent
Given the significant upgrades to \ccake{}~2.0 that include more complicated equations of motion and the move to 3+1D, the run time is significantly longer. Thus, optimizations were required to ensure that event-by-event simulations are still possible.  Below we detail parallelization and GPU-enabling efforts and how they influence the runtime.

\subsubsection{Libraries: \texorpdfstring{\kokkos}{Kokkos} and \texorpdfstring{\cabana}{Cabana}}%
\label{Sec:Hydrodynamics:Numerics:Kokkos}
\noindent
\texttt{\ccake{}~2.0} leverages the \texttt{C++} libraries \kokkos{}  \cite{9485033,CarterEdwards20143202} and \cabana{} \cite{Slattery2022,sam_reeve_2024_13844527} to achieve performance-portable parallelization across CPUs and GPUs.

\kokkos{} provides a high-level abstraction layer over parallel execution and memory management, enabling architecture-agnostic code that can target diverse back-ends such as OpenMP and CUDA. The configuration of execution and memory spaces involves specifying distinct roles for the Host (typically CPU) and Device (GPU, but can be CPU in non-accelerated environments), with data and computation mapped accordingly. \kokkos{} manages memory layout and placement across devices through View objects---independent, multidimensional arrays that are implicitly mapped to the appropriate memory space---allowing fine-grained control over access patterns while maintaining portability between host and device memory. When device-resident data must be accessed or updated on the host, such as in diagnostic routines or during implicit operations, we use \texttt{Kokkos::deep$\_$copy} to transfer data between memory spaces. This can become a performance bottleneck in certain routines, such as the online EoS inverter, where deep copies are triggered at each time step. In contrast, offline inversion strategies avoid this overhead by pre-computing look-up tables, resulting in significantly faster execution. We discuss this more in detail in Fig. \ref{fig:Gubsertest}, and Fig. \ref{fig:pixy_comparison} with corresponding details in Sec.  \ref{Sec:Hydrodynamics:Theory:EquationOfState}. 

\begin{figure}[ht!]
    \centering
    \includegraphics[width=0.85\linewidth,clip]{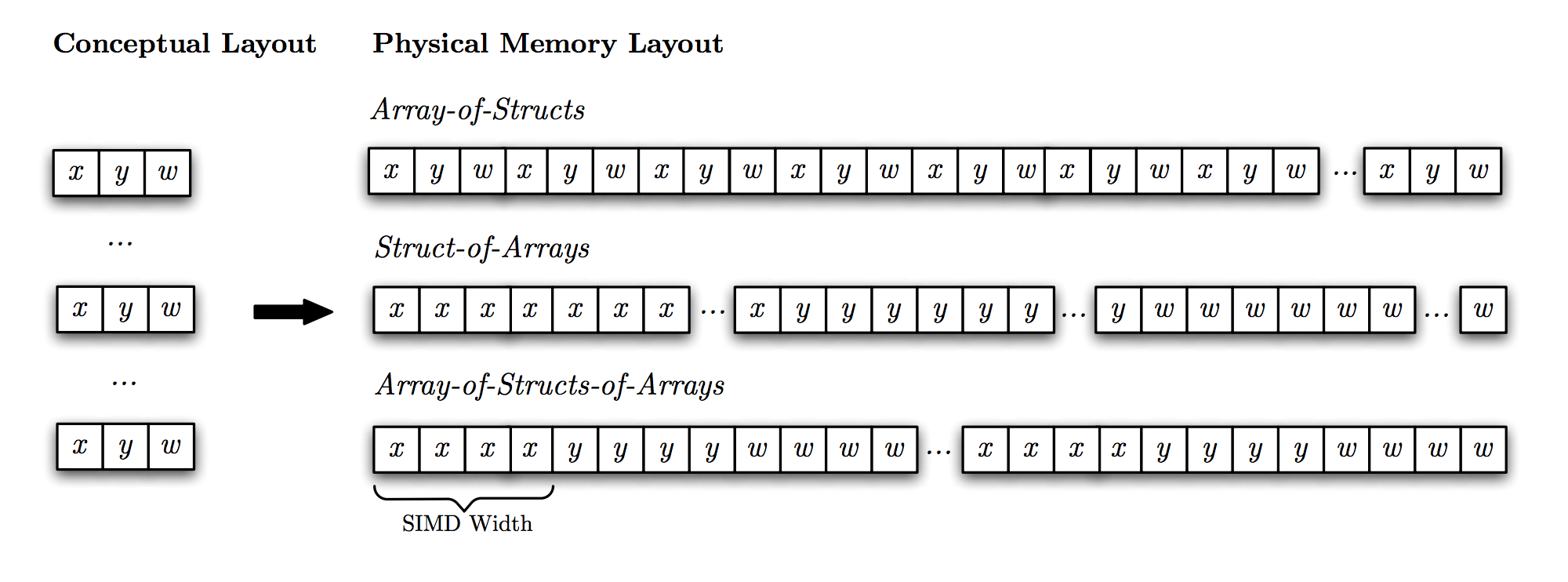}
    \caption{A visual representation that demonstrates how AoSoA type data is stored in memory \cite{doi:10.1177/10943420211022829}.}
    \label{fig:open_angle}
\end{figure}

\cabana{} builds on \kokkos{} and provides specialized support for particle-based simulations. It offers optimized data structures---specifically the Array-of-Structs-of-Arrays (AoSoA) layout---which inherit \texttt{Kokkos::View} memory abstraction features and improve memory access efficiency on modern hardware architectures. AoSoA is a hybrid of the traditional Array-of-Structs (AoS) and Struct-of-Arrays (SoA) layouts. As shown in Fig.~\ref{fig:open_angle}, AoS refers to a memory layout where each particle is represented as a struct containing particle properties, for instance, and these structs are stored sequentially in an array. This layout facilitates access to multiple properties of a single particle but impedes vectorization, which is critical for high-performance simulations involving large particle counts. In contrast, SoA stores each property of all particles in its own contiguous array, grouping data by property rather than by particle. This enables efficient vectorized operations across a single property but suffers from poor cache locality when accessing multiple properties of an individual particle. The AoSoA format balances these trade-offs by grouping small blocks of particles into struct-like chunks while preserving SoA-style field access within each block. This layout is particularly well-suited to SIMD (Single Instruction Multiple data) parallelization---a form of vectorized parallelization where a single instruction operates on multiple data elements simultaneously---and memory coalescing on GPUs, leading to high throughput in large-scale particle-based computations. \\

We also leverage the implementation of \texttt{VerletList} in \cabana{} to efficiently identify neighboring particles for SPH interactions. The domain is partitioned into a uniform grid, and the particles are spatially binned. The neighbors of each particle are drawn from nearby cells within a cut-off radius depending on the smoothing length, reducing the search cost from $\mathcal{O}(N^2)$ of a naïve brute-force algorithm to $\mathcal{O}(N)$.

\subsubsection{Time checks}%
\begin{figure}[ht!]
\centering
     \includegraphics[width=0.99\linewidth]{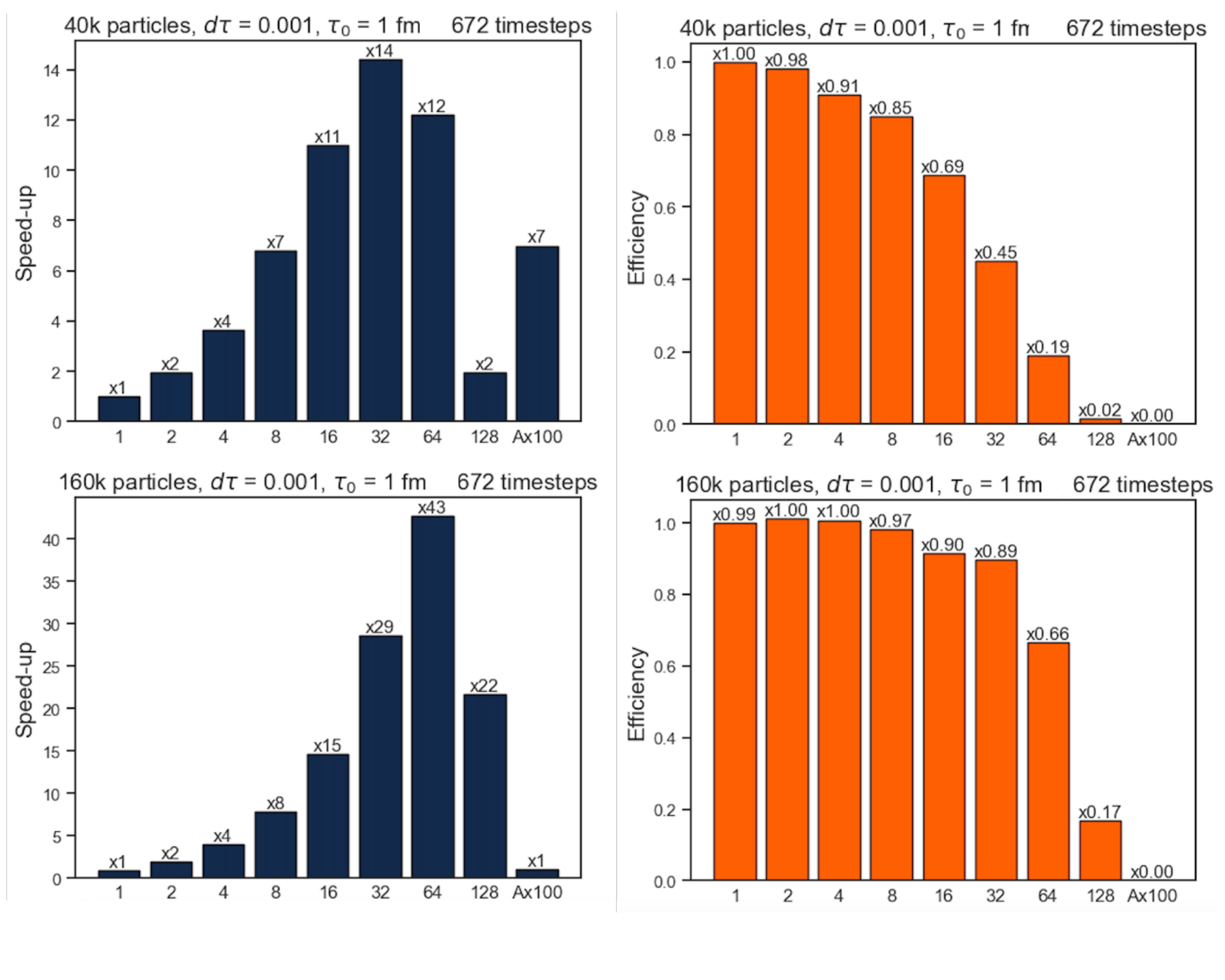}  
    \caption{Parallelization and efficiency time checks for 2+1D viscous Gubser semi-analytical solution. Top row shows a system with only 40k SPH particles (typical for 2+1D simulations), bottoms row shows a system with approximately 4x the number of SPH particles.  }
    \label{fig:gubser_speed_up}
\end{figure}

\label{Sec:Hydrodynamics:SPH:optimization}
\noindent
In Fig.~\ref{fig:gubser_speed_up} we show time checks that have been bench-marked using the semi-analytical Gubser check. We use the Gubser check since we already know the number of SPH particles required to accurately describe the solution, and given that nearly all codes benchmark against these semi-analytical checks it is straightforward for other groups to compare their code against ours. Moreover, we use the offline EoS inverter for these checks to avoid performance bottlenecks with the online inversion routine, which showed a significant reduction in parallelization benefits during testing.
The results were calculated  and averaged over 100 different runs of the Gubser check within \ccake{} to avoid small fluctuations in run time depending on hardware differences on high-performance computing nodes. 
Our Gubser check starts at $\tau_0=1$ fm and is run for 672 timesteps with a step-size in time of $d\tau=0.001$ fm. 
We compare parallelization across different number of cores (along the $x$-axis) versus speed up compared to the single core calculations. We also include NVIDIA A100 GPU in our checks. 
To better understand the importance of parallelization and its relation to the number of SPH particles in the system, we compare runs with 40,000 SPH particles vs. runs with 160,000 SPH particles.

We find that as one increases the number of cores, the code speeds up quite significantly. However, it is not a linear increase. In fact, at some point too many cores actually slows down the run time (this occurs going from 32 to 64 cores for 40k SPH particles and from 64 to 128 cores for 160K SPH particles).  
Generally, systems with more SPH particles see larger speeds up with parallelization. 
This implies that 3+1D simulations that have significantly more SPH particles can benefit the most from parallelization.

However, it is insufficient to run heavy-ion simulations just once to compare to experimental data. Rather, one must run the entire simulation framework (each event) thousands of times to compare to observables. 
Let us give the scenario were a group has access to a certain number of cores: $N_\mathrm{cor}$ and has to choose wisely how to run multiple events. 
Here we assume $X$ is the number of cores that the code is parallelized across. 
Thus, one is actually interested in knowing if which of the following options is the best:
\begin{enumerate}
    \item A given event is run on a single core, but $N_\mathrm{cor}$ events are simultaneously  running in parallel.  
    \item A given event is run in parallel across all $N_\mathrm{cor}$ cores. After that event is completed, the next event is ran across all $N_\mathrm{cor}$ cores. 
    \item A given event is run in parallel across a limited number of cores X. In parallel, other events are also run (each across X cores), allowing a total of $N_\mathrm{cor}/X$ events to be run at a given point in time. 
\end{enumerate}
In order to determine the best option, we now look at the efficiency of the parallelization defined as $\tau_\mathrm{run}/X$.

In Fig.~\ref{fig:gubser_speed_up} our efficiency results are shown, once again looking at 40k SPH particles (top) vs. 160k particles (bottom). One should always compare the results against the single core result (1 on the $x$-axis).  In terms of event-by-event simulations, there is only an advantage of the efficiency is greater than the single core result.  We find that the top figure demonstrates that event-by-event simulations for a small number of SPH particles shows no advantage parallelizing across cores, thus, option 1 is the best. 
However, when there are many SPH particles, there is a small advantage parallelizing across 2-4 cores, but larger number of cores receive a larger penalty. Thus, for very large systems and 3+1D simulations, then a small amount of parallelization can have an advantage (so option 3). 

\subsection{Freeze-out hypersurface construction}%
\label{Sec:Hydrodynamics:FreezeOutHypersurface}
\noindent
The constant $\varepsilon = \varepsilon_\mathrm{FO}$ freeze-out surface is determined by
\begin{equation}
\label{eq:efo_surface}
    \varepsilon_\mathrm{FO} =  \varepsilon \left( x = (x^0,x^1,x^2,x^3)\right)
\end{equation}
To calculate the surface element $\diff \sigma_\mu$ one starts by writing 
\begin{equation}
    \diff \sigma_\mu = |\diff \sigma_\mu| n_\mu,
\end{equation}
where $n_\mu$ is the normal vector to the surface.
Since the surface is given by Eq.~(\ref{eq:efo_surface}), the orthogonal vectors can be calculated by the gradient of this surface. Then, normalizing
\begin{equation}
\label{eq:fonormal}
    n_\mu = \frac{N_\mu}{\sqrt{|N_\mu N^\mu|}}=- \frac{\partial_\mu \varepsilon_\mathrm{FO} (x)}{ \sqrt{|\partial_\mu \varepsilon_\mathrm{FO} (x) \partial^\mu \varepsilon_\mathrm{FO} (x)|}},
\end{equation}
where the $-$ sign is chosen so that the vectors always point ``outwards", and the derivatives are calculated using the SPH method.
And then, the norm $|\diff\sigma_\mu|$, for SPH particle $k$, is \cite{Rone}:
\begin{equation}
    |\diff\sigma_\mu|_k =  \frac{m_k}{\sigma_k |u^\mu_k (n_{\mu})_k|}.
\end{equation}
With these two quantities, it is then possible to construct the freeze-out surface element $d\sigma_\mu$, which is then used in the next stage of the simulation. 
\ccake{} includes the functionality to also output other quantities at the point of freeze-out such as  the thermodynamical variables and velocity fields.

\section{Particlization}%
\label{Sec:Particlization}
\noindent
Once a hydrodynamic event is completed and the hypersurface is calculated, we must convert the fluid surface into particles, which is known as particlization.  
Our prescription for particlization contains both theoretical techniques to convert a fluid into hadrons, taking into account both a thermal distribution and out-of-equilibrium effects, as well as the numerical challenges that must be overcome to accurately describe the final experimental observables. 

The theoretical description of converting particles into hadrons is called Cooper--Frye that has the general format of $f_0+\sum_{\lambda=\eta,\zeta,\dots} \delta f_\lambda$ where the ideal contribution to the distribution function is contained in $f_0$ and it is assumed to be the dominant contribution, then the out-of-equilibrium contribution(s) are contained in $\delta f$ where each type of transport coefficient has its own $\delta f$ term that are summed together. 

In principle, applying our SPH method to the Cooper--Frye formula one could analytically calculate the produced particles as long as one assumes a continuous spectra of particles, i.e., not discrete particle. 
What that implies is that a given fluid cell could produce a fractional contribution to a particle, i.e., there could be 1/2 of a pion. 
From a continuous spectra of particles it is possible to use direct decays (i.e., heavier resonances decays into lighter daughter particles) and calculate all needed collective flow observables, as has been done in a number of previous works e.g. \cite{Kolb:2002ve}. 
However, direct decays do not include $2\leftrightarrow2$ reactions, nor re-scatterings that can affect the overall $p_T$-dependence of the spectra. 
Thus, an alternative approach is to sample discrete particles from an event (the averages for each particle yield is determined from the analytical formulas as well as the underlying distribution to sample from).  
In this way, the sampled event mimics an actual event that produces a similar number particles on an event-by-event basis. 

However, one of the challenges is that a given event $\left\{\mathrm{ev}\right\}$ produces $N^{\left\{\mathrm{ev}\right\}}\sim \mathcal{O}(10^3$--$10^6)$ particles (across the entire coordinate space), which is often not enough to fully reconstruct the spectra and collective flow of that event, especially if one considers identified particles and/or slices in rapidity or $p_T$. 
In experiments, billions of events are collided such that even if a single event's underlying spectra and flow are not completely accurately resolved, we calculate global observables such that these have enough statistics to converge. 
However, in theoretical calculations we do not have enough computational resources to recreate these statistics. 
To avoid numerical artifacts when reconstructing the spectra of $Q_n$ vectors, we then over-sample the event multiple times such that we can accurately represent the spectra and collective flow of that event without introducing numerical error. 
Every time a given event $\left\{\mathrm{ev}\right\}$ is sampled we index the sample with $\left\{s\right\}$. 
Each individual sample is then later ran through a hadronic afterburner, see Sec.~\ref{Sec:Particlization:Rescattering}, and later used to reconstruct the observables for that event. 

Throughout this section, we choose to indicate the purely numerical quantities of the sample and event number in a very specific notation using curly brackets $\left\{\dots\right\}$ because they should not be confused with physical dependencies like the freeze-out hypersurface, coordinates, or the equation of state. 
For the remainder of this section, we will distinguish numerical quantities related to sampling by surrounding them by curly brackets.

In the following, we first describe the theoretical framework of Cooper--Frye that provides our distributions from which we sample in Sec.~\ref{Sec:Particlization:CooperFrye}. Additionally, that section includes the continuous spectra approach that his analytical within SPH. 
Then we walk through our sampling procedure in Sec.~\ref{Sec:Particlization:Sampling} with specific focus on how we enforce BSQ conserved charges globally in a given sample of an event in Sec.~\ref{Sec:Particlization:Sampling:BSQ}. 
Finally, we perform convergence tests in Sec.~\ref{Sec:Particlization:SamplingTests} to determine the number of particles required, which we find depends on the type of particle we are observing.

\subsection{Cooper--Frye (continuous)}%
\label{Sec:Particlization:CooperFrye}
\noindent
The particle distribution for each species i can be calculated from the freeze-out surface using the Cooper--Frye prescription \cite{Cooper:1974mv} 
\begin{equation}
\label{eq:cooperfrye}
E \frac{\diff ^3 N_i}{\diff k^3} \left(x^\mu,k^\mu\right) =\int_{\sigma} k^\nu \diff ^3 \sigma_\nu (f_{0i}(x^\mu,k) + \delta f_i(x^\mu,k^\mu)),
\end{equation}
where $\diff ^3\sigma_\nu$ the infinitesimal freeze-out surface element, $f_{0i}$ the equilibrium distribution is:
\begin{equation}
   f_{0i}({\vec{k}})=  \frac{g_i}{(2\pi)^3} \frac{1}{\exp\left((u(x)\cdot k- \sum_q q_i\mu_q(x))/T\right) \pm 1 },
\end{equation}
with  $+1$ for fermions and to $-1$ for bosons,  $u$  is the four-velocity, $g_i $ the spin degeneracy and $q_i=B_i,S_i,Q_i$ the conserved charge carried by the particle species  $i$.
The $\delta f_i$ term represents the out-of-equilibrium corrections. For the sampling procedure, we take into account only the shear contributions \cite{Noronha-Hostler:2014dqa}.

\subsection{Sampling}%
\label{Sec:Particlization:Sampling}
\noindent
To couple the hydrodynamic simulation to a hadronic transport model, it is not enough to just calculate the particle distributions, but rather one requires information about individual particles. This is done by a Monte Carlo sampling procedure, where the Cooper--Frye distribution in Eq.~(\ref{eq:cooperfrye}) is interpreted as the probability distribution of species $i$ in the phase-space. We generalize the algorithm  \cite{McNelis:2019auj,Shen:2014vra} to take into account the sampling and global conservation of multiple charges.

The sampling algorithm interprets the integrand in Eq.~(\ref{eq:cooperfrye}) as a phase-space probability density associated with a fluid cell. This probability is used to randomly produce particles which are consistent with the properties of the fluid cell. 
Here we outline the steps in our implementation of the algorithm.

\begin{enumerate}
    \item \emph{Pre-sampling:} 
        A known numerical issue is that the Cooper--Frye distribution may have negative contributions at specific $p_T$ or $\phi$. These negative contributions occur for two reasons: a.) spacelike surface element $|p^\nu d^3 \sigma_\nu| <0$ which indicates that a particle is moving into the fluid, or b.) $f_{0i}+\sum_\lambda \delta f_{i,\lambda} <0 $ (with $\lambda=\{\pi^{\mu\nu},\Pi,J^\mu\}$) which may occur at large $p_T$ when out-of-equilibrium corrections are too large. One way to handle this issue is through the modification of the Cooper--Frye formula with a step function: 
        \begin{equation}
        \label{eq:mod-cooperfrye}
        E \frac{\diff ^3 N_i}{\diff k^3} \left(x^\mu,k^\mu\right) =\int_{\sigma} k^\nu \diff ^3 \sigma_\nu f_{i}(x^\mu,k) \Theta (f_{i}(x^\mu,k)) \Theta(k^\nu \diff ^3 \sigma_\nu).
        \end{equation}
        The physical interpretation of the step function is that only outgoing particles with a positive probability distribution are being sampled. This step is done prior to sampling.
    \item 
        The average number of particles emitted (per cell) is given by:
        \begin{equation}
        \label{eq:sampledN}
        \Delta N_a = \sum_{i=\pi,K,\dots}\int_{\sigma} k^\nu \diff ^3 \sigma_\nu f_{i}(x^\mu,k) \Theta (f_{i}(x^\mu,k)) \Theta(k^\nu \diff ^3 \sigma_\nu).
        \end{equation}
        Given that hadronization is quantum mechanical and we require discrete
        particle distributions, the actual number of particles emitted from a
        freeze–out cell will fluctuate from event to event.  
        Thus, for any specific Monte Carlo sampling realization $\{s\}$ of an
        event $\{ev\}$, the sampled particle number from cell $a$,
        $\Delta N_a^{\{s\}}$, will generally differ from the hydrodynamic mean
        value $\Delta N_a$. Therefore, we draw the number of emitted particles from a Poisson
        distribution with mean $\Delta N_a$:
        \begin{align}
            P(N)=\frac{(\Delta N_a)^N}{N!}\,.
            \label{eq:poisson}
        \end{align}
        
        The superscript $\{s\}$ labels an individual Monte Carlo sampling of
        the same hydrodynamic background.  
        The hydrodynamic quantity $\Delta N_a$ is a smooth, deterministic mean,
        whereas a microscopic event must contain an integer number of particles
        that fluctuates around this mean value.  
        Accordingly, for each sampling realization $\{s\}$ we draw
        \begin{equation}
        \Delta N_a^{\{s\}} \sim \mathrm{Poisson}(\Delta N_a).
        \end{equation}
        This procedure generates the correct event-by-event multiplicity
        fluctuations, while the averaging over many samples satisfies
        $\langle \Delta N_a^{\{s\}} \rangle = \Delta N_a$, as discussed
        later in the section.
        
        Here $\Delta N_{a}$ is the total mean number of particles emitted from
        cell $a$, defined as the sum over hadron species,
        \begin{equation}
        \Delta N_{a} \equiv \sum_{i=\pi,K,\dots} \Delta N_{a,i},
        \end{equation}
        with the species--dependent contributions given by
        \begin{equation}
         \Delta N_{a,i}
         = \int_{\sigma}
           k^\nu d^3\sigma_\nu \,
           f_i(x^\mu,k)\,
           \Theta(f_i(x^\mu,k))\,\Theta(k^\nu d^3\sigma_\nu).
        \end{equation}
        where individual particles across a range of momentum are considered.
        The spacelike surface elements, regions with $p \cdot d^3\sigma < 0$ make the integral harder to evaluate, reducing the sampling efficiency. Therefore, the general strategy is to perform rejection sampling whenever the original function is not trivial to sample.
        
        For the hadron number, we start by writing 
        \begin{align}
        \int_{\sigma} k^\nu \diff ^3 \sigma_\nu f_{i}(x^\mu,k^\mu) \Theta (f_{i}(x^\mu,k^\mu)) \Theta(k^\nu \diff ^3 \sigma_\nu) \leq 2|d^3\sigma|\int_k (u(x) \cdot k) f_{0i}\,.
        \label{eq:reduced_yield}
        \end{align}
        Which then defines the envelope (upper bound) for the mean particle yield from cell $a$ for species $i$ as
        \begin{align}
        \label{eq:upperN}
        \Delta N_{a,i}^{\mathrm{max}}
        &\equiv
        2\,|d^3\sigma_a|\ \int_k (u(x) \cdot k) f_{0i}\\
        &= \frac{|d^3\sigma_a|g_i}{\pi^2 \hbar^3} \int^\infty_0 \frac{k^2_{\text{LRF}} \,  dk_{\text{LRF}} }{\exp\left[\dfrac{\sqrt{k^2_{\text{LRF}} + m_i^2}- \sum_q q_i\mu_q(x))}{T(x)}\right]\pm 1},
        \end{align}
        with  $+1$ for fermions and to $-1$ for bosons,
        which replaces the full Cooper--Frye integrand (including step functions and regulated viscous corrections) by its maximal possible value which satisfies from Eq.~(\ref{eq:reduced_yield}) that
        \begin{align}
           \Delta N_{a,i} \le \Delta N^{\mathrm{max}}_{a,i} 
        \end{align}
        for all momenta. Then, we sample the Poisson distribution in Eq.~(\ref{eq:poisson}) with the maximum mean $\Delta N_{a,i}^{\mathrm{max}}$, replacing the original hadron number, giving us an overestimated number of hadrons.
        By setting this upper bound from Eq.~(\ref{eq:upperN}) or ``the envelope", the number of sampled hadrons is guaranteed to be larger than the true count from Eq.~(\ref{eq:sampledN}) for all momenta and viscous corrections. 
        This makes it safe to over sample and then filter which particles to keep later, once we have the four momentum sampled, using an acceptance weight. 
    \item 
        For a given, individual sampled particle $\left\{i\right\}$ , one must next determine its local momentum distribution in the local rest frame from the true local momentum distribution:
        \begin{align}
        Q_i(\vec{k})
          = \frac{2|d^3\sigma| \, f_{0i}(\vec{k})}
            {\Delta N_{a,i,\mathrm{max}}},
        \end{align}
        which includes a non-trivial dependence on momentum and mass. 
        In addition, it is not easy to sample randomly from the above distribution directly. 
        The trick then is to use an auxiliary random variable $\vec{r}$ with a known and easy-to-sample probability density $R_i(\vec{r})$. 
        Momenta are then generated from the proposed distribution $R_i(\vec{r})$,
        using some coordinate transformation from $\vec{r}$ to $\vec{k}$ such that
        \begin{equation}
          Q_i(\vec{k})\,d^3k
          = C \times R_i(\vec{r})\,d^3r
          \times w_i(\vec{k}),
        \end{equation}
        where $C$ is a normalization constant. Here $C$ is chosen such that 
        $C\,R_i(\vec{r}) \ge Q_i(\vec{k})$ for all $\vec{k}$,
        ensuring that the acceptance probability 
        $w_i(\vec{k}) = Q_i(\vec{k})/[C R_i(\vec{r})]$
        lies in the interval $[0,1]$ required by the acceptance--rejection method.
        A sample will be accepted with probability $w_i(\vec{k})$; otherwise it is redrawn until acceptance. 
        
        Typically, $R_i(\vec{r})$ is chosen depending on the particle being sampled \cite{Pang:2018zzo} :
        \begin{itemize}
        \item \textbf{Pions.--}     
            Since pions are the lightest known hadrons, the most efficient function is the massless Boltzmann distribution,
            \begin{align}
               R_i(\vec{r}) \propto \exp{(-k/T)} \,k^2,
            \end{align}
            which is sampled using the trick from \cite{Pang:2018zzo}.
            The acceptance weight correcting the massless proposal to the true Bose distribution generalized for multiple charges is
            \begin{equation}
              w_\pi(\vec{k})
              = \frac{e^{k/T}}{e^{{E/T}- \sum_q q_i\mu_q /T} - 1},
              \qquad E = \sqrt{k^2 + m_i^2}.
              \label{eq:weight_pi_k}
            \end{equation}
            Due to the low pion mass, the denominator from Eq.~(\ref{eq:weight_pi_k}) can be larger than one for low-momentum/high-chemical potentials. To handle this issue, we check if $w_\pi(\vec{k}) \geq 1$ for any $k=|\vec{k}|$ and the particular $(m_i/T,\sum_q q_i\mu_q /T) $ combination considered; if true, we renormalize the weight by the maximum value allowed for this specific combination. The ``allowed region'' is determined by a curve calculated numerically, looping over all possible momenta for each pair $(m_i/T,\Omega_i  =\sum_q q_i\mu_q /T) $, and the maximum value for the pair is calculated numerically beforehand and interpolated during sampling. The maximum value and the allowed region for pions can be found on the left-hand side of Fig.~\ref{fig:sampling_weight}.
        
        \item \textbf{Heavy Hadrons.--}
            For heavy hadrons, we sample from the Boltzmann distribution in spherical coordinates
            \begin{align}
               R_i(\vec{r}) \propto \exp{( \sum_q q_i\mu_q /T  - E_i/T)} \,k^2,
            \end{align}
            which is sampled using the same procedure of \cite{Pang:2018zzo,McNelis:2019auj}.
            The acceptance weight correcting generalized for multiple charges is
            \begin{equation}
              w_i(\vec{k})
              =  \frac{k}{E_i} \frac{e^{{E_i/T}- \sum_q q_i\mu_q /T}}{e^{{E_i/T}- \sum_q q_i\mu_q /T} + \pm 1},
              \qquad E_i = \sqrt{k^2 + m_i^2}.
              \label{eq:weight_heavy_hadrons_k}
            \end{equation}
            with  $+1$ for fermions and to $-1$ for bosons.
            For baryons, the weight is guaranteed to be below or equal to one. For mesons, at very high chemical potentials, it is possible to have a weight greater than one. For this, we proceed in the same way as done for the pions, and the maximum value for the pair is calculated numerically beforehand and interpolated during sampling. The maximum value and allowed region for pions can be found on the right-hand side of Fig.~\ref{fig:sampling_weight}.
            \end{itemize}
            
            \begin{figure}[ht!]
            \centering
                 \includegraphics[width=0.45\linewidth]{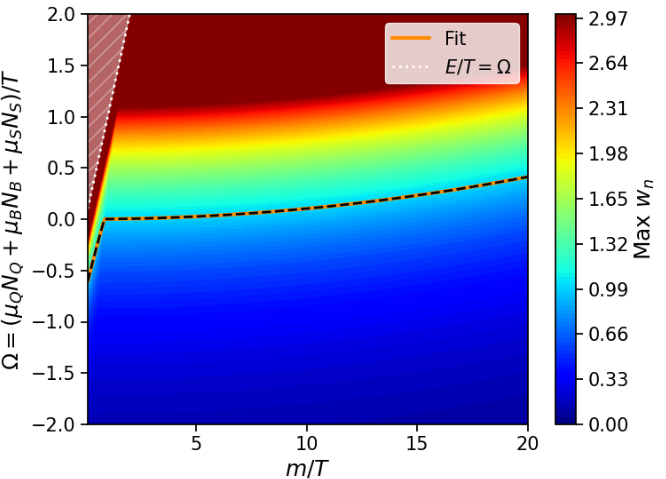} 
                 \includegraphics[width=0.49\linewidth]{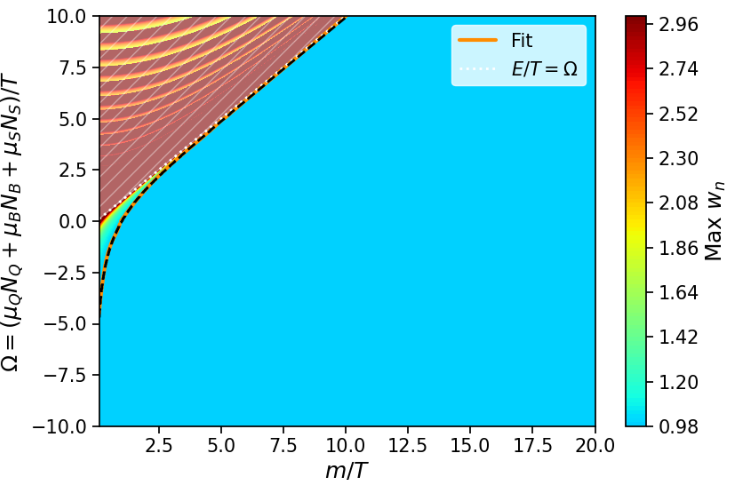}
                \caption{Maximum weight value for any momenta $k$ in the $(m/T, \Omega = \sum_q q_i\mu_q /T)$ plane.
                }
                \label{fig:sampling_weight}
            \end{figure}
    \item
        Once we have the momentum, we can decide to accept or discard the particle, to recover the original Cooper--Frye distribution. 
        The acceptance weight is defined by 
        \begin{equation}
          w_{\mathrm{keep},i}(k)
          = \underbrace{\frac{k \cdot d^3\sigma \,
          \Theta(k \cdot d^3\sigma)}{(u \cdot k)|d^3\sigma|}}_{w_{d\sigma}}
          \times
          \underbrace{\frac{1}{2}\left(1 + \frac{\delta f_{i}}{f_{0i}}\right)}_{w_{\delta f}}\,.
          \label{eq:weight}
        \end{equation}
        Here $w_{d\sigma}(k)$ is a purely geometric flux factor that depends only on the
        hypersurface normal and the particle momentum, and is therefore
        \emph{species-independent}.
        In contrast, $w_{\delta f,i}(k)$ is \emph{species-dependent}, as both the
        equilibrium distribution $f_{0i}(k)$ and the regulated viscous correction
        $\delta f_{i}(k)$ depend on the hadron mass, statistics, and local
        thermodynamic conditions.
        The product $w_{\mathrm{keep},i}(k)$ satisfies
        $0 \le w_{\mathrm{keep},i}(k) \le 1$ by construction, ensuring a valid
        accept--reject probability.
        Applying this species-dependent keep-weight to each oversampled particle
        yields, on average, reproduces the correct multiplicities and momentum distributions for
        all hadron species, as prescribed by the underlying Cooper--Frye formula.
    \item 
        The last step before we store the data is to compute the lab frame momenta and also set the lab frame position to that of the freeze-out cell. Lastly, the particle list of each event is written to be integrated with \smsh{}. 

\end{enumerate}
\noindent
After Steps $\#1\rightarrow\#5$ have been performed for every hadron species and for every
freeze–out cell on the hypersurface, the sampling routine produces, for each
Monte Carlo event $\{ev\}$, a complete list of particles together with their
lab–frame momenta and positions.  
This list represents a microscopic realization of the Cooper–Frye emission
function for that particular event.  
From these sampled particle lists we can reconstruct the \emph{event–wise
momentum–space population densities}
\begin{equation}
\tilde{n}_i^{\{\mathrm{ev}\}}(k_\perp,\phi,\eta),
\end{equation}
which serve as Monte Carlo estimators of the underlying Cooper–Frye spectra.
These population densities form the basis for computing event–wise multiplicities,
charged–particle yields, and the event–averaged observables discussed in the next
section.

For a given event, one obtains the population densities at freeze-out  across momentum space for a given particle $i$, i.e., $\tilde{n}_i^{\left\{\mathrm{ev}\right\}}(k_\perp,\phi,\eta)$, using the analytical formulism. 
To be clear, the population densities are positive definite quantities where we do not include their corresponding conserved charges, i.e., $\tilde{n}_{\bar{\Lambda}}\geq 0$ even though $\bar{\Lambda}$'s are  anti-baryons. 
One can always reconstruct the net-number densities by summing over all relevant particles multiplied by their conserved charge, i.e., for strangeness this would be:
\begin{equation}
    n_s=\sum_{i} S_i \tilde{n}_i.
\end{equation}
Thus, we denote the population densities with tilde's to distinguish them from number densities of conserved charges. 
Then, if we integrate over the entire freeze-out surface we can determine the average number of particles produced within a given event:
\begin{eqnarray}
    \langle N_i\rangle ^{\left\{\mathrm{ev}\right\}}&\equiv&\int_{0}^{\infty}\int_{0}^{2\pi}\int_{\eta_{min}}^{\eta_{max}}\tilde{n}_i^{\left\{\mathrm{ev}\right\}}(k_\perp,\phi,\eta)dk_\perp d\phi d\eta.
\end{eqnarray}
Using these average particle numbers, we can also determine the average number of all particles produced in a given event:
\begin{equation}
    \langle N_\mathrm{all}\rangle ^{\left\{\mathrm{ev}\right\}}=\sum_i\langle N_i\rangle ^{\left\{\mathrm{ev}\right\}}
\end{equation}
as well as the average number of charged particles in that event with
\begin{equation}
    \langle N_\mathrm{ch}\rangle ^{\left\{\mathrm{ev}\right\}}=\sum_i\Theta\left(|Q_i|-1\right)\langle N_i\rangle ^{\left\{\mathrm{ev}\right\}},
\end{equation}
where the Heaviside function enforces that only particles that carry electric charge are counted.

To calculate observables that connect to data, one must then take a second average over events, i.e.,
\begin{equation}
    \langle  \langle N_\mathrm{ch}\rangle \rangle =\frac{1}{N_{\left\{\mathrm{ev}\right\}}}\sum_{\left\{\mathrm{ev}\right\}}^{N_{\left\{\mathrm{ev}\right\}}} \langle N_\mathrm{ch}\rangle ^{\left\{\mathrm{ev}\right\}}
\end{equation}
where $N_{\left\{\mathrm{ev}\right\}}$ is the number of events we are considering and we use the double angular brackets $\langle\langle\dots\rangle\rangle$ to indicate an average over both the samples of events and then again over all events. 

In a given sample of an event, we obtain a yield $N_i^{\left\{s,\mathrm{ev}\right\}}$ for each individual particle $i$  that is not necessarily equal to the average particle yield for that particle, i.e.,  $N_i^{\left\{s,\mathrm{ev}\right\}}\neq \langle N_i\rangle ^{\left\{\mathrm{ev}\right\}}$.
Instead, $N_i^{\left\{s,\mathrm{ev}\right\}}$ is sampled from a Poisson distribution around $\langle N_i\rangle ^{\left\{\mathrm{ev}\right\}}$  such that if enough samples are taken, the average over the samples should reproduce $\langle N_i\rangle ^{\left\{\mathrm{ev}\right\}}$, i.e., 
\begin{equation}
    \langle N_i\rangle ^{\left\{\mathrm{ev}\right\}}=\frac{1}{N_{\left\{s\right\}}} \sum_{\left\{s\right\}}^{N_{\left\{s\right\}}}N_i^{\left\{s,\mathrm{ev}\right\}}.
\end{equation}
where $N_{\left\{s\right\}}$ is the number of times we have sampled the event (this is generally a fixed quantity that does not fluctuate event-to-event). 
Once $N_i^{\left\{s,\mathrm{ev}\right\}}$ is determined from a given event, the underlying momentum space distribution is sampled $N_i^{\left\{s,\mathrm{ev}\right\}}$ times (indexed by $\left\{j\right\}$) to obtain the momentum coordinates of the $\left\{j\right\}$-th particle of particle type $i$:  $\vec{k}_{i,\left\{j\right\}}=\left(k_\perp,\phi,\eta\right)$.
We then have the differential spectra of particles that can be passed onto a hadronic afterburner. 

\subsubsection{BSQ global charge conservation}%
\label{Sec:Particlization:Sampling:BSQ}
\noindent
The above sampling scheme is well-established in the field of heavy-ion collisions and works well in the limit of high-energies where BSQ charge conservation does not play a strong role for certain observables e.g. all charged particle collective flow.
However, certain observables are more sensitive to charge conservation \cite{Denicol:2018wdp,Gardim:2024nyz,Sousa:2025oqf,Pihan:2024lxw} and at the bare minimum, sampling codes should conserve BSQ charges globally. 
It is not yet as clear how important local charge conservation is so we leave that problem for a future work. 

Previous work has already established an algorithm to locally conserve charges in grid-based codes \cite{Oliinychenko:2019zfk,Oliinychenko:2020cmr}. However, this has not yet been attempted in Lagrangian codes. Thus, we develop here an algorithm for global charge conservation for SPH codes (see Tab. \ref{tab:Samp_Summary} for an overview of calculated vs. sampled quantities). Likely this could be extended to local charge conservation by using clustering algorithms for SPH particles, but we leave this for a future work. 

To begin, we need to establish how to calculate the total \emph{population} yields of a specific set of conserved charges.
We select on a specific set of $B,S,Q$ that may be positive or negative using:
\begin{equation}
    \tilde{N}_{B,S,Q}^{\left\{s,\mathrm{ev}\right\}}\equiv\sum_{i}\delta\left(B_{i}-B\right)\delta\left(S_{i}-S\right)\delta\left(Q_{i}-Q\right) N_i^{\left\{s,\mathrm{ev}\right\}},
\end{equation}
where we sum over all particles in a given sample of a given event. 
Within the summation, we only obtain non-zero values of the $\delta(x_i-x)$ functions when $x=x_i$, i.e., 
\begin{equation}
   \delta(x_i-x)=\left\{\begin{matrix}
        1& x_i=x\\
        0& \text{else}
    \end{matrix}\right.
\end{equation}
such that this sum obtains a specific choice for $B,S,Q$ values (these $B,S,Q$ values may be positive or negative). 
We emphasize here that  $\tilde{N}_{B,S,Q}^{\left\{s,\mathrm{ev}\right\}}$ is positive definite, just as the population number density was. 

Then, in order to obtain the net conserved charges (note these may be positive or negative), we calculate:
\begin{eqnarray}
    N_{B}^{\mathrm{net},\left\{\mathrm{ev}\right\}}&\equiv&\sum_iB_i \tilde{N}_i^{\left\{s,\mathrm{ev}\right\}} \\
    &=&\tilde{N}_{+1,a,a}^{\left\{s,\mathrm{ev}\right\}}-\tilde{N}_{-1,a,a}^{\left\{s,\mathrm{ev}\right\}} \label{eqn:netBcalc}\\
    N_{S}^{\mathrm{net},\left\{\mathrm{ev}\right\}}&\equiv&\sum_iS_i \tilde{N}_i^{\left\{s,\mathrm{ev}\right\}} \\
    &=&3\tilde{N}_{-1,+3,-1}^{\left\{s,\mathrm{ev}\right\}}+2\tilde{N}_{-1,+2,a}^{\left\{s,\mathrm{ev}\right\}}+\tilde{N}_{-1,+1,a}^{\left\{s,\mathrm{ev}\right\}}+\tilde{N}_{0,+1,a}^{\left\{s,\mathrm{ev}\right\}}\nonumber\\
    & &-\tilde{N}_{0,-1,a}^{\left\{s,\mathrm{ev}\right\}}-\tilde{N}_{+1,-1,a}^{\left\{s,\mathrm{ev}\right\}}-2\tilde{N}_{+1,-2,a}^{\left\{s,\mathrm{ev}\right\}}-3\tilde{N}_{+1,-3,-1}^{\left\{s,\mathrm{ev}\right\}}\label{eqn:netS_contributions}\\
    N_{Q}^{\mathrm{net},\left\{\mathrm{ev}\right\}}&\equiv&\sum_iQ_i \tilde{N}_i^{\left\{s,\mathrm{ev}\right\}}\\
     &=&  2\tilde{N}_{+1,a,+2}^{\left\{s,\mathrm{ev}\right\}}+\tilde{N}_{a,a,+1}^{\left\{s,\mathrm{ev}\right\}}-\tilde{N}_{a,a,-1}^{\left\{s,\mathrm{ev}\right\}}-2\tilde{N}_{-1,a,-2}^{\left\{s,\mathrm{ev}\right\}}\label{eqn:netQcalc}
\end{eqnarray}
where we use the symbol $a$ to indicate that all possible values of that conserved charge are considered, e.g., for Strangeness $a=\pm 3$, $\pm 2$, $\pm 1$, or $0$.
Note that $N_{X}^{\mathrm{net},\left\{\mathrm{ev}\right\}}$ where $X=B,S,Q$ does \emph{not} have a superscript of $\left\{s\right\}$ because these are conserved quantities that cannot depend on an individual sample. Rather, they must be exactly identical in every sample. 
Thus, an individual sample can achieve a fixed $N_{X}^{\mathrm{net},\left\{\mathrm{ev}\right\}}$ in a variety of different ways, but the net numbers must be conserved.
For instance, if $N_{B}^{\mathrm{net},\left\{\mathrm{ev}\right\}}=3$ this could be achieved with $\tilde{N}_{+1,a,a}^{\left\{s,\mathrm{ev}\right\}}=3, \tilde{N}_{-1,a,a}^{\left\{s,\mathrm{ev}\right\}}=0$ or $\tilde{N}_{+1,a,a}^{\left\{s,\mathrm{ev}\right\}}=5, \tilde{N}_{-1,a,a}^{\left\{s,\mathrm{ev}\right\}}=2$, which could be different different samples. 

Let us now walk through some examples of our notation to make this clearer. The subscript of $\tilde{N}_{B,S,Q}^{\left\{s,\mathrm{ev}\right\}}$ will separate out particle species with specific quantum numbers e.g. $\tilde{N}_{+1,-3,-1}^{\left\{s,\mathrm{ev}\right\}}$ indicates $\Omega^-$ baryons and their resonances. 
If a subscript $a$ is included, that implies all possible combinations are considered, i.e., $\tilde{N}_{+1,a,a}^{\left\{s,\mathrm{ev}\right\}}$ implies that all baryons are consider (include all strangeness and electric charge combinations) but not anti-baryons.
Using our notation, net-baryons are defined as:
\begin{equation}
    N_B^{\mathrm{net},{\left\{\mathrm{ev}\right\}}}= \tilde{N}_{+1,a,a}^{\left\{s,\mathrm{ev}\right\}}-\tilde{N}_{-1,a,a}^{\left\{s,\mathrm{ev}\right\}}
\end{equation}
where all strangeness and electric charge combinations are considered.

\begin{table}[ht!]
    \centering
    \begin{tabular}{c|cc}
    \hline
       & Calculated  & Sampled \\
       \hline
        $N_{B}^\mathrm{net}$ &  Eq.~(\ref{eqn:netBcalc}) &  \\
        $N_{S}^\mathrm{net}$ &  Eq.~(\ref{eqn:netS_contributions}) &  \\
        $N_{Q}^\mathrm{net}$ &  Eq.~(\ref{eqn:netQcalc}) &  \\
      $N_{+1,a,a}^{{\left\{s,\mathrm{ev}\right\}}}$ &   &  \checkmark\\
       $N_{-1,a,a}^{{\left\{s,\mathrm{ev}\right\}}}$ &  Eq.~(\ref{eqn:antibar_samples}) &  \\
       $N_{SB}^\mathrm{net}$ &  Eq.~(\ref{eqn:NSBcalc}) &  \\
        $N_{0,-1,a}^{{\left\{s,\mathrm{ev}\right\}}}$ &   & \checkmark \\
       $N_{0,+1,a}^{{\left\{s,\mathrm{ev}\right\}}}$ &  Eq.~(\ref{eqn:SMcalc}) &  \\
       $N_{BQ}^\mathrm{net}$ &  Eq.~(\ref{eqn:NBQcalc}) &  \\
        $N_{SMQ}^\mathrm{net}$ &  Eq.~(\ref{eqn:NSMQcalc}) &  \\
        $N_{0,0,+1}^{{\left\{s,\mathrm{ev}\right\}}}$ &   &  \checkmark\\
        $N_{0,0,-1}^{{\left\{s,\mathrm{ev}\right\}}}$ & Eq.~(\ref{eqn:NMnegcalc})  &  \\
        $N_{0,0,0}^{{\left\{s,\mathrm{ev}\right\}}}$ &  &  \checkmark\\
        $N^\mathrm{ch}$ & \checkmark & \\
       \hline
    \end{tabular}
    \caption{Summary table of the calculated vs. sampled quantities in our BSQ sampling technique. One starts at the top and works downwards in terms of calculations and sampling. See also the flow chart in Fig.~\ref{chart:sampling_BSQ}. }
    \label{tab:Samp_Summary}
\end{table}

Our algorithm starts by sampling all baryons $N_{+1,a,a}^{{\left\{s,\mathrm{ev}\right\}}}$ (including all light and strange baryons, but not including any anti-baryons) where the only constraint that we must have is that
\begin{equation}
    N_{+1,a,a}^{{\left\{s,\mathrm{ev}\right\}}}\geq N_B^{\mathrm{net},{\left\{\mathrm{ev}\right\}}}.
\end{equation} 
Then, we determine the number of anti-baryons by taking
\begin{equation}\label{eqn:antibar_samples}
N_{-1,a,a}^{{\left\{s,\mathrm{ev}\right\}}}\equiv N_{+1,a,a}^{{\left\{s,\mathrm{ev}\right\}}}-N_B^{\mathrm{net},{\left\{\mathrm{ev}\right\}}},
\end{equation}
where again this includes all baryons (both light and strange).
The value of $N_{-1,a,a}^{\left\{s,\mathrm{ev}\right\}}$ must be exactly constrained to Eq.~(\ref{eqn:antibar_samples}) to enforce baryon number conservation.

Our next step is to randomly sample the $S=-1$ mesons, i.e., $N_{0,-1,a}^{\left\{s,\mathrm{ev}\right\}}$. 
Since we must require strangeness neutrality (with the one extremely rare exception being if hypernuclei were ever collided), then generally we will always have $N_S^{\mathrm{net},\left\{\mathrm{ev}\right\}}=0$, but we leave in the notation of a generic $N_S^{\mathrm{net},\left\{\mathrm{ev}\right\}}$ to allow for a future extension to local charge conservation.
We have no constraints whatsoever on $N_{0,-1,a}^{\left\{s,\mathrm{ev}\right\}}$ such that it can be any positive integer. 
At this point, we have already ensured that the net-baryon number is conserved, but we need to ensure that strangeness is also conserved. 
To do so, we consider all strangeness contributions to net-strangeness, as was done in Eq.~(\ref{eqn:netS_contributions})
where we have separated out the baryons vs. meson contributions to the net-strangeness. 
Given that we have already sampled over all baryons, we can just calculate the net-strangeness contribution from the baryons, i.e.,
\begin{eqnarray}\label{eqn:NSBcalc}
    N_{SB}^{\mathrm{net},\left\{s,\mathrm{ev}\right\}}&\equiv& -N_{+1,-1,a}^{\left\{s,\mathrm{ev}\right\}}+N_{-1,+1,a}^{\left\{s,\mathrm{ev}\right\}}-2N_{+1,-2,a}^{\left\{s,\mathrm{ev}\right\}}+2N_{-1,+2,a}^{\left\{s,\mathrm{ev}\right\}}\nonumber\\
    &-&3N_{+1,-3,-1}^{\left\{s,\mathrm{ev}\right\}}+3N_{-1,+3,+1}^{\left\{s,\mathrm{ev}\right\}}
\end{eqnarray}
where our net-strange, baryon contribution $N_{SB}^{\mathrm{net},\left\{s,\,\mathrm{ev}\right\}}$ \emph{does} have a superscript $\left\{s\right\}$ because this can fluctuate on an sample to sample basis.
Then, we can calculate how many anti-strange mesons we require using
\begin{equation}\label{eqn:SMcalc}
    N_{0,+1,a}^{\left\{s,\mathrm{ev}\right\}}= N_{SB}^{\mathrm{net},\left\{s,\mathrm{ev}\right\}}-N_{0,-1,a}^{\left\{s,\mathrm{ev}\right\}}-N_S^{\mathrm{net},\left\{\mathrm{ev}\right\}}
\end{equation}
to ensure exact strangeness conservation in a given sample of an event. 

\begin{figure}[ht!]
\centering
\begin{tikzpicture}[node distance=1.6cm and 3cm]
\node (n1a) [startstop] at (-3.2,3.2) {Read-in freeze-out hypersurface $\&$ \\ construct $\tilde{n}_i$ };
\node (n1b) [process, below of=n1a] {Compute $\langle N_i \rangle^{\{\mathrm{ev}\}}$ from $\tilde{n}_i$};
\node (n1c) [process, below of=n1b] {Sample $N_i^{\{s,\,\mathrm{ev}\}} \sim \mathrm{Poisson}$};

\node (n2a) [process] at (-3.2,-2) {Sample $N_{+1,a,a}^{\{s,\,\mathrm{ev}\}}$};
\node (n2b) [process, below of=n2a] {Set $N_{-1,a,a}^{\{s,\,\mathrm{ev}\}}$};
\node (n2c) [decision, below=1cm of n2b] {Check $B^\mathrm{net}$ \\conservation};

\node (n3a) [process]at (2,-6.75) {Sample $N_{0,-1,a}^{\{s,\,\mathrm{ev}\}}$};
\node (n3b) [process, below of=n3a] {Set $N_{0,+1,a}^{\{s,\,\mathrm{ev}\}}$};
\node (n3c) [decision, below=1cm of n3b] {Check $S^\mathrm{net}$ \\ conservation};

\node (n4a) [decision]at (7.3,-11.5) {Conserve $Q^\mathrm{net}$};

\node (n4b) [startstop, below=1cm of n4a] {Sample Remaining Neutrals};

\foreach \i/\j in {n1a/n1b, n1b/n1c, n1c/n2a, n2a/n2b, n2b/n2c, n3a/n3b, n3b/n3c}
    \draw [arrow] (\i) -- (\j);


\draw [arrow] (n2c.east) -- ++(0.6,0) node[above, xshift=0.2cm] {if $true$} -- (n3a.west);

\draw [arrow] (n3c.east) -- ++(0.6,0) node[above, xshift=0.2cm] {if $true$} -- (n4a.west);

\draw [arrow] (n4a) -- ++(0,-1.5) node[right, xshift=0.3cm, yshift=-0.5cm] {if $true$} -- (n4b);
\end{tikzpicture}
\caption{Flow chart for the BSQ sampling algorithm inside the \NuclearConfectionery.}
\label{chart:sampling_BSQ}
\end{figure}
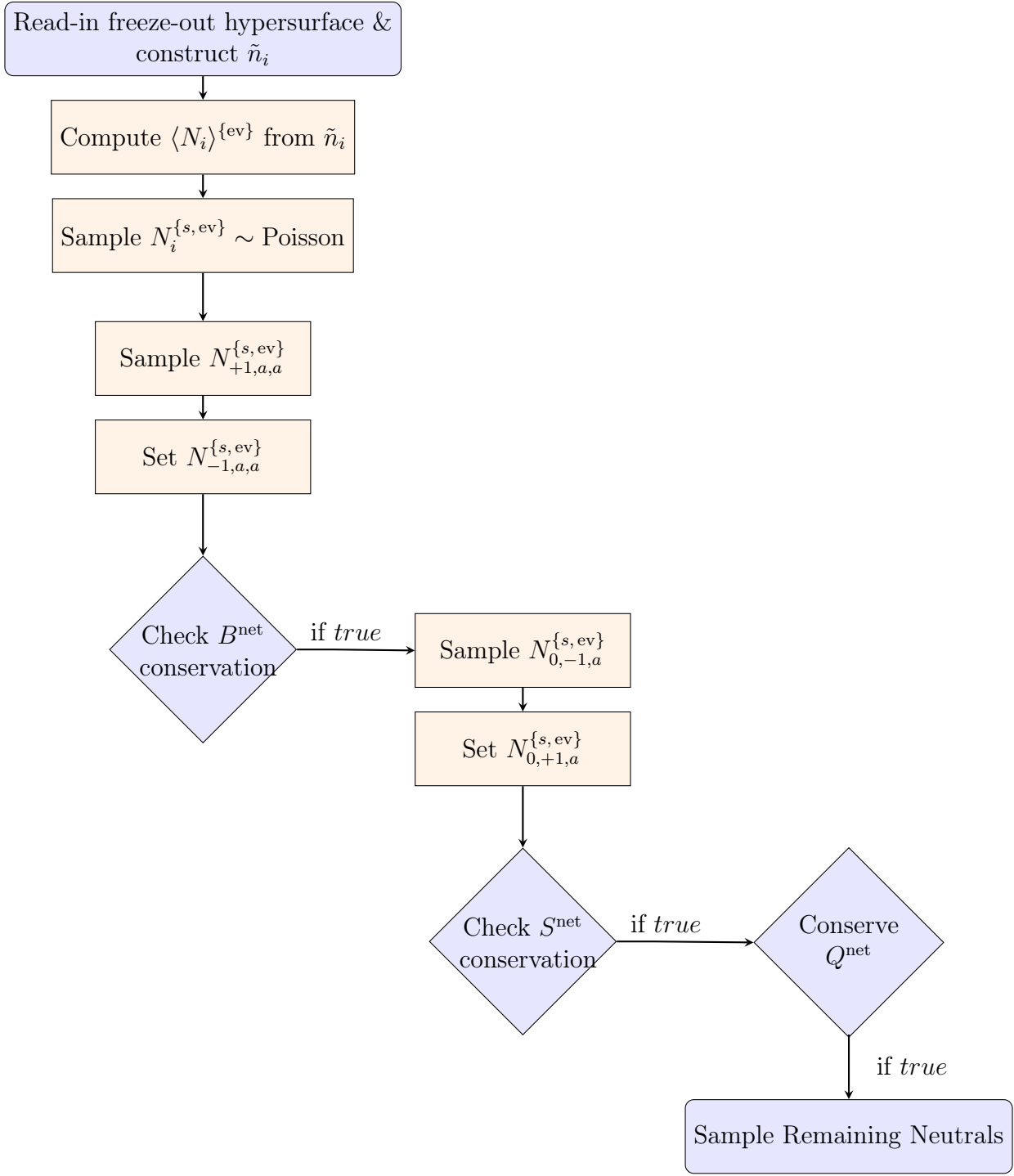

Finally, we must consider electric charge, which is significantly more complicated since we can get (anti)particles that carry anything from $-2,-1,0,+1,+2$ values.
Given that we have sampled all our baryons, we already have knowledge of the following yields:
\begin{eqnarray}\label{eqn:NBQcalc}
    N_{BQ}^{\mathrm{net},\left\{s,\mathrm{ev}\right\}}&\equiv&N_{+1,a,+1}^{\left\{s,\mathrm{ev}\right\}}+N_{-1,a,+1}^{\left\{s,\mathrm{ev}\right\}}-N_{+1,a,-1}^{\left\{s,\mathrm{ev}\right\}}\nonumber\\
    &-&N_{-1,a,-1}^{\left\{s,\mathrm{ev}\right\}}+2N_{+1,a,+2}^{\left\{s,\mathrm{ev}\right\}}-2N_{-1,a,-2}^{\left\{s,\mathrm{ev}\right\}}
\end{eqnarray}
and the strange meson contribution,
\begin{equation}\label{eqn:NSMQcalc}
    N_{SMQ}^{\mathrm{net},\left\{s,\mathrm{ev}\right\}} \equiv N_{0,+1,+1}^{\left\{s,\mathrm{ev}\right\}}-N_{0,-1,-1}^{\left\{s,\mathrm{ev}\right\}}
\end{equation}
from our strangeness sampling procedure. 
Thus, the only remaining unknown contributions to electric charge are from light, charged mesons such that we can write down our net-electric charge contributions as
\begin{eqnarray}
    N_{Q}^{\mathrm{net},\left\{\mathrm{ev}\right\}}&=&  N_{BQ}^{net,\left\{s,\mathrm{ev}\right\}}+N_{SMQ}^{net,\left\{s,\mathrm{ev}\right\}}+N_{0,0,+1}^{\left\{s,\mathrm{ev}\right\}}-N_{0,0,-1}^{\left\{s,\mathrm{ev}\right\}},
\end{eqnarray}
where our only unknowns are the charged light mesons. 
We then arbitrarily choose to sample our positive charged mesons such that  $N_{0,0,+1}^{\left\{s,\mathrm{ev}\right\}}\geq N_{Q}^{net,\left\{\mathrm{ev}\right\}}$.
Then, we calculate our negatively charged mesons,
\begin{equation}\label{eqn:NMnegcalc}
N_{0,0,-1}^{\left\{s,\mathrm{ev}\right\}}=N_{BQ}^{net,\left\{s,\mathrm{ev}\right\}}+N_{SMQ}^{net,\left\{s,\mathrm{ev}\right\}}+N_{0,0,+1}^{\left\{s,\mathrm{ev}\right\}}-N_{Q}^{net,\left\{\mathrm{ev}\right\}}.
\end{equation}
Finally, the only remaining particles are ones that are electrically neutral and either light or have hidden strangeness. The light, neutral particles e.g. $\pi^0$'s, we just sample directly.  
The particles with hidden strangeness like $\phi(s\bar{s})$ are currently sampled directly (just like the $\pi^0$'s), but we plan to study future work where we include finite $\mu_S$ in the Fermi-Dirac distribution. We summarize our notation and how final state observables are obtained in Tab.\ \ref{table:sampling_notation}. Additionally, a flow chart of our algorithm can be see in Fig.~\ref{chart:sampling_BSQ}.

\begin{table}[ht!]
\centering
\begin{tabular}{c|c c}
\hline
{Quantity} & {Description} & {Method of computation}\\
\hline
$\tilde{n}_i$ & Momentum-space particle density & Cooper--Frye 
\\
$\langle N_i \rangle^{\left\{\mathrm{ev}\right\}}$ & Avg. particle yield per event & Integrated over $\tilde{n}_i$ 
\\
$N_i^{\left\{s,\mathrm{ev}\right\}}$ & Sampled yield for species $i$ & Poisson around $\langle N_i \rangle$ 
\\
$N_{B,S,Q}^{\left\{s,\mathrm{ev}\right\}}$ & Yield for specific charge & Summation over sampled particles 
\\
$N_X^{\mathrm{net},\left\{\mathrm{ev}\right\}}$ & Net $B$, $S$, or $Q$ per event & Must be conserved (Input constraint)\\
\hline
\end{tabular}
\caption{A summary of the notation for the quantities used in the sampling algorithm and how each quantity is calculated. }
\label{table:sampling_notation}
\end{table}

\subsubsection{Sampling tests}%
\begin{figure*}
    \centering
    \includegraphics[width=0.9\linewidth]{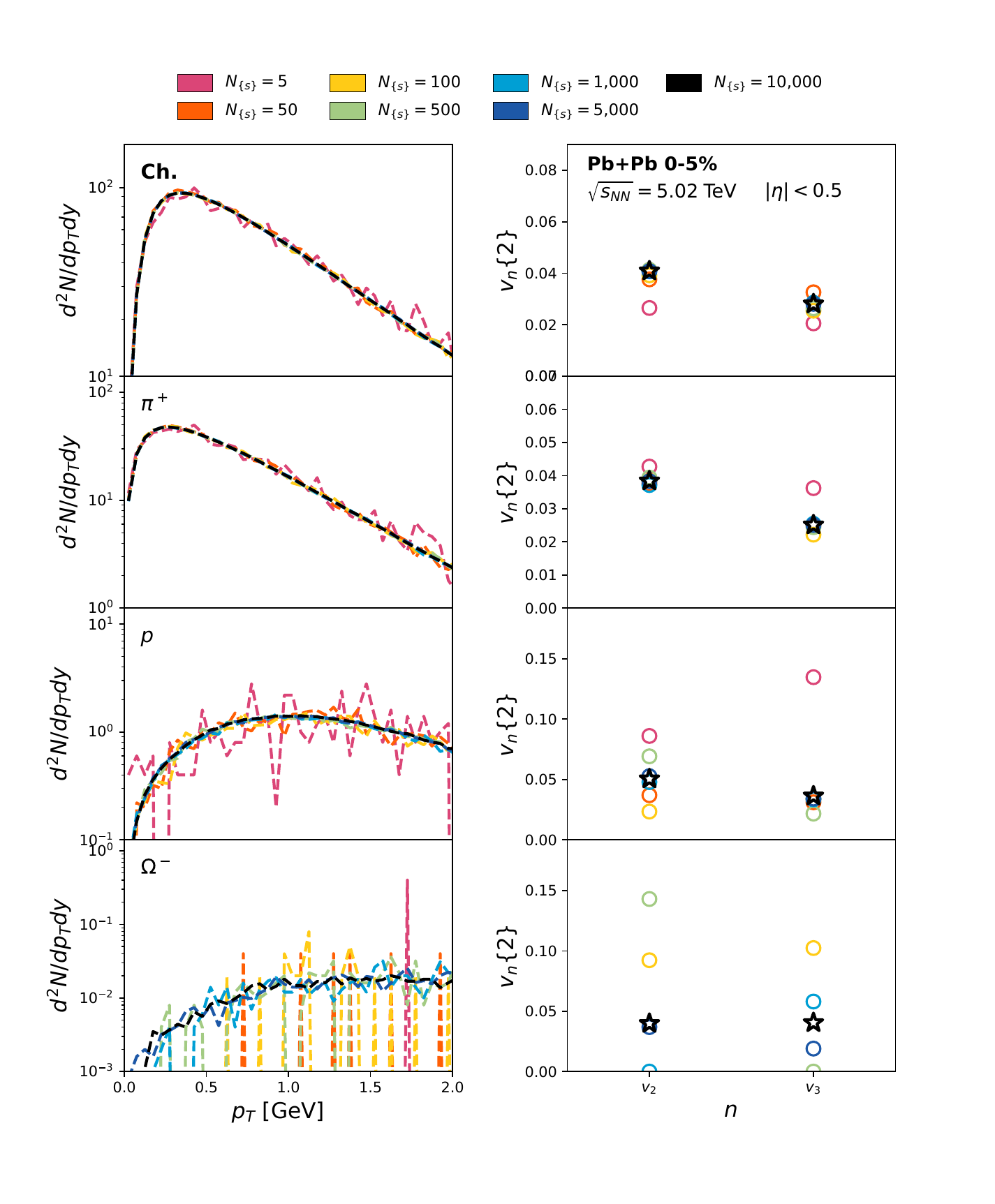}
\caption{Convergence check for a Pb+Pb event, comparing several values of the particle sample number. The convergence is heavily dependent on the particle mass. 
In the rows we compare all charged particles, pions, protons, and Omega baryons. The column compares the particle spectra (left) vs. the integrated particle flow (right) for both elliptical and triangular flow. On the right-hand side the star marker represents the 10,000 samples.
}
    \label{Figure-samplingconvergence}
\end{figure*}

\label{Sec:Particlization:SamplingTests}
\noindent
We developed a sampling subroutine on the hypersurface such that \ccake{} can be connected to \smsh{}. An open question remains as to the number of samples required to accurately reproduce the particle spectra. Here we use the number of samples $N^{\left\{s\right\}}$ as a free parameter to look for convergence, as was done in Fig.~\ref{Figure-samplingconvergence}.

In order to test our numerical sampling method, we take a single event $\left\{\mathrm{ev}\right\}$ in \ccake{}~2.0 for Pb+Pb $\sqrt{s_{NN}}=5.02$ TeV collisions, so \trento{}+\ccake{}. In these simulations, we only run direct decays in \smsh{}. We then keep the hypersurface fixed and resample the hypersurface $N_{\left\{s\right\}}$ times. Every time  a single event $\left\{\mathrm{ev}\right\}$ is sampled, we will indicate it by the index $\left\{s\right\}$. 
A single sample $\left\{s\right\}$ of an event $\left\{\mathrm{ev}\right\}$ produces a large number of particles across a wide range of hadronic species where we sum up all particles for a species $i$ in $N_i^{\left\{s,\mathrm{ev}\right\}}$
\begin{eqnarray}\label{eqn:ntest}
    N_i^{\left\{\mathrm{test},\mathrm{ev}\right\}}&=&\sum_{\left\{s\right\}}^{N_{\left\{s\right\}}} N_i^{\left\{s,\mathrm{ev}\right\}}\\
    &\sim& N_{\left\{s\right\}}\langle N_i\rangle^{\left\{\mathrm{ev}\right\}} \label{eqn:ntest_approx}
\end{eqnarray}
where $i=\pi,K,\Lambda,\dots$, $\left\{\mathrm{ev}\right\}$ indicates a single event, and $\left\{s\right\}$ is a sampled set of particles from that event. 
We define that the number of test particles $N_i^{\left\{\mathrm{test},\,\mathrm{ev}\right\}}$ is the summation of the total number of sampled particles $i$ across all samples of an event $\left\{s\right\}$.
If we have taken enough samples $N_{\left\{s\right\}}$ then  Eq.~(\ref{eqn:ntest_approx}) is a good approximation.

In Fig.~\ref{Figure-samplingconvergence} (left) we compare the differential spectra versus transverse momentum $p_T$ of both all charged particle vs. identified particles, while varying $N_{\left\{s\right\}}$ for each species.  
Here we are looking for an overall convergence such that increasing $N_{\left\{s\right\}}$ does not significantly affect our spectra. 
We find that for all charged particles (top left) we find a very quick convergence with few samples required. Already with $N_{\left\{s\right\}}=50$, we obtain a spectra that is relatively smooth and by $N_{\left\{s\right\}}=500$ it appears that the results have converged such that we see no difference with further samples.  
It is common to use $N_{\left\{s\right\}}\sim 200$, which seems reasonable for all charged particles. 

A large motivation of the development for \ccake{}~2.0 is to study identified particles (PID) such that also want to study the convergence of pions $\pi^+$, protons $p$, and omegas $\Omega^-$. 
Given that pions make up the vast majority of all charged particles, we expect similar results for the $\pi^+$ spectra compared to all charged particles. 
Indeed, in the upper right-hand figure we find that $N_{\left\{s\right\}}=100$ is already quite close to converging for the $\pi^+$ spectra and by $N_{\left\{s\right\}}=500$ we have converged to a smooth result. 

However, heavier particles are significantly suppressed due to their mass such that $\langle N_i\rangle^{\left\{\mathrm{ev}\right\}}$ may be orders of magnitude smaller compared to pions. 
Then, for heavier particles one obtains a significantly smaller $  N_i^{\left\{\mathrm{test},\,\mathrm{ev}\right\}}$ to sample the differential spectra across $p_T$.
For instance, at the Beam Energy Scan the average proton yields per event (so in our language per sample) is on the order of $\langle \langle N_p\rangle\rangle \sim \mathcal{O}(10)$ \cite{STAR:2017sal}. We can compare this to the order of magnitude for the pions that is an order of magnitude larger $\langle \langle N_{\pi}\rangle\rangle\sim\mathcal{O}(100)$. 
With this estimate on hand, we can then make a guess for the sample size that allows for convergence ($ N_{\left\{s\right\}}^\mathrm{con}$) we require, i.e,
\begin{equation}\label{eqn:Ncon}
     N_{\left\{s\right\}}^\mathrm{con}\sim \frac{10^4}{\langle\langle N_i\rangle \rangle}
\end{equation}
where $\langle\langle N_i\rangle \rangle$ is the experimentally obtained average yield for particle $i$ in a given central class and beam energy. 
Since we found that pions required samples that had an order of magnitude of $N_{\left\{s\right\}}\sim \mathcal{O}(100)$, then this implies that a reasonable spectra must have  approximately $N^{\left\{\mathrm{test},\,\mathrm{ev}\right\}}\sim \mathcal{O}(10^4)$.
Thus, given that we produce an order of magnitude fewer protons, we estimate that we will require $N_{\left\{s\right\}}=\mathcal{O}(10^3)$. 
Indeed, when we study the protons we find that $N_{\left\{s\right\}}=1,000$ is quite reasonable for convergence and by $N_{\left\{s\right\}}=5,000$ we can see clear convergence.

We now check the $\Omega^-$ baryons which are even more significantly suppressed because of their mass. 
Very few $\Omega^-$ baryons are produced per event (in fact, it is not guaranteed that every event will produce an $\Omega^-$ baryon) such that their order of magnitude in a given event is $\langle\langle N_{\Omega^-}\rangle\rangle\sim\mathcal{O}(0.1$--$1)$ such that we expect the number of samples that we will require to be $N_{\left\{s\right\}}\sim \mathcal{O}(10^4$--$10^5)$. 
Here we find that even $N_{\left\{s\right\}}=10,000$ that we have not fully converged and one would require more samples to obtain an accurate convergence of the spectra. 
However, our estimate of $N_{\left\{s\right\}}=10,000$ is probably a reasonable lower bound for the number of samples required, being that we appear to be close to obtaining a smooth spectra. 

It is important to not only just to consider the differential spectra but also the integrated flow of the event (specifically elliptical and triangular flow).  In the right column in Fig.~\ref{Figure-samplingconvergence} we perform our convergence test with $N_{\left\{s\right\}}$ for the integrated flow as well. Generally, we find similar values of $N_{\left\{s\right\}}$ for both the spectra and the flow. 

For all charged particles, we find a rather quick convergence over $N_{\left\{s\right\}}$ such that $N_{\left\{s\right\}}\gtrsim 500$ is already a reasonable estimate of the flow (especially for $v_2$). 
A similar result can be found for the pions, which is unsurprising since the all charged particle result is primarily composed of pions. 
However, the integrated flow for protons requires a large number of samples and only taking $N_{\left\{s\right\}}=500$ leads to around a $20\%$ error on the integrated flow. 
Thus, for protons one should consider at least $N_{\left\{s\right\}}\gtrsim 1,000$ samples, which is similar in order of magnitude for what we found for the spectra. 
Finally, we find similar results for the integrated flow of $\Omega^-$ baryons. While the elliptical flow appears to have a near convergence already for $N_{\left\{s\right\}}=5,000$, triangular flow has not yet converged at that point. Thus, for $\Omega^-$ baryons we suggest at least $N_{\left\{s\right\}}=10,000$ samples are required. 

The results here present a specific challenge for low beam energies where PID provides an important insight into the properties of the dense QGP.  
At this point we have only considered single particle measurements but if one wanted to look net particle yields such that one should accurate obtain the results of anti-baryons than we anticipate further challenges because of how suppressed anti-baryons are at low beam energies. 

\begin{table}[ht!]
    \centering
    \begin{tabular}{c|ccc}
    \hline
    & \multicolumn{3}{c}{Estimates for $N_{\left\{s\right\}}^\mathrm{con}$ for $0$--$5\%$ centrality}\\
       PID  &  $\sqrt{s_{NN}}=5.02$ TeV  &  $\sqrt{s_{NN}}=200$--$39$ GeV &  $\sqrt{s_{NN}}=7.7$ GeV\\
       \hline
        ch or $\pi$ & $\mathcal{O}(10)$ & $\mathcal{O}(10^2)$ & $\mathcal{O}(10^2)$ \\
         $p$ or $\Lambda$ & $\mathcal{O}(10^3)$ & $\mathcal{O}(10^3)$ & $\mathcal{O}(10^3)$ \\
         $\bar{p}$ or $\bar{\Lambda}$ & $\mathcal{O}(10^3)$ & $\mathcal{O}(10^3$--$10^4)$ & $\mathcal{O}(10^4$--$10^5)$ \\
         $\Xi$ or $\Omega$ & $\mathcal{O}(10^4)$ & $\mathcal{O}(10^4)$ & $\mathcal{O}(10^4)$ \\
        \hline
    \end{tabular}
    \caption{Estimated number of samples required for convergence for flow and spectra calculations within hydrodynamics for $A$+$A$ collisions based on Eq.~(\ref{eqn:Ncon}) and using hadron yield data from the LHC from the ALICE experiment \cite{ALICE:2016fzo} as well as Beam Energy Scan data from STAR \cite{STAR:2017sal}.
    }
    \label{tab:Est_Ncon}
\end{table}

Using our formula in Eq.~(\ref{eqn:Ncon}) we can refer to experimental data at the LHC and RHIC to make order of magnitude estimates for the number of samples required to converge to reasonable results for the spectra and flow across beam energies. In Tab.\ \ref{tab:Est_Ncon} we show our estimates.  We should note that here we have note studied more differential quantities like fluctuations of net-particles yields (e.g., \cite{STAR:2020tga}), which would like require a much large sample sizes.

\section{Re-scattering (\texorpdfstring{\smsh{}}{SMASH})}%
\label{Sec:Particlization:Rescattering}
\noindent
Following our sampling procedure, we run the hadronic afterburner \smsh{} \cite{SMASH:2016zqf} using the PDG2021+ list from \cite{SanMartin:2023zhv}.
In \cite{SanMartin:2023zhv}, it was found that the hadronic cross sections must be rescaled when more resonances are included and a systematic procedure was developed. 
Within \smsh{} one can run direct decays that includes multi-body decays ($1\rightarrow2$, $1\rightarrow3$, and $1\rightarrow4$) or perform hadronic re-scatterings that include $2\leftrightarrow2$ body interactions. 
For most of this paper, we are only concerned with convergence and time checks such that the hadronic re-scatterings are not relevant to these studies, such that we will primarily focus on direct decays.
That being said, hadronic re-scatterings can have a small but not insignificant effect on a number of experimental observables. Thus, future work that focuses on direct comparisons to experimental data will include hadronic re-scattering as well. 

\section{Conclusions and Outlook}%
\label{Sec:Conclusions}
\noindent
In this work, we have made significant upgrades to the relativistic viscous hydrodynamics code with BSQ conserved charges, \ccake{}, that is solved using the SPH algorithm that is now open-source with the release of this paper. 
The primary changes include the flexibility in dimensions to run in 0+1D, 1+1D, 2+1D, or 3+1D in a generalized coordinate system (Cartesian or Hyperbolic); the inclusion of second-order transport coefficients so one can run with Israel--Stewart, DNMR, or in the future ADNH equations of motion---all three of  which include the BSQ diffusion matrix; enforcing global BSQ conservation at particlization, followed by the hadronic afterburner \smsh{}; the inclusion of source terms that are relevant to jets or dynamical initialization; the ability to calculate causality constraints; and it is rewritten using \kokkos/\cabana{} to allow the flexibility to run on either CPUs or GPUs. 
We have ensured that with all the improvements to \ccake{}~2.0 that it still passes all known (semi)-analytical checks and convergence tests.

After a series of time checks we have determined the optimal methods to run \ccake{}~2.0 on an event-by-event basis. 
There is not a universal answer because it depends on both the dimensionality and $\sqrt{s_{NN}}$ of the system. 
Generally, 2+1D systems work best in serial given that they require fewer SPH particles and already have short runtimes $\mathcal{O}(10^{2}$--$10^{3})$ seconds. 
For 3+1D systems, running in parallel/GPUs can aid run times due to the significant increase in SPH particles.  
In parallel using 2-4 cores appear to be optimal. 
Overall, we have runtimes of $\mathcal{O}(0.1)$ CPU hours for 2+1D systems and $\mathcal{O}(1)$ CPU hours for 3+1D systems (without any parallelization). 
The way we initialize the system can also play a role in runtime. A run with low $\sqrt{s_{NN}}$ in 3+1D is slightly faster using Cartesian coordinates. 
Otherwise, at higher $\sqrt{s_{NN}}$ we find that hyperbolic coordinates are quicker.

Here we benchmark the number of samples required to reproduce spectra for identified particles and flow observables, leading to an important benchmark to compare against for sampling the production of hadrons. 
While $\mathcal{O}(10^1$--$10^2)$ samples of the freeze-out hypersurface in a given event shows convergence well for all-charged particles, we find that identified particles (especially for baryons) require significantly more samples (approximately $\mathcal{O}(10^3$--$10^4)$ for baryons and up to $\mathcal{O}(10^5)$ for antibaryons at low $\sqrt{s_{NN}}$) to accurately reproduce their spectra. 
This finding implies that accurate calculations of PID observables at low $\sqrt{s_{NN}}$ require a significant number of samples per event. 

Here we have demonstrated future use cases for \ccake{}~2.0 such as jets fully coupled to the medium.  Additionally, we demonstrate that we obtain a reasonable nuclear modification factor, reproducing previous results. 
Furthermore, sources terms were also used for dynamical initializations that make \ccake{}~2.0 ready to run a beam energy scan to compare to heavy-ion collision data. 
Obvious future extensions include methods to include critical fluctuations and explorations that search for the QCD critical point. 

\section*{Acknowledgments}%
\noindent
The authors would like to thank Renan Hirayama for his technical assistance with \smsh{}. 
This research was supported by the US-DOE Nuclear Science Grant No. DE-SC0023861,  within the framework of the Saturated Glue (SURGE) Topical Theory Collaboration, and by the National Science Foundation (NSF) within the framework of the MUSES collaboration, under grant number OAC-2103680 and partially by from Fundação de
Amparo à Pesquisa do Estado de São Paulo (grants 2020/15893-4, 2024/08903-4 and 2018/24720-6). F.G. was supported by CNPq (Conselho Nacional de Desenvolvimento Científico) through  307806/2025-1. We also acknowledge support from the Illinois Campus Cluster, a computing resource that is operated by the Illinois Campus Cluster Program (ICCP) in conjunction with the National Center for Supercomputing Applications (NCSA), which is supported by funds from the University of Illinois at Urbana-Champaign.










\clearpage
\bibliographystyle{elsarticle-num} 
\bibliography{inspire,NOTinspire}

\newpage

\appendix
\section{Modified EoS to permit lower \texorpdfstring{$\varepsilon_{\mathrm{min}}$}{εₘᵢₙ}}
\label{NewEoS}

\subsection{Modifying the conformal-diagonal EoS}
\noindent
The conformal-diagonal EoS is defined as
\begin{equation}\label{eqn:pressureConDiag}
        p_\mathrm{cd}(T, \vec{\mu} )
        = A_0 T_0^4 \left[ \left( \frac{T}{T_0} \right)^4
             +\sum_{q=B,S,Q} \left( \frac{\mu_q}{\mu_{q,0}} \right)^4\right] ,
    \end{equation}
This EoS guarantees the existence of an exact solution to the inversion problem as long as the following condition is satisfied:
\begin{equation}
    \varepsilon_0 \geq \varepsilon_\mathrm{min}(\vec{n}_0)
    \equiv \frac{3}{4\cdot 2^{2/3} (A_0 T_0^4)^{1/3}} \sum_{q=B,S,Q}\left( \mu_{q,0} \left| n_{q,0} \right| \right)^{4/3},
\end{equation}
where
\begin{align}
    A_0 &\equiv p_{T,0}/T_{\mathrm{scale}}^4, \nonumber \\
    T_0 &\equiv 1 \text{ fm}^{-1}, \label{eq:cd_parameter_constraints} \\
    \mu_{q,0} &\equiv T_0 \mu_{q,{\max}}\l(\frac{A_0}{p_{X,\max} - p_{T,0}}\r)^{1/4}, \nonumber \\
    T_{\mathrm{scale}} &\equiv 1.1\, T_\mathrm{FO}\,,
\end{align}
where $T_\mathrm{FO}$ is fixed by the point where the $\varepsilon = \varepsilon_\mathrm{FO}$ hypersurface intersects the $T$ axis in the phase diagram (at $\vec{\mu} = \vec{0}$) and
\begin{align*}
    p_{T,0} &\equiv p_\mathrm{table}(T_{\mathrm{scale}}, \vec{0}), \\
    p_{B,\max} &\equiv p_\mathrm{table}(T_{\mathrm{scale}}, \mu_{B,{\max}}, 0, 0), \\
    p_{S,\max} &\equiv p_\mathrm{table}(T_{\mathrm{scale}}, 0, \mu_{S,{\max}}, 0), \label{EoS_constraints_definitions} \\
    p_{Q,\max} &\equiv p_\mathrm{table}(T_{\mathrm{scale}}, 0, 0, \mu_{Q,{\max}}),
\end{align*}
The problem may occasionally arise that this inequality is not satisfied for a given combination of input densities.  One way to handle this situation is simply to decrease the values of the $\mu_{q,0}$ until the $\varepsilon_\mathrm{min}(\vec{n}_0)$ can be presumed sufficiently small to cover all practical scenarios of interest.

On the other hand, one may want to fix the $\mu_{q,0}$ scales independently as was done above, in which case there is no additional flexibility for reducing $\varepsilon_\mathrm{min}$.  In this case, we can introduce a new, more flexible EoS (also conformal) with the following form:

\begin{equation}\label{eqn:pressureModifiedConformal}
        p_\mathrm{mc}(T, \vec{\mu} )
        = A_0 T_0^4 \left[ \left( \frac{T}{T_0} \right)^4
             + \sum_{q=B,S,Q} \left( \frac{\mu_q}{\mu_{q,0}} \right)^4
             + \sum_{q=B,S,Q} {\mathcal{c}}_q \left( \frac{T}{T_0} \right)^2\left( \frac{\mu_q}{\mu_{q,0}} \right)^2 \right] ,
    \end{equation}
    where ${\mathcal{c}}_q>0$.  The other thermodynamic quantities (e.g., densities and susceptibilities) are found immediately:
    \begin{align}
        s(T, \vec{\mu} ) &= 2 A_0 T \l( 2 T^2 + \sum_{q=B,S,Q} {\mathcal{c}}_q \l(\frac{\mu_q}{\mu_{q,0}}\r)^2 \r), \\
        n_q(T, \vec{\mu} ) &= \frac{2 A_0 \mu_q}{\mu_{q,0}^4}\l( 2\mu_q^2 +  {\mathcal{c}}_q T^2 \mu_{q,0}^2\r), \\
        \chi_{TT} &= 2A_0\l( 6 T^2 + {\mathcal{c}}_q \l(\frac{\mu_q}{\mu_{q,0}}\r)^2 \r), \\
        \chi_{Tq} &= \frac{4 A_0 {\mathcal{c}}_X T \mu_X}{\mu_{q,0}^2}, \\
        \chi_{qq'} &= 
        \begin{cases}
            \frac{2A_0}{\mu_{q,0}^4} \l( 6 \mu_q^2 + {\mathcal{c}}_q T^2 \mu_{q,0}^2 \r) & q = q' \\
            0 & q \neq q'.
        \end{cases}
    \end{align}
    One checks easily that $\varepsilon = -p + s T + \sum_{q}n_q \mu_q = 3p$, as must hold for any conformal EoS.  Note also that the new EoS reduces to the conformal-diagonal EoS if we set all ${\mathcal{c}_q} \to 0$.

\subsection{Computing the minimum energy density at finite charge density}
\noindent
    The problem we need to solve is again $\varepsilon(T,\vec\mu) = \varepsilon_0$, $\vec{n}(T,\vec\mu) = \vec{n}_0$.  Introducing the variables $\mu_q = \chi_q T$, these conditions are written as
    \begin{align}
        \chi_q \l( 2 \chi_q^2 + {\mathcal{c}}_q \beta_q^2 \r) &= \frac{\beta_q^4}{2 A_0} \frac{n_{q,0}}{T^3}, \\
         1 + \sum_{q=B,S,Q} \l( \frac{\chi_q^4}{\beta_q^4} + {\mathcal{c}}_q \frac{\chi_q^2}{\beta_q^2} \r) &= \frac{\varepsilon_0}{3 A_0 T^4},
    \end{align}
    where $\beta_q \equiv \mu_{q,0}/T_0$.  The density conditions are all cubic equations of the general form $\chi_q^3 + \frac{1}{2} {\mathcal{c}}_q \beta_q^2 \chi_q - \frac{\beta_q^4}{4 A_0} \frac{n_{q,0}}{T^3} = 0$.  Using standard results about the solutions of the cubic equation, one can show from the fact that
    \[{\mathcal{c}}_q^3 + \frac{27 \beta_X^2 n_{q,0}^2}{8 A_0^2 T^6} > 0\]
    that there is only one real solution for $\chi_q$.  This solution can be written in a closed form which depends on $T$:
    \begin{align}
        \chi_q(T) &= \frac{2^{1/3}\delta_q^{2/3}(T) - 2\cdot3^{1/3}\alpha_{q,1}}{6^{2/3} \delta_q^{1/3}(T)},
    \end{align}
    where
    \begin{align}
        \alpha_{q,1} &\equiv \frac{1}{2} {\mathcal{c}}_q \beta_q^2, \quad \alpha_{q,2}(T) \equiv \frac{\beta_q^4}{4 A_0} \frac{n_{q,0}}{T^3},\nonumber \\
        \delta_X(T) &\equiv \sqrt{12\alpha_{q,1}^3+81 \alpha_{q,2}^2(T)} + 9 \alpha_{q,2}(T).
    \end{align}
    Note that $\alpha_{q,1} > 0$ and therefore $\delta_q > 0$ as long as ${\mathcal{c}}_q > 0$.
    The remaining equation for $T$ is then obtained by substituting back the $\chi_X$ solutions into the energy condition.  This latter condition cannot be solved for $T$ in closed form, so it must be solved numerically instead.  The energy condition becomes
    \begin{align}
        1 + \sum_{q=B,S,Q} \l({\mathcal{c}}_q \frac{\chi_q^2(T)}{\beta_q^2} + \frac{\chi_q^4(T)}{\beta_q^4} \r) = \frac{\varepsilon_0}{3 A_0 T^4}.
    \end{align}
    The minimum energy density $\varepsilon_\mathrm{min}(\vec{n}_0)$ occurs when $T \to 0$, corresponding to
    \[\varepsilon_\mathrm{min}(\vec{n}_0) = 3A_0 \sum_{q=B,S,Q} \l( \frac{\mu_{q,0} |n_{q,0}|}{4 A_0 T_0} \r)^{4/3},\]
    which is the same condition as was found above for the conformal-diagonal EoS.  The minimum energy may therefore be adjusted by modifying the values of $T_0$ and the $\mu_{q,0}$ as before.

\subsection{Fixing parameters of the modified EoS}
\noindent
    Finally, in order to reduce the value of $\varepsilon_\mathrm{min}$ in practice, we rescale the values of the $\mu_{q,0}$ above to
    \begin{align}
    \mu_{q,0} &\equiv \frac{T_0 \mu_{q,{\max}}}{f} \l(\frac{A_0}{p_{q,\max} - p_{T,0}}\r)^{1/4}, \nonumber
    \end{align}
    where $f \geq 1$ is the resulting factor by which $\varepsilon_\mathrm{min}$ is to be reduced.  Note that there is no reason \textit{a priori} that all the $\mu_{q,0}$ should be scaled by the same factor $f$, and the question of whether it may be advantageous to scale them by different factors is deferred to future work.

    We fix the remaining parameters in the new EoS by matching to the tabulated EoS at $T_{\mathrm{scale}}$ and $\mu_q = 0$:
    \begin{align}
        A_0 &\equiv p_{T,0}/T_{\mathrm{scale}}^4, \nonumber \\
        T_0 &\equiv 1 \text{ fm}^{-1}, \label{eq:mcd_parameter_constraints} \\
        {\mathcal{c}}_q &\equiv \frac{\mu_{q,0}^2}{2 A_0 f^2 T_{\mathrm{scale}}^2} \chi_{qq,\max} \,,\nonumber
    \end{align}
    where $\chi_{qq,\max} \equiv \chi_{qq}( T, \vec{0} )$.  The original conformal-diagonal EoS corresponds instead to choosing $f=1$ and ${\mathcal{c}}_q = 0$.
    
 \end{document}